\documentclass[a4paper,11pt]{article}

\usepackage{jheppub}
\usepackage[T1]{fontenc}
\usepackage[utf8]{inputenc}

\usepackage{tikz}
\usetikzlibrary{math}
\usetikzlibrary{calc}
\usetikzlibrary{trees}
\usetikzlibrary{decorations.pathmorphing}
\usetikzlibrary{decorations.markings}
\usetikzlibrary{external}

\usepackage{dsfont, bm}

\newcommand{\beq}{\begin{equation}}
\newcommand{\eeq}{\end{equation}}

\newcommand{\cN}{{\mathcal N}}
\newcommand{\cO}{{\mathcal O}}
\newcommand{\cM}{{\mathcal M}}

\def\deltaQ{\;\delta^{(16)} (Q)}
\def\spa#1.#2{\left\langle#1\,#2\right\rangle}
\def\spb#1.#2{\left[#1\,#2\right]}

\def\sqrtmQSq{\sqrt{-q^2}}
\newcommand{\vect}{\bm}
\def\relf{\sigma}
\def\relfbar{y}

\def\dlog{\mathrm{d}\!\log}

\def\II{\mathrm{II}}
\def\X{\mathrm{X}}

\def\RT{\mathrm{III}}
\def\xRT{\overline{\RT}}
\def\IX{\mathrm{IX}}
\def\xIX{\overline{\IX}}
\def\XI{\mathrm{XI}}
\def\xXI{\overline{\XI}}
\def\H{\mathrm{H}}
\def\xH{\overline{\H}}
\def\IYtop{\mathrm{IY}}
\def\YItop{\overline{\IYtop}}
\def\IYbot#1{\mathrm {I\raisebox{\depth}{\scalebox{-#1}{Y}}}}
\def\YIbot#1{\overline{\IYbot{#1}}}

\newcommand{\mushroomtop}[1]{
  \begin{tikzpicture}[#1]
  \coordinate (t1) at (0ex,0ex);
  \coordinate (t2) at (0.5ex,0ex);
  \coordinate (b1) at (0ex,-0.5ex);
  \coordinate (b2) at (0.5ex,-0.5ex);
  \draw[line width=0.25mm] (0.75ex,0)arc(0:180:0.5ex);
  \draw[line width=0.1mm] ($(t1) + (-0.35ex,0)$) -- ($ (t2) + (0.35ex,0) $);
  \draw[line width=0.1mm] ($(b1) + (-0.25ex,0)$) -- ($ (b2) + (0.25ex,0) $);
\draw[line width=0.25mm] (t1) -- (b1);
\draw[line width=0.25mm] (t2) -- (b2);
\end{tikzpicture}
}

\newcommand{\mushroombot}[1]{
  \begin{tikzpicture}[#1]
  \coordinate (t1) at (0ex,0ex);
  \coordinate (t2) at (0.5ex,0ex);
  \coordinate (b1) at (0ex,-0.5ex);
  \coordinate (b2) at (0.5ex,-0.5ex);
  \draw[line width=0.25mm] (0.75ex,-0.5ex)arc(0:-180:0.5ex);
  \draw[line width=0.1mm] ($(t1) + (-0.25ex,0)$) -- ($ (t2) + (0.25ex,0) $);
  \draw[line width=0.1mm] ($(b1) + (-0.35ex,0)$) -- ($ (b2) + (0.35ex,0) $);
\draw[line width=0.25mm] (t1) -- (b1);
\draw[line width=0.25mm] (t2) -- (b2);
\end{tikzpicture}
}

\clearpage{}

\usepackage{bbold}

\usepackage{graphicx,epsf}
\usepackage{amsmath,amsfonts,amssymb,amsbsy}
\usepackage{mathtools}

\usepackage{xcolor}

\usepackage{multirow}
\usepackage{stackrel}

\usepackage{placeins}
\usepackage{slashed} 

\usepackage{booktabs}
\usepackage{adjustbox}

\usepackage{subcaption}

\def\oneloop{{1 \mbox{-} \rm loop}}
\def\twoloop{{2 \mbox{-} \rm loop}}
\def\tree{{\rm tree}}

\newcommand{\EulerGamma}{\gamma_\mathrm{E}}
\renewcommand{\imath}{\mathrm{i}}

\newcommand{\pot}{(p)}

\usepackage{slashed}
\newcommand{\cutindex}{\slashed}

\begin{document}
\title{Extremal black hole scattering at $\cO(G^3)$:\\ graviton dominance, eikonal exponentiation, \\and differential equations}

\author[1]{Julio~Parra-Martinez,}
\affiliation[1]{Mani L. Bhaumik Institute for Theoretical Physics,\\
	UCLA Department of Physics and Astronomy,\\
	Los Angeles, CA 90095, USA}
\emailAdd{jparra@physics.ucla.edu}

\author[2]{Michael S. Ruf,}
\emailAdd{michael.ruf@physik.uni-freiburg.de}
\affiliation[2]{Physikalisches Institut, Albert-Ludwigs Universit\"at Freiburg, \\
	D-79104 Freiburg, Germany}

\author[3]{Mao Zeng}
\affiliation[3]{Institut f\"ur Theoretische Physik, Eidgen\"ossische Technische Hochschule Z\"urich, \\
Wolfgang-Pauli-Strasse 27, 8093 Z\"urich, Switzerland}
\emailAdd{mzeng@phys.ethz.ch}

\abstract{
  We use $\mathcal N=8$ supergravity as a toy model for understanding the dynamics of black hole binary systems via the scattering amplitudes approach. We compute the conservative part of the classical scattering angle of two extremal (half-BPS) black holes with minimal charge misalignment at $\cO(G^3)$ using the eikonal approximation and effective field theory, finding agreement between both methods.
  We construct the massive loop integrands by Kaluza-Klein reduction of the known $D$-dimensional massless integrands.
  To carry out integration we formulate a novel method for calculating the post-Minkowskian expansion with exact velocity dependence, by solving velocity differential equations for the Feynman integrals subject to modified boundary conditions that isolate conservative contributions from the potential region. 
 Motivated by a recent result for universality in massless scattering, we compare the scattering angle to the result found by Bern et.\ al.\ in Einstein gravity and find that they coincide in the high-energy limit, suggesting graviton dominance at this order.
}

\preprint{FR-PHENO-2020-007, UCLA/TEP/2020/103}

\maketitle

\section{Introduction}
The direct detection of gravitational waves at LIGO/VIRGO \cite{Abbott:2016blz, TheLIGOScientific:2017qsa} has started an exciting new age of gravitational wave astronomy. Scattering amplitudes have emerged as the latest tool in computing the gravitational dynamics of binary systems in the perturbative regime. In contrast to the traditional post-Newtonian expansion, which is a simultaneous expansion in the Newton constant, $G$, and a relative velocity, $v$,  relativistic scattering amplitudes can naturally lead to results up to a fixed order in $G$ but to all orders in velocity, known as the post-Minkowskian (PM) expansion \cite{Bertotti:1956pxu,Kerr:1959zlt,Bertotti:1960wuq,Portilla:1979xx, Westpfahl:1979gu,Portilla:1980uz,Bel:1981be,Westpfahl:1985tsl, Ledvinka:2008tk, Bjerrum-Bohr:2013bxa, Bjerrum-Bohr:2018xdl, Damour:2016gwp, Damour:2017zjx, Cheung:2018wkq, Kosower:2018adc, Maybee:2019jus, Bern:2019nnu, Bern:2019crd, Antonelli:2019ytb, KoemansCollado:2019lnh, Cristofoli:2020uzm}. A recent highlight is the result of Bern et al. \cite{Bern:2019nnu,Bern:2019crd} for the conservative dynamics of black hole binary systems at $\cO(G^3)$, i.e.\ the third-post-Minkowskian (3PM) order. The result points to many interesting questions, some of which are explored in the present paper.
\begin{enumerate} \item The scattering angle for massive particles in
    Refs.~\cite{Bern:2019nnu,Bern:2019crd} contains a term that diverges in the
    high-energy limit.  Ref.~\cite{Bern:2020gjj} recently found that the high-energy 
    limit of massless scattering is universal at $\cO(G^3)$, i.e., it is independent of the number
    of supersymmetries and dominated by graviton exchange. Does the
    aforementioned term in the massive scattering angle appear in the presence of supersymmetry, and does it
    exhibit universality in the high-energy limit?
  \item The computation of the
    scattering angle in Ref.~\cite{Bern:2019nnu,Bern:2019crd} proceeds by first
    extracting a classical potential using a non-relativistic effective field
    theory (EFT) \cite{Cheung:2018wkq}, then calculating the scattering
    trajectory by solving the classical equations of motion. However, there is
    a well-known alternative method: the eikonal approximation 
    \cite{Glauber, Levy:1969cr, Soldate:1986mk, tHooft:1987vrq, Amati:1987wq, Amati:1987uf, Muzinich:1987in,
    Amati:1990xe, Kabat:1992tb, Amati:2007ak, Laenen:2008gt,
    Melville:2013qca, Akhoury:2013yua, Divecchia:2019myk, DiVecchia:2019kta, Kulaxizi:2019tkd,
    KoemansCollado:2019ggb, Bern:2020gjj, Abreu:2020lyk, Cristofoli:2020uzm, Bern:2020buy}, 
    which calculates the classical scattering angle from suitable Fourier
transforms of quantum scattering amplitudes. Do the two methods give
    equivalent results at $\cO(G^3)$ for the scattering of massive scalar
    particles?
    \item Refs.~\cite{Bern:2019nnu,Bern:2019crd} resum the velocity
    dependence of the $\cO(G^3)$ result by first calculating the velocity expansion
    to the 7th-post-Newtonian order, i.e.\ $\cO(G^3 v^{10})$ around the static
    limit, then fitting the series to an ansatz, which is shown to be unique.
    Can we instead directly obtain exact velocity dependence, as is common in
    the calculation of relativistic scattering amplitudes?
\end{enumerate}
The answers to the above three questions are all \emph{yes}, as we will show
using a calculation of two-loop, i.e.\ $\mathcal O(G^3)$, scattering of
extremal black holes in $\mathcal N=8$ supergravity \cite{Cremmer:1978km,Cremmer:1978ds,Cremmer:1979up}.
The last question about exact velocity dependence is especially of current interest due to two reasons.
First, a direct calculation without a series expansion to high orders can be computationally more efficient.
Second, Ref.~\cite{Damour:2019lcq} raised questions about the velocity resummation of
Refs.~\cite{Bern:2019nnu,Bern:2019crd} in the case of Einstein gravity. Since
then, the correctness of the latter result has been verified at high orders in
the velocity expansion \cite{Blumlein:2020znm, Bini:2020wpo}, an alternative
method for resummation of the velocity series has been used with identical results
\cite{Blumlein:2019bqq}, and the unitarity cut construction of the loop integrand
has been checked against direct Feynman diagram computations
\cite{Cheung:2020gyp}. Still, a direct relativistic calculation that bypasses
velocity resummation will be a valuable additional confirmation of the result,
and will provide a way to streamline future calculations at $\cO(G^3)$ and
beyond.

The study of classical gravitational scattering in $\mathcal N=8$ supergravity
was initiated in a beautiful paper by Caron-Huot and Zahraee
\cite{Caron-Huot:2018ape}, which we build upon. The large set of symmetries of
this theory provides important simplifications, which make it the perfect
theoretical laboratory to study various conceptual questions and test the
technology to be applied at higher orders in perturbation theory.  This is
familiar to how precision QCD practitioners have often sharpened their
axes with simpler calculations in $\mathcal{N}=4$ super-Yang--Mills theory before honing in on the beast. A
lot is known about  $\mathcal N=8$ supergravity, in particular the complete loop
integrands for the quantum four-point amplitude are available through five
loops
\cite{Green:1982sw,Bern:1998ug,Bern:2007hh,Bern:2008pv,Bern:2009kd,Bern:2012uf,Bern:2017ucb,Bern:2018jmv,Edison:2019ovj}.
These were constructed using the unitarity method
\cite{Bern:1994zx,Bern:1994cg,Bern:1997sc,Bern:2004cz,Britto:2004nc} and the
different incarnations of the double copy
\cite{Kawai:1985xq,Bern:2008qj,Bern:2010ue,Bern:2017yxu}.  These results, being
valid in $D$-dimensions, can be used to easily obtain massive integrands via
Kaluza-Klein reduction, as we will do in this paper.\footnote{See also \cite{KoemansCollado:2019lnh} for a recent application of
KK reduction in the context of the eikonal approximation.}

Moving on to integration, we will obtain the part of the amplitude relevant for
classical conservative dynamics using the method of regions
\cite{Beneke:1997zp}. In particular, integration in the ``soft region''
produces the correct small momentum transfer expansion of the amplitude
\cite{Neill:2013wsa, Cristofoli:2020uzm}, up to contact terms that are irrelevant for long-range classical physics at any order in $G$.
However, conservative classical dynamics actually arises from the ``potential
region'' which is a sub-region contained in the soft region \cite{Cheung:2018wkq}.
Strictly speaking, the potential region is defined in the near-static limit and
produces an expansion of the Feynman integrals as a series in small
velocity. But since the velocity series can be resummed to all orders,
the resummed result will be also referred to as the amplitude evaluated in the potential region. In addition to isolating conservative effects, evaluating in the potential region also simplifies the integrals considerably.

Refs.~\cite{Bern:2019nnu, Bern:2019crd} exploits the fact that infrared (IR)
divergences cancel when matching the EFT against full theory, and circumvents
the evaluation of IR divergent integrals. In this paper, all IR divergent
integrals will be evaluated explicitly in dimensional regularization (which
serves as both UV and IR regulators). This will allows us to check against the
predictions from eikonal exponentiation, which expresses the divergent amplitude in an exponentiated form. Additionally, we will evaluate all
integrals relativistically with full dependence on velocity, without
constructing and resumming a velocity series. This is made possible by
employing the method of differential equations for Feynman integrals
\cite{Kotikov:1990kg,Bern:1993kr,Remiddi:1997ny,Gehrmann:1999as}, with a crucial new ingredient being the use of modified
boundary conditions that isolate the contributions from the potential region.
While Ref.~\cite{Bern:2019crd} already presented a precursor of our
differential equations method as an alternative to the ``expansion-resummation
method'', it was only successfully applied to a subset of the needed integrals
that do not involve infrared divergences due to ``iteration'' of graviton
exchanges. This paper will use a finer control of boundary conditions to
evaluate all integrals using differential equations. We also perform soft
expansions prior to the construction of differential equations, resulting in
dramatic speedups in computation. Another improvement is that we
transform the differential equations into Henn's canonical form
\cite{Henn:2013pwa, Henn:2014qga}. In this form, the differential equations have a simple analytic structure, and can be easily solved to higher orders in the $\epsilon$ expansion. (See also \cite{Ablinger:2018zwz} for advances in automated solution of generic univariate differential equations that are solvable by iterated integrals.)

In the context of $\cN=8$ supergravity, Ref.~\cite{Caron-Huot:2018ape} put
forward a tantalizing conjecture: that the energy levels of a pair of black
holes in such theory retain hydrogen-like degeneracies to all orders in
perturbation theory. This is tantamount to the classical black hole binary orbits
being integrable and showing no precession. Two pieces of evidence were
provided in support of this conjecture: first, the absence of precession for
the full $\cO(G^2)$ dynamics, which directly follows from an analog of the
``no-triangle'' hypothesis
\cite{Bern:2005bb,BjerrumBohr:2005xx,BjerrumBohr:2006yw,Bern:2006kd,Bern:2007xj,BjerrumBohr:2008vc} for massive scattering;
and second, various all-orders-in-$G$ calculations in the probe limit for different
charge configurations. It is known that $\cO(G^3)$ (or any odd power of $G$)
corrections to the conservative dynamics cannot yield precession \cite{Kalin:2019rwq,Kalin:2019inp}. Instead we will use the
scattering angle at $\cO(G^3)$ to test this conjecture, and see that it
deviates from the integrable Newtonian result at this order.  We will extract the scattering
angle both from appropriate derivatives of the ``eikonal
phase'' and via the EFT techniques of Refs.~\cite{Bern:2019nnu,
Bern:2019crd}, finding agreement between both methods.

Although we perform our calculations in $\cN=8$ supergravity, we expect the techniques here developed to be directly applicable to Einstein gravity as well. Such application is beyond the scope of the present paper and we leave it for future work.

This paper is organized as follows: In Section \ref{sec:setup} we setup our conventions, we review some basic features of extremal black holes in $\cN =8$ supergravity, and discuss the different limits that will be used in the paper. In Section \ref{sec:integrand} we construct the tree-level four-point amplitude, as well as the one- and two-loop massive integrands from the known massless integrands via Kaluza-Klein reduction and truncation to the appropriate sector. In Section~\ref{sec:int} we briefly discuss the integration regions involved in our problem, and introduce our new integration method based on differential equations, which is applied to calculate the full one- and two-loop amplitudes in the potential region. In Section~\ref{sec:amplitudes} we assemble the scattering amplitudes. In Section~\ref{sec:eikonal} we review the eikonal method, and use it to calculate the order $G^{n \leq 3}$ eikonal phase, while checking exponentiation. Then we use the eikonal phase to calculate the gravitational scattering angle and we compare its high-energy limit with the result of Refs.~\cite{Bern:2019nnu, Bern:2019crd} in Einstein gravity. In Section~\ref{sec:eft} we cross check our results via the EFT method of Ref.~\cite{Cheung:2018wkq}, and we comment on the advantages of this approach. We calculate the conservative Hamiltonian by matching, and find the scattering angle by solving the classical equations of motion. In Section~\ref{sec:conclusions} we present our conclusions. We include two appendices: Appendix~\ref{sec:3dintegrals} contains some technical details about the computation of integrals in the near-static limit and Appendix~\ref{sec:twoloopdetails} collects the solution to our two-loop differential equations.
The results are provided in computer-readable format in several ancillary files (see comments at the beginning of each file for detailed descriptions).

 \section{Kinematics and setup}
\label{sec:setup}
We model the dynamics of two half-BPS black holes in $\cN = 8$ supergravity \cite{Andrianopoli:1997wi,Arcioni:1998mn} by considering the scattering of two massive point particles in half-BPS multiplets, which interact via the massless supergravity multiplet. We use an all-outgoing convention for the external momenta $p_i$, and the masses of the particles are
\begin{equation}
  p_1^2 = p_4^2 = m_1^2\,, \qquad p_2^2 = p_3^2 = m_2^2\,.
\end{equation}
We will parametrize the scattering amplitudes in terms of the usual  invariants $s=(p_1+p_2)^2$, $t=(p_1+p_4)^2= q^2$ and $u=(p_1+p_3)^2$, where we introduced the four-momentum transfer $q = p_1+p_4$ for later convenience.  As is common in the study of scattering amplitudes we will cross the incoming particles to the final state, so that all particles are outgoing. The physical scattering configuration corresponds to the region $s>(m_1+m_2)^2, t=q^2<0$ and $u<0$.\footnote{We use a mostly minus metric.}

The half-BPS multiplet in $\cN=8$ supergravity contains massive states with spin $0\leq S \leq 2$. In this work we will focus on particular scalar components, $\phi$ and $\bar\phi$, with $S=0$ and leave the study of spinning states for later work. The interactions between different half-BPS particles are mediated by the massless supergraviton multiplet. In addition to gravitons the $\cN = 8$ supergraviton multiplet contains 28 (vector) graviphotons, $A_{IJ}$, and 70 scalars $\phi_{IJKL}$, as well as fermions which will not be important for our discussion. Black holes in $\cN =8$ supergravity interact with the graviphotons and scalars with charges $C_{IJ}$ given by an $8\times8$ matrix. Here $I,J,\ldots$ are $\mathrm{SU}(8)$ $R$-symmetry indices and the vectors a scalars are in  $\mathrm{SU}(8)$ representations of the appropriate dimension. We will not print the Lagrangian here, because it is lengthy. For our purposes, however, all scattering amplitudes could be built from the three-particle amplitudes:
\begin{align}
  M^\tree_{3}(1_{\vphantom{\bar\phi}\phi},2_{\bar\phi},3_{h}) &=  16\pi G \, ( \varepsilon_3 \cdot p_1 )^2\,, \\
  M^\tree_{3}(1_{\vphantom{\bar\phi}\phi},2_{\bar\phi},3_{A_{IJ}}) &=  8\pi G \sqrt{2} \, ( \varepsilon_3 \cdot p_1 ) \, C_{IJ}\,, \\
  M^\tree_{3}(1_{\vphantom{\bar\phi}\phi},2_{\bar\phi},3_{\phi_{IJKL}}) &= 16 \pi G \,(C_{IJ}\,C_{KL} - C_{IK}\,C_{JL} + C_{IL}\,C_{JK}) \,,
\end{align}
using factorization and unitarity, as done in Ref.~\cite{Caron-Huot:2018ape}. Here $\varepsilon$ are polarization vectors. 

In general the charges, $C_{IJ}$ are complex and the black holes are dyonic. The charges are also central charges of the supersymmetry algebra, and the BPS condition requires their magnitude to be equal to the mass
\beq
C^{IK} C_{KJ} = m^2\,\delta^I{}_J \,.
\eeq
When studying a pair of black holes we need only consider the relative phases in their BPS charges. These are parameterized by three angles along which the charges might be misaligned
\beq
C_1 = m_1  \begin{pmatrix}	    0        & 1_{4\times4} \\
                              -1_{4\times4}  &      0 
			    \end{pmatrix}\,, \qquad 
C_1 = m_2 \begin{pmatrix}    0   &  \Phi \\
                          -\Phi  &    0 
			\end{pmatrix} \,,
\eeq
with 
$\Phi = \text{diag}(e^{\imath\phi_1},e^{\imath\phi_2},e^{\imath\phi_3},e^{\imath\phi_4})$ and  $\sum_i \phi_i = 0$.
For the two- and one-angle cases, however, there always exist a duality frame where the magnetic charges are zero. 
We point the reader to  Ref.~\cite{Caron-Huot:2018ape} for a more detailed discussion of the charges.

Although we will construct the full (quantum) loop integrands for the scattering amplitudes of these black holes, we are ultimately interested in their classical conservative dynamics. In the classical limit of hyperbolic scattering, the orbital angular momentum of the black hole binary system is much larger than $\hbar$. Thus, the \emph{classical} limit of the scattering amplitudes simply corresponds to the large angular momentum limit $J\gg 1$ (in natural units), which establishes a hierarchy of scales
\begin{equation}
s,|u|, m_1^2,m_2^2 \sim J^2 |t| \gg  |t| = |q|^2\,.
\label{eq:scalehierarchy}
\end{equation}
As a result, we are interested in calculating scattering amplitudes in the limit of small $q$, or more precisely as an expansion in small $q$. From a heuristic calculation in the Newtonian limit, the leading-order scattering angle $\theta$ is of the order $Gm/(v r) \sim Gmq / v$, where $m$ and $r$  are the total mass and relative transverse distance of the system. So for generic values of $v$, the quantity $Gmq$ is of order $\theta$,
and for each additional order of $G$, we need to expand the amplitude up to one additional power of $q$ to obtain corrections to the scattering angle of order $\theta^L$, where $L$ is the loop order. Terms that are more subleading in $q$ at the same power of $G$ are quantum corrections that vanish classically. In summary, at $\cO(G^n)$, we only need to expand the scattering amplitude of massive particles up to $\cO(|q|^{n-2})$ in the small-$q$ expansion \cite{Neill:2013wsa}, in order to extract the classical dynamics. In practice this will imply, among other things, that when we calculate an amplitude some loop integrals can be discarded before any calculation, if they are beyond the classical order.

Furthermore, we will only be interested in the conservative dynamics, so we will restrict the components of the momentum transfer $q=(q^0,\vect q)$ to scale as
\begin{equation}
  |\vect q| \gg q^0\,,
\end{equation}
so that the graviton multiplet mediates instantaneous long-range interactions. Note that the latter expansion involves an additional small parameter, a velocity $|\vect v| = q^0/|\vect q| \ll 1$, on top of the classical limit $J \gg1$. We will refer to this expansion as the \emph{near-static} limit, and we delay a more detailed discussion to Section \ref{sec:int}. 

Finally, in comparing our results to Einstein gravity, it will be useful to take the \emph{high-energy} or ultra-relativistic limit in which the black holes are highly boosted. This simply makes the hierarchy of scales in Eq.~\eqref{eq:scalehierarchy} more strict
\begin{equation}
s,|u| \gg m_1^2,m_2^2 \sim J^2 |t| \gg  |t|\,.
\label{eq:scalehierarchyHE}
\end{equation}
In this context, it will be useful to introduce the variable
\begin{equation}\label{eq:LorentzFactor}
  \sigma = \cosh\eta = \frac{s-m_1^2-m_2^2}{2m_1m_2} = \frac{p_1\cdot p_2}{m_1m_2}\,,
\end{equation}
which is simply the relativistic factor of particle 1 in the rest-frame of particle 2 (or vice versa).
In terms of this variable the high-energy limit simply corresponds to taking $\sigma \gg 1$. Note that in our setup it is important that we take the classical limit first, before taking the high-energy limit, so that $J\gg \sigma$. This is equivalent to having the hierarchy of scales in Eq.~\eqref{eq:scalehierarchyHE}. The opposite limit, $J\ll \sigma$, is closely connected to the regime of massless high-energy scattering considered in Ref.~\cite{Bern:2020gjj}.

In summary, we will be interested in the three limits
  \begin{align}
    \text{Generic classical limit:}   &\qquad J\gg 1\,,\\
    \text{near-static classical limit:}   &\qquad J\gg 1, \quad |\vect v|\ll 1\,,\\
    \text{high-energy classical limit:} &\qquad J\gg 1 \quad \text{then} \quad \sigma \gg 1\,,
  \end{align}
  expressed here in terms of their corresponding dimensionless expansion parameters. 
 \section{Integrands from Kaluza-Klein reduction}
\label{sec:integrand}

In this section we construct the tree amplitude and loop integrands for the scattering of the two black holes via Kaluza-Klein (KK) reduction. Ref.~\cite{Caron-Huot:2018ape} studied the case of three-angle misalignment in the BPS charges. While such case is the most rich and interesting, we will focus on the single-angle misalignment case, which is the one we can access via KK  reduction from the existing integrands. Let us explain this in more detail: we consider Type IIA supergravity in $D=10$ and perform KK reduction on a six-torus of radius $R$. When dimensionally reducing the massless integrand we will identify the massive black holes with KK gravitons, with  ten-dimensional momenta $k_i$ and masses arising from the components of momenta outside of $D=4$. The supersymmetry algebra in higher dimensions, upon reduction, identifies the extra-dimensional momenta as BPS charges (see e.g. Appendix B of Ref.~\cite{Caron-Huot:2018ape}). There is only one relative angle between the extra dimensional momenta, so the dimensional reduction only provides the result for one-angle misalignment. Because of this, one might perform a rotation to set the momenta along all but two directions to zero and effectively reduce from $D=6$. Henceforth, for simplicity, we shall then pretend we are reducing from six dimensions. Then we can write the momenta of the four particles as
\beq
k_1 = \begin{pmatrix} p_1 \\ 0 \\ m_1 \end{pmatrix}\,, \qquad
k_2 = \begin{pmatrix} p_2 \\ m_2 \sin\phi \\ m_2 \cos\phi \end{pmatrix}\,, \qquad
k_3 = \begin{pmatrix} p_3 \\ -m_2 \sin\phi \\ -m_2 \cos\phi \end{pmatrix}\,, \qquad
k_4 = \begin{pmatrix} p_4 \\ 0 \\ -m_1 \end{pmatrix} \,.
\label{eq:6dto4d}
\eeq
The compactness of the extra dimensions requires the extra dimensional momenta to be discrete and of order $\sim R^{-1}$. We will choose the masses $m_1$ and $m_2$ to correspond to the two lightest KK modes, $\phi_1, \phi_2$. Depending on the momentum in the extra dimensions the massless six-dimensional scalar, $\phi$, will reduce to either of these.

We will see momentarily that the massless integrands for maximal supergravity, have two simplifying features which imply that we just need a few basic rules to perform the KK reduction. First, the loop integrands are proportional to the tree amplitude to all orders. This follows from the supersymmetry Ward identity \cite{Bern:1998ug}, and implies that the polarization dependence is trivial and factors out of the integrand. Second, through two loops, the integrands are composed only of scalar loop integrals, so we only need to understand how to KK reduce propagators. 

Let us first discuss the rules for reducing the massless loop integrand. The integration over loop momentum reduces as
\beq
\mathrm{d}^6\ell \longrightarrow \frac{1}{(2\pi R)^2}\sum_{\ell^4,\ell^5\in 2\pi R \mathbb{Z}} \mathrm{d}^4\ell\,,
\eeq
where the factors of $(2\pi R)^2$ simply relate the $D=6$ and $D=4$ Newton's constant $G=G^{6D} (2\pi R)^{-2}$, and the sum is over all possible ways to assign a KK momentum to each leg in a given diagram, subject to the constraint of momentum conservation in the extra dimensions. Since we are choosing our external legs to be two particular KK modes, this means in practice that we should sum over all the ways the external massive particles could route inside the diagram. We are interested in the diagrams that feature the exchange of massless particle in the graviton multiplet, so we will truncate the full massive integrand to this sector. We delay a discussion about the consistency of this truncation to the end of this section. The truncation to massless exchange, together with momentum conservation imposes an additional rule when routing the external particles through the diagram, namely that lines corresponding to different KK modes cannot cross at a three-point vertex. 

\begin{figure}[tb]
  \centering
	\begin{subfigure}[b]{0.3\textwidth}
		\centering
		\includegraphics{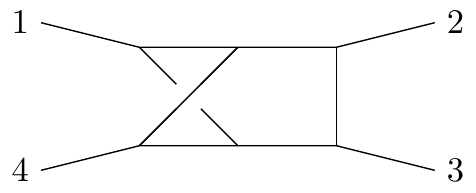}
		\caption{}
	\end{subfigure}
        \hfill
	\begin{subfigure}[b]{0.3\textwidth}
		\centering
		\includegraphics{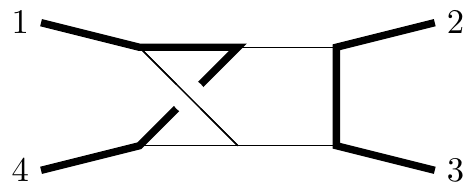}
		\caption{}
	\end{subfigure}
        \hfill
	\begin{subfigure}[b]{0.3\textwidth}
		\centering
		\includegraphics{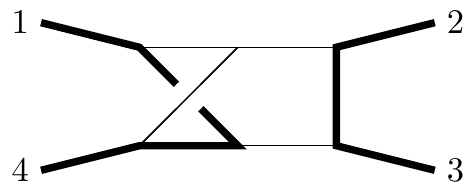}
		\caption{}
	\end{subfigure}
	\caption{Example of KK reduction with massless exchange. The diagram in (a) reduces to the pair of diagrams in (b) and (c). The thin and  thick lines denote massless and massive momenta respectively.}
  \label{fig:exampleKKred}
\end{figure}

As an example, consider the massless non-planar double-box integral in Fig.~\ref{fig:exampleKKred}(a). It is easy to see that there are two alternative ways to route the mass/extra-dimensional momenta through the diagram, shown in Fig.~\ref{fig:exampleKKred}(b) and (c). So this massless integral will yield two contributions to the massive integrand. In contrast, there are also examples in which there there is no way to route the masses. We will find several of these when constructing the two loop integrand.

Finally, using the identifications in Eq.~\eqref{eq:6dto4d} we find that the reduction of the external invariants is given by the following simple replacement rules 
\beq
  s \rightarrow s - |m_1+m_2e^{\imath\phi}|^2\,, \qquad
  t \rightarrow t\,, \qquad 
  u \rightarrow u - |m_1-m_2e^{\imath\phi}|^2\,,
\eeq
and similarly for loop momenta
\beq
(\ell+k_i)^2 \rightarrow (\ell+p_i)^2 -m_i^2\,,
\eeq
which follows from the orthogonality of the four-dimensional loop momentum  and the extra-dimensional components of the external momentum.

\subsection{Tree level amplitude}
As a warmup let's start with the tree level amplitude. We will write it as
\beq
M_4^{\tree}(1,2,3,4) = 8\pi G^{6D}  \frac{\mathcal{K}}{stu}\,,
\eeq
where $\mathcal{K}$ is the four point matrix element of the supersymmetric $t_8t_8R^4$ operator (see e.g. Ref.~\cite{Green:1987mn}, Eq.~(9.A.18)). In four dimensions $\mathcal{K}=\spb3.4^4/\spa1.2^4 \deltaQ$, where $Q$ is the on-shell super-momentum \cite{Nair:1988bq}.
For simplicity we choose the incoming and outgoing states to be complex conjugate scalars $\phi$ and  $\bar\phi$ in the graviton multiplet. The corresponding component of $\mathcal{K}$ is simply $s^4$ and the $D$-dimensional scalar amplitude is
\beq
M_4^{\tree}(1_{\vphantom{\bar\phi}\phi},2_{\vphantom{\bar\phi}\phi},3_{\bar\phi},4_{\bar\phi}) =  8\pi G^{6D}\, \frac{s^3}{tu} \,.
\eeq
Using our rules for dimensional reduction we find the result
\beq
M_{4}^{\tree}(1_{\vphantom{\bar\phi_1}\phi_1},2_{\vphantom{\bar\phi_2}\phi_2},3_{\bar\phi_2},4_{\bar\phi_1}) =  8\pi G \frac{(s-|m_1+m_2e^{\imath\phi}|^2)^3}{t(u - |m_1-m_2e^{\imath\phi}|^2)} \,.
\label{eq:treeres}
\eeq
Although this is the full amplitude we want to restrict to the massless exchange sector. We can partial fraction \eqref{eq:treeres} as 
\beq
M_{4}^{\tree}(1_{\vphantom{\bar\phi_1}\phi_1},2_{\vphantom{\bar\phi_2}\phi_2},3_{\bar\phi_2},4_{\bar\phi_1}) =  8\pi G \frac{(s-|m_1+m_2e^{\imath\phi}|^2)^2}{-t} + \text{massive exchange}\,,
\label{eq:treeresmasslessexchange}
\eeq
which using $s-|m_1+m_2e^{\imath\phi}|^2 = 2m_1m_2 (\cosh\eta - \cos\phi)$, where $\eta$ is the relative rapidity, $\eta= \text{arccosh}(\relf)$, agrees with Eq.~(3.18) of Ref.~\cite{Caron-Huot:2018ape}, restricted to the one-angle case.

\subsection{One-loop integrand}

\begin{figure}
	\centering
	\begin{subfigure}[b]{0.4\textwidth}
\centering
		\includegraphics{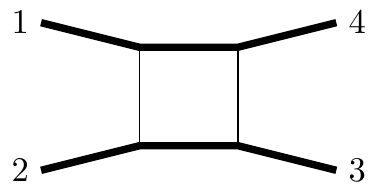}
		\caption{Box: $I_{\rm II}$}
	\end{subfigure}
	\begin{subfigure}[b]{0.4\textwidth}
\centering
		\includegraphics{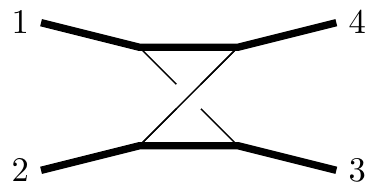}
		\caption{Non-planar box: $I_{\rm X}$}
	\end{subfigure}
	\caption{One-loop topologies.} \label{fig:oneloop}
\end{figure}

The one-loop massless integrand was constructed long ago in Refs.~\cite{Green:1982sw,Bern:1998ug},
\begin{align}
  M_4^{\oneloop}(1,2,3,4) = -\imath 8\pi G^{6D} stu M_4^{\rm tree}(1,2,3,4) \, \big(\, I_{1234}^{(1)} + I_{1342}^{(1)} + I_{1423}^{(1)} \,)\,,
\end{align}
where all the integrals are one-loop boxes with the specified ordering of the external legs.
Using the reduction rules described above
\beq
stuM_{4}^{\rm tree}(1_{\vphantom{\bar\phi}\phi},2_{\vphantom{\bar\phi}\phi},3_{\bar\phi},4_{\bar\phi})  \rightarrow   8\pi G (s-|m_1+m_2e^{\imath\phi}|^2)^4\,,
\eeq
and KK reduction maps the massless integrals to massive integrals as follows,
\beq
  I_{1234}^{(1)}  \rightarrow I_{\rm II} \,, \qquad
  I_{1342}^{(1)}  \rightarrow 0\,, \qquad
  I_{1423}^{(1)}  \rightarrow I_{\rm X}\,.
\eeq
Where the integrals are shown in Fig.~\ref{fig:oneloop}, and $0$ indicates that there is no way to route the momenta so the reduction yields zero. Putting all together we find the one-loop integrand
\beq
M_{4}^{\oneloop}(1_{\vphantom{\bar\phi_1}\phi_1},2_{\vphantom{\bar\phi_2}\phi_2},3_{\bar\phi_2},4_{\bar\phi_1}) = -\imath(8\pi G)^2\,(s-|m_1+m_2e^{\imath\phi}|^2)^4\, \big(\, I_{\rm II} + I_{\rm X}\,)\,,
  \label{eq:1lintegrand}
\eeq
where we have truncated to the massless exchange sector.
This matches the result in Eqs.~(3.33) and~(3.34) of Ref.~\cite{Caron-Huot:2018ape}.

\subsection{Two-loop integrand}
\begin{figure}[bt]
  \centering
	\begin{subfigure}[b]{0.4\textwidth}
\centering
		\includegraphics{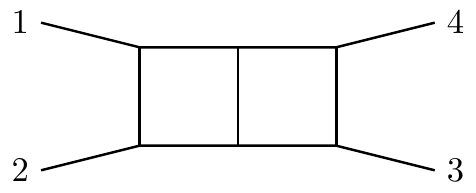}
		\caption{Double box: $I^{(2)\, \rm P}_{1234}$}
	\end{subfigure}
	\begin{subfigure}[b]{0.4\textwidth}
\centering
		\includegraphics{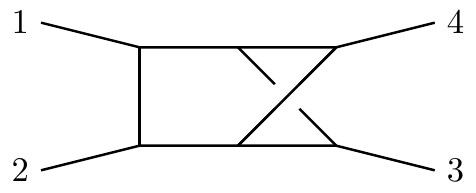}
		\caption{Non-planar double-box: $I^{(2)\, \rm NP}_{1234}$}
	\end{subfigure}
  \caption{Massless two-loop topologies}
  \label{fig:twoloopmassless}
\end{figure}

\begin{figure}
	\centering
	\begin{subfigure}[b]{0.3\textwidth}
\centering
		\includegraphics{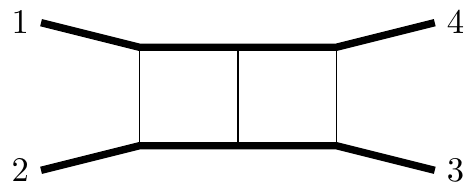}
		\caption{$\RT$}
	\end{subfigure}
	\hfill
	\begin{subfigure}[b]{0.3\textwidth}
\centering
		\includegraphics{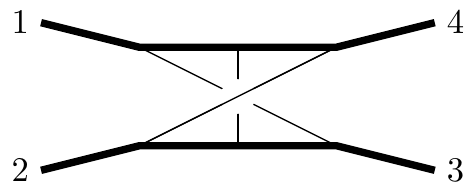}
		\caption{$\xRT$}
	\end{subfigure}
	\hfill
        \begin{subfigure}[b]{0.3\textwidth}
\centering
		\includegraphics{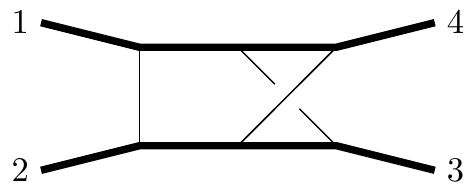}
		\caption{$\IX$}
	\end{subfigure}\\
	\begin{subfigure}[b]{0.3\textwidth}
\centering
		\includegraphics{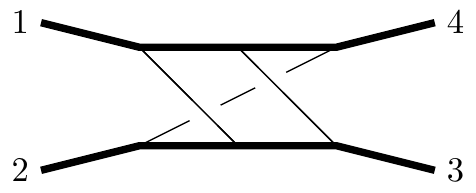}
		\caption{$\xIX$}
	\end{subfigure}
	\hfill
	\begin{subfigure}[b]{0.3\textwidth}
\centering
		\includegraphics{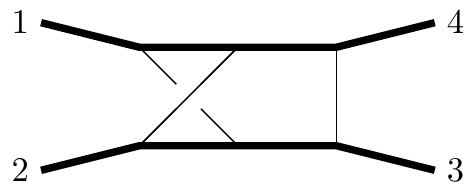}
		\caption{$\XI$}
	\end{subfigure}
	\hfill
	\begin{subfigure}[b]{0.3\textwidth}
\centering
		\includegraphics{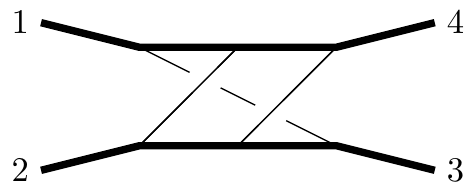}
		\caption{$\xXI$}
	\end{subfigure}
	\caption{Two loop integrals that are of the double-box type.}\label{fig:twoloophor}
\end{figure}

\begin{figure}
	\centering
	\begin{subfigure}[b]{0.3\textwidth}
\centering
		\includegraphics{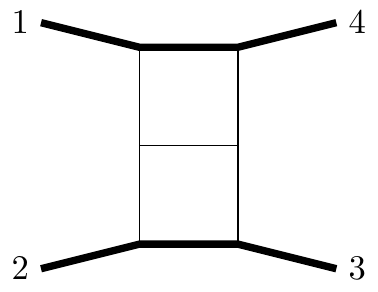}
		\caption{$\H$}
	\end{subfigure}
	\begin{subfigure}[b]{0.3\textwidth}
\centering
		\includegraphics{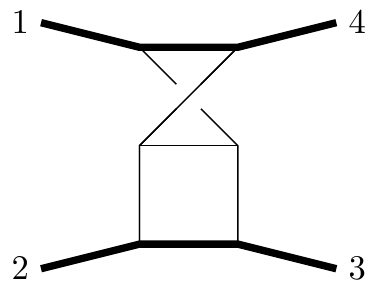}
		\caption{$\xH$}
	\end{subfigure}
	\begin{subfigure}[b]{0.3\textwidth}
\centering
		\includegraphics{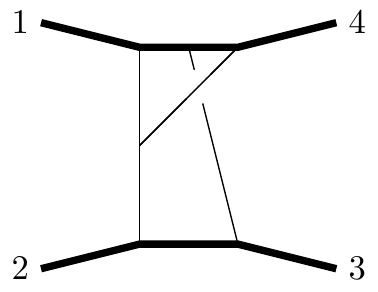}
		\caption{$\YItop$}
	\end{subfigure}\\
	\begin{subfigure}[b]{0.3\textwidth}
\centering
		\includegraphics{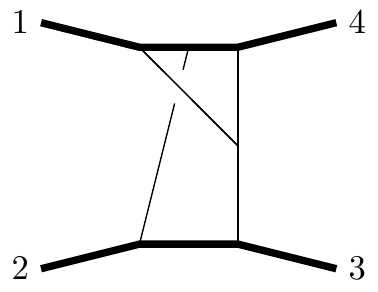}
		\caption{$\IYtop$}
	\end{subfigure}
	\begin{subfigure}[b]{0.3\textwidth}
\centering
		\includegraphics{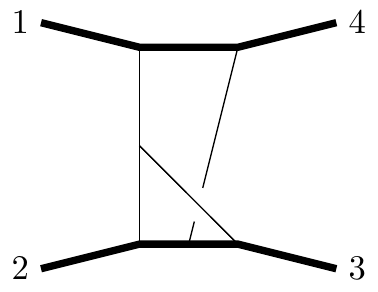}
		\caption{$\YIbot1$}
	\end{subfigure}
	\begin{subfigure}[b]{0.3\textwidth}
\centering
		\includegraphics{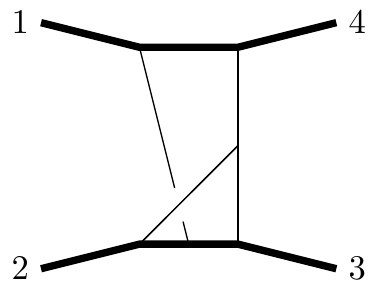}
		\caption{$\IYbot1$}
	\end{subfigure}
	\caption{Two loop integrals that are not of the double-box type.}\label{fig:twoloopvert}
\end{figure}

\begin{figure}
	\centering
	\raisebox{2pt}{\begin{subfigure}[b]{0.225\textwidth}
\centering
		\includegraphics{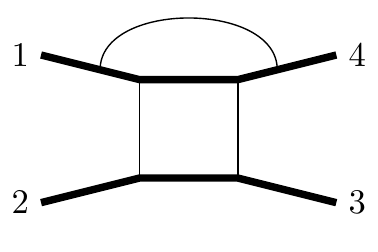}
		\caption{$\mushroomtop{scale=1.5}$}
	    \end{subfigure}} 
	\hfill
	\begin{subfigure}[b]{0.225\textwidth}
\centering
		\includegraphics{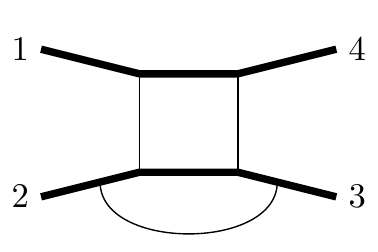}
		\caption{$\mushroombot{scale=1.5}$}
	\end{subfigure}
	\hfill
	\raisebox{2pt}{\begin{subfigure}[b]{0.225\textwidth}
\centering
		\includegraphics{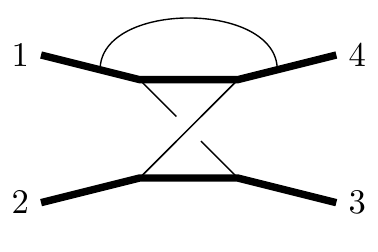}
		\caption{$\overline{\mushroomtop{scale=1.5}}$}
	    \end{subfigure}}
	    \hfill
	\begin{subfigure}[b]{0.225\textwidth}
\centering
		\includegraphics{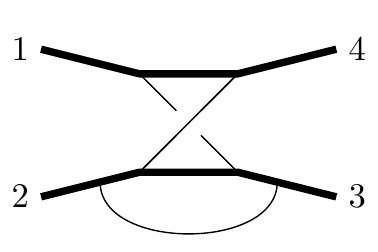}
		\caption{$\overline{\mushroombot{scale=1.5}}$}
	\end{subfigure}
	\caption{Two-loop integrals that include a self-interaction.}\label{fig:twoloopself}
\end{figure}

The massless two loop integrand was constructed in Ref.~\cite{Bern:1998ug} using the unitarity method. It takes the remarkably simple form
\begin{align}
\begin{aligned}
  M_4^{\twoloop}(1,2,3,4) &= -(8\pi G^{6D})^2 stu M_4^{\rm tree}(1,2,3,4) \\ & \times \big( s^2\,I_{1234}^{(2)\,\rm P}  + s^2\,I_{3421}^{(2)\,\rm P} + s^2\,I_{1234}^{(2)\,\rm NP}  + s^2\,I_{3421}^{(2)\,\rm NP}  + \text{cyclic} \big)\,,
\end{aligned}
\end{align}
where `` $+$ cyclic''  means adding the two other cyclic permutations of $(2,3,4)$ and the integrals, which are all scalar, are shown in Fig.~\ref{fig:twoloopmassless}. It is easy to find how the integrals map under the dimensional reduction. The planar integrals reduce as follows
\begin{align}
\begin{aligned}
  I_{1234}^{(2)\,\rm P} & \rightarrow I_{\rm III} \,, \qquad 
  & I_{1342}^{(2)\,\rm P} & \rightarrow 0 \,, \qquad                  
  & I_{1423}^{(2)\,\rm P} & \rightarrow I_{\overline{\rm H}} + I_{\overline{\mushroomtop{}}} +  I_{\overline{\mushroombot{}}}\,, \\
  I_{3421}^{(2)\,\rm P} & \rightarrow 0 \,, \qquad               
  & I_{4231}^{(2)\,\rm P} & \rightarrow I_{\overline{\rm III}} \,, \qquad
  & I_{2341}^{(2)\,\rm P} & \rightarrow I_{\rm H} + I_{\mushroomtop{}} +  I_{\mushroombot{}}\,,
\end{aligned}
\end{align}
where $I_{\RT}$ is the double-box integral in Fig.~\ref{fig:twoloophor}(a), $I_{\H}$ is the $\H$ integral in Fig.~\ref{fig:twoloopvert}(a), $I_{\mushroomtop{}},I_{\mushroombot{}}$ are the self-energy diagrams in Fig.~\ref{fig:twoloopself}(a-b), and the integrals with a bar denote their crossed versions, obtained by exchanging $p_2\leftrightarrow -p_3$, which are also shown in the same figures. It is interesting to note that the $\H$ and self-energy diagrams come from the dimensional reduction of the same massless diagrams.  The non-planar integrals reduce as follows
\begin{align}
\begin{aligned}
   I_{1234}^{(2)\,\rm NP} & \rightarrow I_{\XI} \,, \qquad
 & I_{1342}^{(2)\,\rm NP} & \rightarrow I_{\xXI} \,, \qquad
 & I_{1423}^{(2)\,\rm NP} & \rightarrow I_{\vphantom{\YIbot{0.73}}\IYbot{0.73}} + I_{\YIbot{0.73}}\,, \\
   I_{3421}^{(2)\,\rm NP} & \rightarrow I_{\IX} \,, \qquad
   & I_{4231}^{(2)\,\rm NP} & \rightarrow I_{\xIX} \,, \qquad
   & I_{2341}^{(2)\,\rm NP} & \rightarrow I_{\vphantom{\YItop}\IYtop} + I_{\YItop}\,,
\end{aligned}
\end{align}
where we will refer to $I_{\IX}$ and $I_{\XI}$ as non-planar double-boxes, and the rest of the integrals are shown in Figs.~\ref{fig:twoloophor} and~\ref{fig:twoloopvert}.
The KK reduced two-loop integrand is then given by
\begin{align} 
  &M_{4}^{\twoloop}(1_{\vphantom{\bar\phi_1}\phi_1},2_{\vphantom{\bar\phi_2}\phi_2},3_{\bar\phi_2},4_{\bar\phi_1}) = (8\pi G)^3 (s-|m_1+m_2e^{\imath\phi}|^2)^4\, \nonumber \\
  & \hspace{0.5cm}  \times \bigg[ (s-|m_1+m_2e^{\imath\phi}|^2)^2(I_{\RT} +I_{\XI} + I_{\IX}) + (u-|m_1-m_2e^{\imath\phi}|^2)^2(I_{\xRT} +I_{\xXI} + I_{\xIX})  \nonumber\\
  & \hspace{1.25cm} + t^2 (I_{\rm H} + I_{\mushroomtop{}} +  I_{\mushroombot{}} + I_{\IYtop} + I_{\IYbot{0.73}} + I_{\xH} + I_{\overline{\mushroomtop{}}} +  I_{\overline{\mushroombot{}}} + I_{\YItop} + I_{\YIbot{0.73}})\, \bigg]\,. 
\label{eq:twoloopintegrand}
\end{align}

\subsection{Comments on the consistency of the integrands}
Finally, let us make some brief comments about the consistency of the integrands we have constructed in this section. We have focused on the sector of the theory where KK modes with masses $m_1, m_2$ of order $R^{-1}$ exchange massless particles. This is not a consistent truncation, however, since there is no parametric separation between the masses of the KK modes which are all of order $R^{-1}$.\footnote{We thank Chia-Hsien Shen for discussions related to this point.} This manifests itself in various ways. For instance, the tree amplitude in Eq.~\ref{eq:treeres}, features the exchange of massive particle of mass $\sim m_1-m_2$, which is required by crossing symmetry. Similarly, at loop-level, it is known that the sum of the box and crossed-box integrals contains a mass singularity (see e.g.~\cite{Beenakker:1988jr}). Consequently, the amplitude in Eq.~\eqref{eq:1lintegrand} has collinear divergences in violation of the theorem of Ref.~\cite{Akhoury:2011kq} which precludes them in quantum gravity.\footnote{This stands in contrast to Einstein gravity, whose quantum one-loop amplitude was shown in Ref.~\cite{Bern:2019crd} to lack collinear divergences} All of these issues are manifestations of the well known fact that there is no consistent quantum theory of a finite number of massive particles coupled to maximal supergravity. In attempting to fix these problems, one is bound to discover the tower of KK modes, which arise from a consistent massless theory in higher dimensions. In spite of these comments, our truncated theory has a well-defined classical Lagrangian and is a useful toy model to explore the questions we are interested in in this paper, so we will ignore all of these issues henceforth.

 \section{Integration via velocity differential equations}
\label{sec:int}
In the previous section we have constructed the full quantum integrand for the four-point amplitude through two loops.
In this section we will calculate the integrals using the method of regions~\cite{Beneke:1997zp, Smirnov:2004ym} to extract the contributions which are relevant for the conservative dynamics.
After briefly reviewing the various regions involved the problem, we introduce a new method to calculate the integrals in the potential region, using single-scale fully relativistic differential equations with modified boundary conditions.
We illustrate the method using several examples at one and two loops.

\subsection{Regions and power counting}

Following the discussion in Ref.~\cite{Bern:2019crd}, we consider an internal graviton line with four-momentum $\ell = (\omega, \vect \ell)$, whose components can scale as
\begin{align}
{\rm hard}: \quad (\omega, \vect \ell) &\sim (m,m) \,,\\
{\rm soft}: \quad (\omega, \vect \ell) &\sim  (|\vect q|,|\vect q|) \sim J^{-1}\,(m |\vect v|,m |\vect v|) \,, \\
{\rm potential}: \quad  (\omega, \vect \ell) &\sim (|\vect q| |\vect v|, |\vect q|) \sim J^{-1}\,(m |\vect v|^2,m |\vect v|) \,, \\
{\rm radiation} : \quad  (\omega, \vect \ell) &\sim (|\vect q| |\vect v|, |\vect q| |\vect v|) \sim J^{-1}\,(m |\vect v|^2, m |\vect v|^2)\,,
  \label{eq:modes}
\end{align}
where we take as reference scale $m = m_1 + m_2$, and each scaling defines a \emph{region}.
Note that the four different regions are defined using two small parameters $|\vect{q}|$ (or $J^{-1}$) and the velocity
$|\vect v|$, which define the classical and non-relativistic limit respectively. 
Of the four regions, only the potential region contains off-shell modes, which can be integrated out and yield the conservative part of the dynamics. Their contributions can be captured by a non-relativistic EFT which was introduced and put to use in Refs.~\cite{Cheung:2018wkq, Bern:2019nnu, Bern:2019crd}, and we will utilize in Section \ref{sec:eft}. 

The method of regions~\cite{Beneke:1997zp, Smirnov:2004ym} instructs us to expand the integrand using the scaling corresponding to a given region, and then integrate over the whole space of loop momenta in dimensional regularization. 
Our goal is to calculate the contributions from the potential region.

\subsection{Outline of the new method}
Ref.~\cite{Bern:2019nnu} introduced a ``non-relativistic integration'' method by which one must first expand in velocity before expanding in $|\vect q|$. This produces simple integrals akin to those appearing in NRGR \cite{Goldberger:2004jt,Gilmore:2008gq} at the cost of breaking manifest relativistic invariance in the first step. As explained above the potential region is defined by a double expansion, and we might chose to expand in the opposite order, first in small $|\vect q|$, and then in velocity.  The expansion in small $|\vect q|$ is just the expansion in the \emph{soft region} where all graviton momentum components are uniformly small (of order $|\vect q|$). The result of this expansion is a power series in $|\vect q|$ truncated at an appropriate order, with each term in the expansion given by fully relativistic \emph{soft integrals} with linearized matter propagators. To simplify the expressions, we will apply the well-known method of integration-by-parts reduction \cite{Chetyrkin:1981qh} to these soft integrals to rewrite them as a linear combination of \emph{master integrals}. Then we will construct differential equations \cite{Kotikov:1990kg,Bern:1993kr,Remiddi:1997ny,Gehrmann:1999as} in the canonical form \cite{Henn:2013pwa, Henn:2014qga} for these master integrals. The choice of a \emph{basis} of the master integrals will be an important technical point to be discussed later. The selection of pure basis integrals is also facilitated by automated tools \cite{Prausa:2017ltv,Gituliar:2017vzm}.

The upside of the soft expansion is that it keeps the integrals fully relativistic, but here we are only interested in the contributions from the potential region. Thus, in a second step we should \emph{re-expand} the integrals in the potential region where graviton momenta are dominated by spatial components, since the potential region isolates conservative classical effects \cite{Cheung:2018wkq, Bern:2019nnu, Bern:2019crd}. After the expansion in the potential region, each term in the previous small-$q$ expansion would be rewritten as a Taylor series in the velocity  (ratio of spatial to time components) of external momenta. Unlike the first step, which gives the expansion in small $|q|$ to some finite order, in the second step the velocity expansion can be performed to all orders by using method of differential equations for the soft master integrals. A key observation is that we can construct differential equations for the soft integrals directly before re-expanding in the potential region, as the re-expansion does not change the differential equations, but changes the boundary conditions near the static limit. Thus, it suffices to expand the soft master integrals to leading order in velocity in the potential region, to obtain the boundary conditions that allow us to uniquely solve the differential equations and determine the integrals to all orders in velocity.\footnote{The true values of the soft integrals, which will be useful for future calculations beyond conservative classical dynamics, can be obtained by solving differential equations subject to the boundary conditions of the ``full'' soft integrals near the static limit or another suitable kinematic limit.}

Let us now explain each of these steps in more detail.

\subsubsection{Soft expansion using special variables}

\begin{figure}\centering \includegraphics{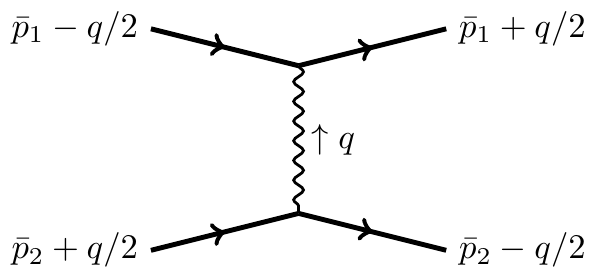}
  \caption{Kinematic setup with special variables.}\label{fig:TreeKin}
\end{figure}

In order to carry out the procedure outline above, it will be useful to parametrize the external kinematics as\footnote{To our knowledge this parameterization was introduced in \cite{Landshoff:1989ig}. Notice that in our convention all external $p_i^\mu$ are outgoing, but $\bar p_i$ can be either incoming or outgoing.}
\begin{align}
p_1=-(\bar{p}_1-q/2) \,,\quad  p_4 = \bar{p}_1 + q/2 \,,\\
p_2=-(\bar{p}_2+q/2) \,,\quad  p_3 = \bar{p}_2 - q/2 \,,
\end{align}
as displayed in Fig.~\ref{fig:TreeKin}. Note that $s=(p_1+p_2)^2=(\bar{p}_1+\bar{p}_2)^2$, so the physical region is still given by $s>(m_1+m_2)^2$.
By construction the $\bar{p}_i$'s are orthogonal to the momentum transfer $q=(p_1+p_4)$,
\begin{align}
  &p_1^2 - p_4^2 = -2 \, \bar{p}_1\cdot q = 0\,,\\
  &p_2^2 - p_3^2 = 2 \, \bar{p}_2\cdot q = 0\,.
\end{align}
We would like expand the full topologies in the \emph{soft region}, which in these variables is characterized by the following hierarchy of scales
\begin{equation}
|\ell|\sim |q|\ll |\bar{p}_i|,m_i,\sqrt{s}\,,
\end{equation}
where $\ell$ stands for any combination of graviton momenta ($\ell_1,\ell_2,\ell_1\pm\ell_2,\cdots$), or equivalently
\begin{align}
  (\bar p_i^0, \vect{\bar p}_i) & \sim m (1,1), \\
  (q^0\,, \vect q) & \sim (|\vect q|, |\vect q|), \\
  \quad (\ell^0\,, \vect \ell) & \sim (|\vect q|, |\vect q|).
\end{align}
The massless graviton propagators typically take the form
\begin{equation}
  \frac{1}{\ell^2}\,, \quad \frac{1}{(\ell-q)^2}\,,
\end{equation}
so they have uniform power counting $| q|^{-2}$ in the small-$|q|$ limit, without further expansion terms.
Meanwhile, the momentum of each matter propagator is the sum of an external matter momentum $\bar{p}_i\pm\frac{1}{2}q$ and the momentum $\ell$ injected by gravitons (here $\ell$ is generally some linear combination of one or more graviton momenta). We have to expand these matter propagators in the soft region,
\begin{equation}
  \frac{1}{\left(\ell+\bar{p}_i\pm\frac{1}{2}q\right)^2-m_i^2}
  = \frac{1}{2\bar{p}_i\cdot\ell} - \frac{\ell^2\pm\ell\cdot q}{(2 \bar{p}_i\cdot\ell )^2}+\cdots \,,
  \label{eq:linearize}
\end{equation}
so all massive propagators are replaced by ``eikonal'' propagators that are linear in loop momenta.
We can further define normalized external momenta,
\begin{equation}
  u_1^\mu = \frac {\bar{p}_1^\mu} {\bar m_1}, \quad u_2^\mu = \frac {\bar{p}_2^\mu} {\bar m_2},
  \label{eq:u1u2def}
\end{equation}
with 
\begin{equation}
  \bar m_1^2 = \bar{p}_1^2 = m_1^2 - \frac {q^2} 4, \quad \bar m_2^2 =  \bar{p}_2^2 = m_2^2 - \frac {q^2} 4\,. \label{eq:mbar}
\end{equation}
We can then rewrite the denominators of Eq.~\eqref{eq:linearize} by following Eq.~\eqref{eq:u1u2def} and factoring out the scale associated to $\bar{p}_i$ from the propagators,
\begin{equation}
  \frac{1}{2\bar{p}_i\cdot\ell}=\frac{1}{(2u_i\cdot \ell) \sqrt{m_i^2 - q^2/4}}
  =  \frac {1} {2 u_i \cdot \ell} \left( \frac 1 {m_i} + \frac{q^2}{8m_i^3} + \frac{3 q^4}{128 m_i^5} + \cdots \right),
  \label{eq:pToU}
\end{equation}
where the relevant kinematic factor is again expanded in small $|q|$. This choice of variables, has the advantage that each order in the expansion is homogeneous in $|q|$, due to the absence of products between external and graviton momenta in the numerators.

In summary, in the soft region the graviton propagators remain unexpanded, while the matter propagators have the form $1/(2 u_i \cdot \ell)$, generally raised to higher powers when we look at terms beyond the leading order in the expansion.
Thus, we can write down the following power counting rules applicable at any loop order, before we actually carry out the expansion in the soft region,
\begin{align}
  \begin{split}
  \text{Graviton propagator:} &\quad \sim \frac 1 {|q|^2},  \\
  \text{Matter propagator:}   &\quad \sim \frac 1 {|q|},  \\
  \text{Integration measure per loop:} &\quad \mathrm{d}^4 \ell \sim |q|^4. 
\end{split}
\label{eq:softpowercounting}
\end{align}

At successively higher orders in the expansion Eq.~\eqref{eq:linearize}, we encounter integrals with propagators raised to higher powers as well as higher-degree polynomials in the numerators. Fortunately, all such integrals can be reduced to a finite number of master integrals via integration by parts \cite{Chetyrkin:1981qh} automated by the Laporta algorithm \cite{Laporta:2001dd, Laporta:1996mq}, and we use the \texttt{FIRE6} software package \cite{Smirnov:2019qkx} to perform the calculation. This allows the soft expansion result to be expressed in terms of a small number of master integrals, whose values will be calculated by the method of differential equations.

\subsubsection{Velocity differential equations for soft integrals}

Next we want to integrate the master integrals, which we will do by the method of differential equations. Importantly, by virtue of the normalization \eqref{eq:u1u2def} we have 
\begin{equation}
u_1^2 = u_2^2 = 1, \quad u_1 \cdot q = u_2 \cdot q = 0 . \label{eq:u1u2dotproducts}
\end{equation}
Hence, after the soft expansion, the only dimensionful scale of the integrals is $q^2$. The dependence on $q^2$ of each integral can be easily fixed by dimensional analysis, and the integrals only depend non-trivially on the following dimensionless parameter,
\begin{equation}
  \relfbar = u_1 \cdot u_2 . \label{eq:ydef}
\end{equation}
Hence our differential equations will depend on this single variable, $\relfbar$, which is related to the relativistic Lorentz factor in Eq.~\eqref{eq:LorentzFactor},
\begin{equation}
\relfbar = \relf + \frac{\relf  (m_1^2+m_2^2) +2 m_1 m_2}{8 m_1^2 m_2^2}\, q^2  + \cO(q^4) . \label{eq:yGamma}
\end{equation}
We give this relation to the next-to-leading order in $q^2$ since it will be used later to convert amplitude results in $\relfbar$ to results in $\relf$.

We will construct the differential equations by taking derivatives of the master integrals.  The choice of a \emph{basis} master integrals is not unique; we choose a \emph{pure basis} in which each master integral has an $\epsilon$ expansion where each term is a generalized polylogarithms \cite{Goncharov:2001iea, Goncharov:2010jf, Duhr:2012fh} of uniform transcendentality. This is largely just a technical point, because at the order of $\epsilon$ expansion needed, the integrals in this paper do not contain any functions more complicated than logarithms (which are a special case of generalized polylogarithms). However, this will yield simple differential equations. A possible form of the differential operator $\partial_{\relfbar}$, rewritten as derivatives against normalized external momenta $u_i^\mu$, is
\begin{equation} \frac {\mathrm{d}}{\mathrm{d}\relfbar} =\frac{1}{\relfbar^2-1}\left(\relfbar u_1^\mu-u_2^\mu\right) \frac{\partial}{\partial u_1^\mu}\,. \label{eq:yDerivativeInU}
\end{equation}
The original form $\mathrm{d}/\mathrm{d}\relfbar$ is fine for differentiating the explicit $\relfbar$-dependent factors in the normalization of the master integrals, but the RHS of Eq.~\eqref{eq:yDerivativeInU} is needed to differentiate the propagators and numerators expressed in terms of external and internal momenta. After differentiating any of the pure integrals with respect to $y$, the result can be IBP-reduced back to the basis of master integrals. We will rationalize the square root $\sqrt{\relfbar^2-1}$ using the change of variable
\begin{equation}
  \relfbar = \frac{1+x^2}{2x}, \quad \sqrt{y^2-1} = \frac{1-x^2}{2x},\quad y \geq 1, \quad 0<x \leq 1,
  \label{eq:xvar}
\end{equation}
under which
\begin{equation}
\frac {\mathrm{d}}{\mathrm{d}\relfbar} = \frac{2 x^2}{x^2-1} \frac{\mathrm{d}}{\mathrm{d}x}\, .
\end{equation}
In terms of these variables, the physical region in our scattering processes is given by $1<\relfbar<\infty$, i.e.\ $0<x<1$.

Our differential equations will take the canonical form \cite{Henn:2013pwa, Henn:2014qga}
\begin{equation}
  \mathrm{d} \vec f = \epsilon \sum_i A_i \, \dlog \alpha_i(x) \vec f\,,
  \label{eq:diffeqform}
\end{equation}
where $A_i$ are numerical matrices and each $\alpha_i(x)$, called a \emph{symbol letter}, is a rational functions in $x$, and $\epsilon=(4-D)/2$ is the dimensional regularization parameter.\footnote{Henn's canonical form can also be used for finite integrals without a dimensional regulator, see \cite{Caron-Huot:2014lda,Herrmann:2019upk}.} The set of  the $\alpha_i$ is called the symbol alphabet, in the formalism of Ref.~\cite{Goncharov:2010jf} which uses ``symbols'' to elucidate functional identities between generalized polylogarithms.

These differential equations can be easily solved, given appropriate boundary conditions. While we could use them to calculate the full soft integrals, we will use them to directly extract the values of the integrals evaluated in the potential region. By expanding in the potential region and summing the expansion to all orders, we have localized the loop integration on the poles of matter propagators. We are essentially dealing with a version of cut integrals (see e.g. Refs.~\cite{Kosower:2011ty, CaronHuot:2012ab, Abreu:2017ptx, Bosma:2017ens, Sogaard:2014ila, Primo:2016ebd}), which satisfy the same IBP relation and differential equations as original uncut integrals. This is the reason why the only changes are in the boundary conditions, obtained in the near-static limit $y\to 1$ by re-expanding the master integrals in the potential region.

\subsubsection{Static boundary conditions from re-expansion in the potential region}
\label{sec:potentialexpansion}

We are ready to write down the power counting of momentum components in the potential region, in terms of a small velocity parameter $v$. Since we have first expanded in the soft region and transitioned to normalized external momenta in Eq.~\eqref{eq:pToU}, we will write down the power counting for $u_i^\mu$ instead of $p_i^\mu$, and for graviton momenta $\ell^\mu$,
\begin{align}
    u_i^\mu &= (u_i^0, \vect u_i)  \sim (1, |\vect v|), \label{eq:potentialU} \\
   \ell^\mu \, &= (\omega, \vect \ell)  \sim |\vect q| (|\vect v|, 1). \label{eq:potentialL}
\end{align}
The factor of $|q|$ is unimportant in our two-step expansion procedure, where the integrals are already homogeneous in $q^2$ (i.e.\ proportional to a definite power of $q^2$ without further corrections) after the soft expansion is carried out.

Now we can expand graviton and matter propagators. Recall that graviton propagators $\sim 1/ \ell^2$ are unchanged in the soft expansion. Their expansion in the potential region is
\begin{equation}
  \frac 1 {\ell^2} = \frac 1 {\omega^2 - \vect \ell^2} = - \left(
  \frac 1 {\vect \ell^2}
  + \frac {\omega^2} {(\vect \ell^2)^2}
  + \frac {\omega^4} {(\vect \ell^2)^3}
  + \cdots \right) . \label{eq:gravPotRegion}
\end{equation}
On the other hand, matter propagators of the form~\eqref{eq:pToU} are homogeneous in $v$ and the expansion consists of a single term,
\begin{equation}
\frac {1} {2 u_i \cdot \ell} = \frac {1} {2 \left( u_i^0\, \omega - \vect u_i \vect \ell \, \right)} \label{eq:matterPotRegion} \, .
\end{equation}
The power counting rules for propagators and integration measure in the potential region are
\begin{align}
  \text{Graviton propagator:} & \quad \sim 1, \label{eq:powerGravPot} \\
  \text{Matter propagator:} & \quad \sim \frac{1}{|\vect v|}, \label{eq:powerMatterPot} \\
  \text{Integration measure:} & \quad \mathrm{d}^4 \ell \sim |\vect v|. \label{eq:powerIntPot}
\end{align}
We will only need to expand to leading order in $|\vect v|$, since we only wish to obtain the value of the integrals at one point, to supply a boundary condition.

The expanded integrals can be evaluated by residues by performing contour integration over the graviton energies $\omega$. Such energy integrals can be ambiguous until one applies a proper prescription \cite{Cheung:2018wkq,Bern:2019crd}. Such a prescription is effectively part of the \emph{definition} of the potential region which separates it from the larger soft region. Refs.~\cite{Cheung:2018wkq,Bern:2019crd} presented the prescription in the absence of double poles, i.e.\ squared matter propagators, but we will show in our examples that when the energy integral prescription is formulated in terms of residues, double poles can be treated in a natural manner and cause no difficulty. As explained in Ref.~\cite{Bern:2019crd}, this prescription generally implies that an integral in the potential region with less than one massive propagators per loop is necessarily zero. Finally, the resulting $D-1$-dimensional integrals can be easily evaluated using traditional methods, and provide the desired boundary conditions to solve our soft integrals in the potential region. 

 \subsection{One-loop integrals}
\label{sec:oneloop}

Next we will illustrate the method above with some simple one-loop examples. We will evaluate all the box-type integrals, which appear in the one-loop $\cN=8$ integrand in Eq.~\eqref{eq:1lintegrand} with scalar numerator.\subsubsection{Box integral}
The box integral with two opposite masses has been evaluated in Ref.~\cite{Beenakker:1988jr} in dimensional regularization up to order $\epsilon^0$. It has also been discussed in detail in Ref.~\cite{Bern:2019crd}.
\begin{figure}[h]
	\centering
	\includegraphics{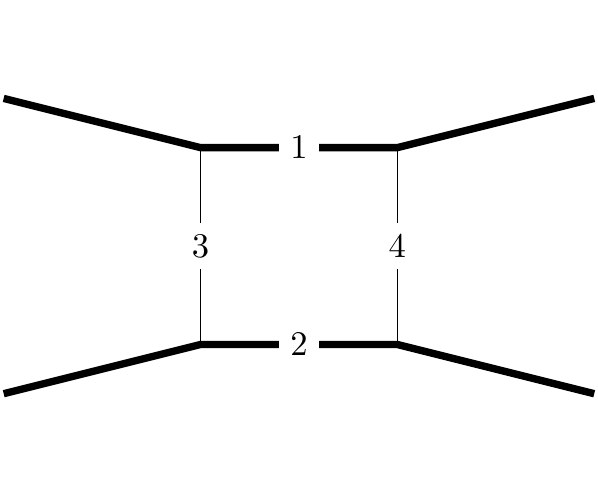}
	\caption{Top level topology at one-loop. Indices correspond to the propagators listed in eq.~\eqref{eq:BoxPropsFull}.}
	\label{fig:BoxTopo}
\end{figure}
As shown in Fig.~\ref{fig:BoxTopo}, a generic integral in the box topology is of the form
\begin{align} 
I_{i_1, i_2, i_3, i_4} = \int\frac{\mathrm{d}^{D}\ell}{(2\pi)^{D}}\frac{1}{\tilde\rho_1^{i_1} \tilde\rho_2^{i_2} \tilde\rho_3^{i_3} \tilde\rho_4^{i_4}}\,, \label{eq:fullBox}
\end{align} 
where the propagator denominators are explicitly 
\begin{align}
\tilde\rho_1=(\ell-p_1)^2-m_1^2\,,\quad
\tilde\rho_2=(\ell+p_2)^2-m_2^2\,,\quad
\tilde\rho_3=\ell^2\,,\quad
\tilde\rho_4=(\ell-q)^2\,,\label{eq:BoxPropsFull}
\end{align}
and the scalar box integral is $I_{\II} = I_{1,1,1,1}$. 
The crossed box integral topologies are related to the box integral ($\X$) by the replacement $u_1 \rightarrow u_1$, $u_2 \rightarrow -u_2$.

\subsubsection*{Integration-by-parts reduction of soft integrals}
Using the soft power counting rules explained in the previous section we see that the box integrals are $\mathcal{O}(|q|^{-2})$. Thus, classical power counting requires expanding the integral to subleading powers. The box propagators reduce in the soft expansion to
\begin{equation}
\rho_1 = 2u_1\cdot\ell\,,\quad
\rho_2 = -2u_2\cdot\ell\, , \quad
\rho_3 = \ell^2\, , \quad
\rho_4 = (\ell-q)^2\, ,
\label{eq:softBoxRhos}
\end{equation} 
which upon expansion of the integral will generally appear raised to integer powers. The numerators appearing in the expansion are polynomials in $\rho_i$, so each order in the soft expansion is a sum of integrals of the form
\begin{equation}
G_{i_1, i_2, i_3, i_4} = \int \frac{\mathrm{d}^D \ell \, e^{\EulerGamma \epsilon}}{\imath \pi^{D/2}} \frac {1} {\rho_1^{i_1} \rho_2^{i_2} \rho_3^{i_3} \rho_4^{i_4}}, \label{eq:GnotationBox}
\end{equation}
with each such integral multiplied by a rational function of the external kinematic variables $m_i^2$, $q^2$, and $\relfbar$. As we already mentioned, $q^2$ is the only dimensionful scale in such integrals. Whenever $i_1$ or $i_2$ is non-positive, the integral will become scaleless and vanish in dimensional regularization.\footnote{Physically speaking, this is because the soft expansion only captures the part of the amplitude that is non-analytic in $q^2$ and relevant for long-range classical physics.} Here and in the following, when writing such soft integrals we adopt the convention of Ref.~\cite{Smirnov:2012gma}, in which we remove an overall factor of
\begin{equation}
  \frac{\imath}{(4\pi)^2}(\bar\mu^2)^\epsilon \equiv \frac{\imath}{(4\pi)^2}\left(4\pi e^{-\EulerGamma} \mu^2\right)^{\epsilon}\,,
  \label{eq:loopfactor}
\end{equation}
per loop, where $\epsilon=(4-D)/2$ and $\mu$ is the dimensional regularization scale. In other words, we write the integration measure for each loop as $ \mathrm{d}^D \ell/(\imath \pi^{D/2})$, and multiply by a factor of Eq.~\eqref{eq:loopfactor} per loop in the end to recover results defined with the more common normalization $\mathrm{d}^D \ell/(2\pi)^D$.

\begin{figure}[tb!]
	\begin{subfigure}[b]{0.3\linewidth} 
		\centering
		\includegraphics{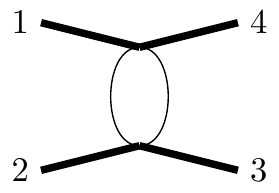}
		\caption{$f_1$} 
	\end{subfigure}
	\begin{subfigure}[b]{0.3\linewidth}
		\centering	\includegraphics{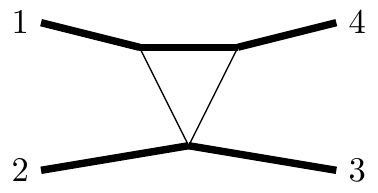}
		\caption{$f_2$}   
	\end{subfigure}
	\begin{subfigure}[b]{0.3\linewidth}
		\centering 
		\includegraphics{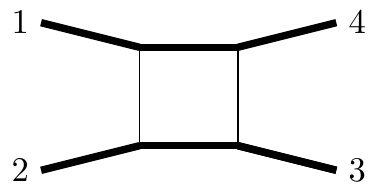}
		\caption{$f_3$} 
	\end{subfigure}
	\caption{Topologies for the box master integrals.}
        \label{fig:BoxMastersSoft}
\end{figure}

Using integration-by-parts reduction, all such integrals are rewritten as linear combinations of the following three \emph{master integrals}\footnote{In contrast the full box system has 10 master integrals, see e.g. Ref.~\cite{Bern:2019crd}.}
\begin{equation}
f_1=  \epsilon (-q^2)\, G_{0,0,2,1}\,,\quad
f_2=  \epsilon^2\sqrtmQSq \, G_{1,0,1,1}\,, \quad
f_3=  \epsilon^2\sqrt{\relfbar^2-1}\, (-q^2)\, G_{1,1,1,1}\,. \label{eq:boxPureBasis}
\end{equation}
whose corresponding topologies are depicted in Fig.~\ref{fig:BoxMastersSoft}. So all integrals given by Eq.~\eqref{eq:GnotationBox} span not an infinite-dimensional, but a three-dimensional vector space. The above integrals are all proportional to $(-q^2)^{-\epsilon}$ times a $q$-independent function of the dimensionless parameter $y$.
The basis does not involve the other triangle integral $G_{0,1,1,1}$ with a (linearized) matter propagator on the bottom -- this is because with the linearized propagator denominators $2 u_i \cdot \ell$ the two triangle integrals are identical and we may freely choose either one as part of the basis of master integrals.

Starting from the original box integral Eq.~\eqref{eq:fullBox} with $a_k=1$, we expand the propagators as in Eqs.~\eqref{eq:linearize} and \eqref{eq:pToU}, and perform integration-by-parts reduction to obtain the small-$q$ expansion of the box integral in terms of the three master integrals in Eq.~\eqref{eq:boxPureBasis},
\begin{align}
  I_{\II} &= \frac{\imath}{(4\pi)^2}(\bar\mu^2)^\epsilon \bigg[
\frac{1}{\epsilon ^2 \bar m_1 \bar m_2 \sqrt{\relfbar^2-1}} \frac{1}{(-q^2)} f_3 \nonumber\\
&\quad +  \frac{\left(\bar m_1+\bar m_2\right)}{\epsilon\bar m_1^2 \bar m_2^2(\relfbar-1) } \frac{1} {\sqrtmQSq} f_2\nonumber \\
&\quad - \frac{(1+2 \epsilon) \left(2 \bar m_2 \bar m_1 \relfbar+\bar m_1^2+\bar m_2^2\right)}{8 \epsilon ^2 \bar m_1^3 \bar m_2^3 \left(\relfbar^2-1\right)^{3/2}} f_3
+ \frac{(1 + 2 \epsilon) \left[ \left(\bar m_1^2+\bar m_2^2\right) \relfbar+2 \bar m_1 \bar m_2 \right]}{8 \epsilon \bar m_1^3 \bar m_2^3 (\relfbar^2-1)} f_1\bigg], \nonumber \\ \label{eq:boxExpandedReduced}
\end{align}
where the 1st, 2nd, and 3rd lines are of order $|q|^{-2}$, $|q|^{-1}$, and $|q|^0$, respectively\footnote{This is true up to the factors of $|q|$ hidden in the definition of $\relfbar$ and $\bar m_i$.}. The bubble integral $f_1$ will be eventually set to zero because we will evaluate the integrals in the potential region.
\subsubsection*{Differential equations for soft integrals}
Now we will construct differential equations for the three pure master integrals in Eq.~\eqref{eq:boxPureBasis}. The original form of the differential operator $\mathrm{d}/\mathrm{d}\relfbar$ is used for differentiating the explicit $\relfbar$-dependent factors in Eq.~\eqref{eq:boxPureBasis}, such as $\sqrt{\relfbar^2-1}$, and the RHS of Eq.~\eqref{eq:yDerivativeInU} is used to differentiate the propagators in Eqs.~\eqref{eq:softBoxRhos} and \eqref{eq:GnotationBox}. After differentiating any of the three pure integrals with respect to $y$, the result is a sum of integrals of the form Eq.~\eqref{eq:GnotationBox}, and can be IBP-reduced back to the basis Eq.~\eqref{eq:boxPureBasis}. After IBP-reduction we use the change of variables from $y$ to $x$ in Eq.~\eqref{eq:xvar}, to rationalize the square roots. The resulting differential equation is
\begin{equation}
\frac{\mathrm{d}\vec{f}} {\mathrm{d} x} =\epsilon\frac{A}{x} \vec{f} , 
\end{equation}
where the matrix $A$ is explicitly given by
\begin{equation} 
A=  
	\begin{pmatrix} 
		0 & 0 & 0 \\ 
		0 & 0 & 0 \\ 
		1 & 0 & 0 
	\end{pmatrix}\,.
\end{equation}
This can be written in the form \eqref{eq:diffeqform}
\begin{equation}
\mathrm{d}\vec{f} = \epsilon A_1 \, \dlog(x)\vec{f}\,,
\label{eq:1loopSoftDE}
\end{equation}
so we recognize $x$ as the only \emph{symbol letter} for the integrals relevant at one loop.

\subsubsection*{Static boundary conditions from re-expansion in the potential region}

Finally, we need to obtain the appropriate boundary conditions to solve the differential equation \eqref{eq:1loopSoftDE} in the potential region. As explained above, we proceed by expanding the pure basis of master integrals Eq.~\eqref{eq:boxPureBasis} in the near-static limit $|\vect v|\ll 1$, using the rules in Section~\ref{sec:potentialexpansion}. After expanding in $|\vect v|$, each order in the series consists of a sum of integrals of the form
\begin{equation}
\int \mathrm{d}^{D-1} \vect \ell \int_{-\infty}^{\infty} \mathrm{d} \omega \, \frac{\mathcal N(\omega, \vect \ell, u_i^0, \vect u_i)} { \left( \vect \ell^2-\imath0 \right)^{i_1} \left[ \left( \vect \ell - \vect q \right)^2 - \imath0 \right]^{i_2} \left( 2 \vect u_1 \vect \ell - 2u_1^0 \, \omega- \imath0 \right)^{i_3} \left( -2 \vect u_2 \vect \ell + 2u_2^0\, \omega- \imath0 \right)^{i_4}}, \label{eq:integralsAfterVexpansion}
\end{equation}
with some polynomial numerator $\mathcal N$.

These integrals can be evaluated by performing integration over energy $\omega$ by residues.  We work in a frame where the momentum transfer $q^\mu$ has no energy component, so the energy of the two graviton lines are $\omega$ and $-\omega$, respectively.  For convenience, we can further boost our frame so that particle 1 is at rest\footnote{To be precise, particle 1 is only at rest up to $\mathcal{O}(q^2)$, as $u_1$ only coincides with the four-velocity of particle 1 at leading order in $q$.} and $u_2$ moves in $z$-direction
\begin{equation}
u_1=(1,0,0,0)\,,\quad u_2=(\sqrt{1+v^2},0,0,v\,)\,. \label{eq:u1u2framechoice}
\end{equation}
The $\relfbar$ variable defined in Eq.~\eqref{eq:ydef} is related to the above parametrization by $v=\sqrt{\relfbar^2-1}$.

We symmetrize over the energy components of the two graviton lines, and rewrite Eq.~\eqref{eq:integralsAfterVexpansion} using the transformation
\begin{equation}
  \int_{-\infty}^{\infty} \mathrm{d} \omega \, \mathcal I(\omega) \rightarrow
  \int_{-\infty}^{\infty} \mathrm{d} \omega \, \frac 1 2 \left[ \mathcal I(\omega) + \mathcal I(-\omega) \right] \, . \label{eq:oneloopsym}
\end{equation}
Then we perform the $\omega$ integral by closing the contour either in the upper half plane or the lower half plane, and pick up contributions from poles at finite values of $\omega$, discarding poles at infinity, i.e.\ neglecting possible non-zero contributions from the arc of a semi-circle contour whose radius tends to infinity. After the $\omega$ integral in Eq.~\eqref{eq:integralsAfterVexpansion} is carried out in this way, we are left with the spatial integral $\mathrm{d}^{D-1} \vect\ell$, and the only denominators left are massless quadratic propagators in three dimensions and linear propagators 
\begin{equation}
	\frac 1 {\vect{\ell}^{2} - \imath0}\,,
        \frac 1 {(\vect{\ell} - \vect{q}) ^{2} - \imath0}\,,
        \frac 1 {-2\ell^z - \imath0} \,.
\end{equation}
The resulting spatial integrals only depend on a single scale $\vect{q}^{\,2}$,
and are related to standard propagator integrals.

The bubble integral $f_1$ in Eq.~\eqref{eq:boxPureBasis} trivially vanishes in the potential region, because there are no poles at finite values of $\omega$ and poles at infinity are discarded in our integration prescription. Using the power counting rules in the potential region, Eqs.~\eqref{eq:powerGravPot} to \eqref{eq:powerIntPot}, we can see that $f_2$ and $f_3$, i.e.\ triangle and box integrals with appropriate prefactors that ensure a canonical form of differential equations, both start at $\cO(v^0)$ in the velocity expansion. For example, $f_3$ has a prefactor $\sqrt{\relfbar^2-1} = v$, two matter propagators giving $\cO(1/v^2)$, and an integration measure of $\cO(v)$, so overall $f_3$ is of $\cO(v^0)$. This is not surprising, since it is well known that integrals of unit leading singularity can have at most logarithmic singularities in any kinematic limit. To obtain $f_2$ and $f_3$ evaluated in the potential region at the leading order in $v$, we keep only the leading term in Eq.~\eqref{eq:gravPotRegion} for each graviton propagator, and then use Eq.~\eqref{eq:oneloopsym} to perform the energy integral, leaving spatial integrals 
\begin{align} 
f_1^{\pot} \big|_{\relfbar=1} &\equiv 0\,,\\
f_2^{\pot} \big|_{\relfbar=1} & = -\frac{\sqrt{\pi}}{2} \epsilon^2 {\sqrtmQSq} \int\frac{\mathrm{d}^{D-1}\ell \, e^{\EulerGamma \epsilon}}{ \pi^{(D-1)/2}}\frac{1}{\left(\vect{\ell}^2-\imath0\right)\left[(\vect{\ell}-\vect{q})^2-\imath0\right]}\,,
\label{eq:linbub}\\
f_3^{\pot} \big|_{\relfbar=1} & =  \sqrt{\pi} \epsilon^2 {(-q^2)} \int\frac{\mathrm{d}^{D-1}\ell \, e^{\EulerGamma \epsilon}}{ \pi^{(D-1)/2}}\frac{1}{ (\vect{\ell}^2 - \imath 0) [(\vect{\ell}-\vect{q})^2 - \imath 0] }\frac{1}{(-2 \ell^z - \imath 0)}\,.
\label{eq:lintri}
\end{align} 
The bubble integral vanishes as the propagator does not have any energy dependence in the potential limit. 
The $(D-1)$-dimensional integrals are calculated in Appendix \ref{sec:3dintegrals} and given in Eqs.~\eqref{eq:Bub3D} and \eqref{eq:Tri3D}. The result for the static limit is then
\begin{align}
f_1^{\pot} \big|_{\relfbar=1} & = 0, \label{eq:g1pot} \\
f_2^{\pot} \big|_{\relfbar=1} & = -\epsilon^2 (-q^2)^{-\epsilon} e^{\EulerGamma \epsilon}
\frac{
	\sqrt{\pi} \, \Gamma \left(\frac{1}{2}-\epsilon \right)^2 \Gamma \left(\epsilon
	+\frac{1}{2}\right)}{2 \Gamma(1-2 \epsilon )}\,,    \label{eq:g2pot}\\
f_3^{\pot} \big|_{\relfbar=1} &= \epsilon^2 (-q^2)^{-\epsilon} e^{\EulerGamma \epsilon}
        \frac{\imath \pi}{2} \frac{\Gamma (-\epsilon )^2 \Gamma (1 + \epsilon)}{\Gamma (-2 \epsilon )}\,. \label{eq:g3pot}
\end{align} 
Solving the differential equation \eqref{eq:1loopSoftDE} shows that Eqs.~\eqref{eq:g1pot}--\eqref{eq:g3pot} in fact are correct to all orders in $v$, i.e.\ for any values of $y\geq 1$, so they are the final expressions for the pure basis Eq.~\eqref{eq:boxPureBasis} as evaluated in the potential regions to all orders in velocity,
\begin{align}
  f_1^{\pot} = f_1^{\pot} \big|_{\relfbar=1}\,,\quad  f_2^{\pot} = f_2^{\pot} \big|_{\relfbar=1}\,, \quad f_3^{\pot} = f_3^{\pot}\big|_{\relfbar=1} \,.
\end{align} 
Looking forward to the next sections, we will find the solutions to differential equations to be more non-trivial for two-loop integrals.

\subsubsection*{Result}
Substituting the results Eqs.~\eqref{eq:g1pot}--\eqref{eq:g3pot} into Eq.~\eqref{eq:boxExpandedReduced} and taking into account Eqs.~\eqref{eq:mbar} and \eqref{eq:yGamma}, we obtain the box integral evaluated in the potential region to all order in velocity, given as a small-$|q|$ expansion, 
\begin{align}
  I_{\II}^{\pot} &= \frac{\imath}{(4\pi)^2} \left(\frac{-q^2}{\bar\mu^2}\right)^{-\epsilon}
\bigg\{
\frac{1}{(-q^2)} \frac {\imath \pi} {2 m_1 m_2 \sqrt{\relf^2-1}}
\frac{e^{\epsilon \EulerGamma}\Gamma (-\epsilon )^2 \Gamma (1 + \epsilon)}{\Gamma (-2 \epsilon )}\nonumber \\
&\quad - \frac{1}{\sqrtmQSq} \frac{ \epsilon(m_1 + m_2)}{m_1^2 m_2^2(\relf-1) }
\frac{\sqrt{\pi} \,e^{\epsilon \EulerGamma} \Gamma \left(\frac{1}{2}-\epsilon \right)^2 \Gamma \left(\epsilon+\frac{1}{2}\right)} {2 \Gamma(1-2 \epsilon )}\nonumber \\
&\quad - \frac{ \imath \pi \epsilon \left(m_1^2+m_2^2+2 m_1 m_2 \relf\right)}{8 m_1^3 m_2^3 \left(\relf^2-1\right)^{3/2}}
\frac{e^{\epsilon \EulerGamma}\Gamma (-\epsilon )^2 \Gamma (1 + \epsilon)}{\Gamma (-2 \epsilon )}
\nonumber \\
&\quad + \mathcal O \left( \sqrtmQSq \right)
\bigg\}  \, , \quad \relf>1 \,.
\label{eq:IIres}
\end{align}

\subsubsection{Crossed box integral}

We end with a discussion of the crossed box integrals. As mentioned above, the unexpanded crossed integral is related to the box integral by the crossing replacement $u_1 \rightarrow u_1$, $u_2 \rightarrow -u_2$. Therefore, the same soft differential equations \eqref{eq:1loopSoftDE} are satisfied by these integrals, and one only needs to be careful about the boundary conditions.

The specific choice of reference frame Eq.~\eqref{eq:u1u2framechoice} is changed by crossing into
\begin{equation}
u_1=(1,0,0,0)\,,\quad u_2=(-\sqrt{1+v^2},0,0,-v\,)\,. \label{eq:u1u2framechoiceCrossed}
\end{equation}
In terms of Lorentz invariants, this is $\relfbar \rightarrow -\relfbar$. However, our results for the box integral at $\relfbar>1$ cannot be analytically continued to negative values of $\relfbar$, because the energy integration prescription produces non-analytic behavior in $\relfbar$. For example, when performing the energy integration for $f_3$ in Eq.~\eqref{eq:boxPureBasis} in the potential region, the two poles lie on the same side of the contour when $\relfbar<0$, and the contour integration gives zero. The correct result for crossed integrals in the static limit (analogous to Eqs.~\eqref{eq:g1pot}--\eqref{eq:g3pot} for the box) is
\begin{align}
f_1^{\pot} \big|_{\relfbar=-1} & = 0, \label{eq:g1potx} \\
f_2^{\pot} \big|_{\relfbar=-1} & = -\epsilon^2 (-q^2)^{-\epsilon} e^{\EulerGamma \epsilon}
\frac{
	\sqrt{\pi} \, \Gamma \left(\frac{1}{2}-\epsilon \right)^2 \Gamma \left(\epsilon
	+\frac{1}{2}\right)}{2 \Gamma(1-2 \epsilon )}\,,    \label{eq:g2potx}\\
f_3^{\pot} \big|_{\relfbar=-1} &= 0 \, . \label{eq:g3potx}
\end{align}
Again, the above equations are derived from the static limit but are actually valid to all orders in velocity, because the velocity differential equations have trivial solutions at one loop. 

\subsubsection*{Result}

To obtain the small-$|q|$ expansion of the crossed box, we also need to make the $\relfbar \rightarrow -\relfbar$ replacement in the coefficients of $f_i$ master integrals in Eq.~\eqref{eq:boxExpandedReduced}. The end result for the small-$|q|$ expansion of the crossed box integral is
\begin{align}
  I_{\X}^{\pot} 
&= \frac{\imath}{(4\pi)^2} \left(\frac{-q^2}{\bar\mu^2}\right)^{-\epsilon}
\bigg\{
 \frac{1}{\sqrtmQSq} \frac{ \epsilon(m_1 + m_2)}{m_1^2 m_2^2(\relf+1) }
\frac{\sqrt{\pi} \, e^{\epsilon \EulerGamma} \Gamma \left(\frac{1}{2}-\epsilon \right)^2 \Gamma \left(\epsilon+\frac{1}{2}\right)} {2 \Gamma(1-2 \epsilon )} \nonumber \\
&\quad + \mathcal O \left( \sqrtmQSq \right)
\bigg\}  \, , \quad \relf> 1 \, .
\label{eq:Xres}
\end{align}
 
 \subsection{Two-loop integrals}
\label{sec:twoloop}
Next we will evaluate the two-loop integrals needed for the two-loop integrand in Eq.~\eqref{eq:twoloopintegrand}.
A simple application of the soft power-counting rules in Eq.~\eqref{eq:softpowercounting} reveals that all the double-box-type integrals in the second line of Eq.~\eqref{eq:twoloopintegrand} contribute in the classical limit with leading power $\cO(q^{-2})$ so they need to be expanded to subleading powers. On the other hand, of the integrals in the third line of Eq.~\eqref{eq:twoloopintegrand}, only the $\H$ and $\xH$ integrals at leading power survive the $(q^2)^2$ suppression of the numerator, and the rest do not contribute in the classical limit.\footnote{The ``mushroom'' integrals $I_{\overline{\mushroomtop{}}}$ and $I_{\overline{\mushroombot{}}}$ vanish identically when evaluated in the potential region \cite{Bern:2019nnu, Bern:2019crd}, so cannot contribute even without the $(q^2)^2$ suppression from the integrand numerator.}

We will describe in detail the computation of the double-box ($\RT$) and $\H$-type integrals, and present results for the rest of the integrals. As usual we will strip from our integrals a factor of \eqref{eq:loopfactor} per loop in intermediate steps, to be restored at the end.

\subsubsection{Double-box ($\RT$)}
\begin{figure}[tb!]  
	\centering
\includegraphics{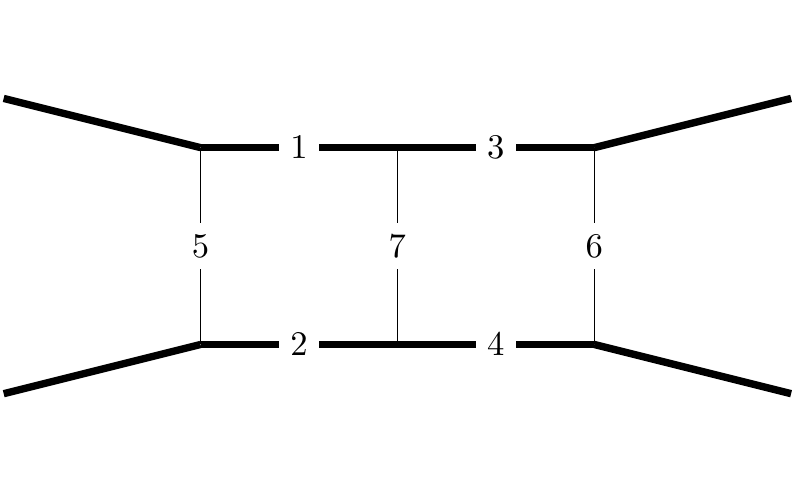}
\caption{III topology. Indices correspond to the propagators listed in eq.~\eqref{eq:TwoLoopPropsFull}.}\label{fig:TopLevelTopos}
\end{figure}  
We first consider generic integrals of the form
\begin{equation}
  I_{i_1,i_2,\dots, i_9}=\int\frac{\mathrm{d}^{D}\ell_1}{(2\pi)^D}\int\frac{\mathrm{d}^{D}\ell_2}{(2\pi)^D}\frac{1}{\tilde \rho_1^{i_1}\tilde \rho_2^{i_2}\cdots\tilde \rho_9^{i_9}}\,.
\end{equation}
Where the propagators are
\begin{alignat}{3}
&\tilde \rho_1=(\ell_1-p_1)^2-m_1^2\,,\qquad\quad &&\tilde \rho_2=(\ell_1+p_2)^2-m_2^2\,,\qquad\quad && \tilde \rho_3=(\ell_2-p_4)^2-m_1^2\,,\nonumber\\
&\tilde \rho_4=(\ell_2+p_3)^2-m_2^2\,,\qquad\quad&&\tilde \rho_5=\ell_1^2\,,\qquad\quad&&\tilde \rho_6=\ell_2^2 \,,\nonumber\\
&\tilde \rho_7=(\ell_1 + \ell_2 - q)^2\,,\qquad\quad&&\tilde \rho_8=(\ell_1 - q)^2\,,\qquad\quad&&\tilde \rho_9=(\ell_2 - q)^2\,.\label{eq:TwoLoopPropsFull}
\end{alignat}
The double-box ($\RT$) topology can be embedded in this family of integrals, as shown in Fig.~\ref{fig:TopLevelTopos}, so that the scalar box integral is given by $I_{\RT} = I_{1,1,1,1,1,1,1,0,0}$. Later we will see that the $\H$ topology can also be embedded in the same family. We note that the equal-mass double-box integral has been evaluated in Refs.~\cite{Smirnov:2001cm, Henn:2013woa} without expansion in the soft or potential region, but the case of generic masses has not been discussed in the literature.

\subsubsection*{Soft expansion and differential equations}
\begin{figure}[tbh!]  
	\centering
	\begin{subfigure}[b]{0.3\linewidth}
		\centering 
		\includegraphics{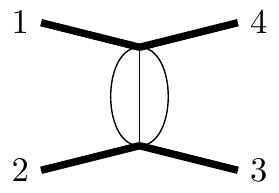}
		\caption{$f_{\RT,1}$} 
	\end{subfigure}
	\begin{subfigure}[b]{0.3\linewidth}
		\centering	\includegraphics{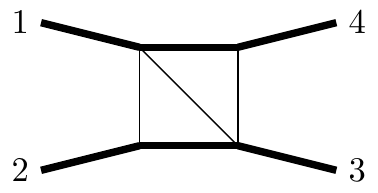}
		\caption{$f_{\RT,2},f_{\RT,3},f_{\RT,4}$}
	\end{subfigure}
	\begin{subfigure}[b]{0.3\linewidth}
		\centering
		\includegraphics{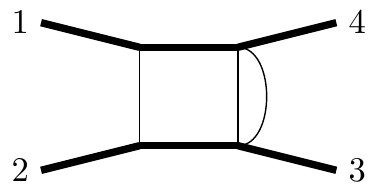}
		\caption{$f_{\RT,5}$}
	\end{subfigure}
	\begin{subfigure}[b]{0.3\linewidth}
		\centering 
		\includegraphics{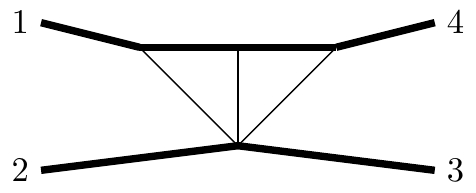}
		\caption{$f_{\RT,6}$}
	\end{subfigure}\qquad\qquad
	\begin{subfigure}[b]{0.3\linewidth}
		\centering	\includegraphics{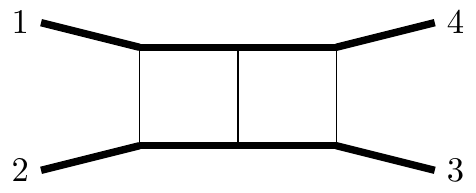}
		\caption{$f_{\RT,7}$}
	\end{subfigure}
	\caption{Even-$|q|$ topologies relevant for the double-box master integrals.}\label{fig:RTTopologiesEven}
\end{figure}

\begin{figure}[tbh!]  
	\centering
	\begin{subfigure}[b]{0.3\linewidth}
		\centering 
		\includegraphics{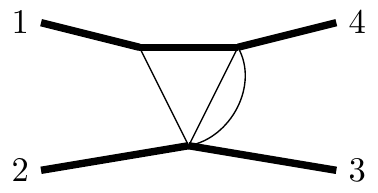}
		\caption{$f_{\mathrm{\RT},8}$} 
	\end{subfigure}
	\begin{subfigure}[b]{0.3\linewidth}
		\centering	\includegraphics{./figures/DE_SlashBox}
		\caption{$f_{\mathrm{\RT},9}$}
	\end{subfigure}
	\begin{subfigure}[b]{0.3\linewidth}
		\centering
		\includegraphics{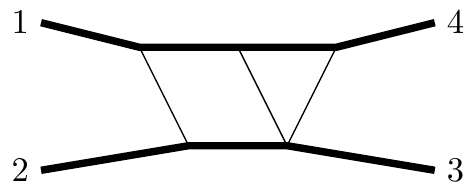}
		\caption{$f_{\mathrm{\RT},10}$}
	\end{subfigure}
	\caption{Odd-$|q|$ topologies relevant for the double-box master integrals.}\label{fig:RTTopologiesOdd}
\end{figure}

In the soft region, we construct an expansion of the integrand around small $|\ell_i| \sim |q|$. In the expansion, only the leading order parts of $\tilde \rho_i$, denoted by $\rho_i$ and given by
\begin{alignat}{3}
&\rho_1=2\, \ell_1\cdot u_1\,,\qquad\qquad &&\rho_2=-2\, \ell_1 \cdot u_2\,,\qquad\qquad && \rho_3=-2\, \ell_2\cdot u_1\,,\nonumber\\
&\rho_4=2\, \ell_2\cdot u_2\,,\qquad\qquad&&\rho_5=\ell_1^2\,,\qquad\qquad&&\rho_6=\ell_2^2 \,,\nonumber\\
&\rho_7=(\ell_1 + \ell_2 - q)^2\,,\qquad\qquad&&\rho_8=(\ell_1 - q)^2\,,\qquad\qquad&&\rho_9=(\ell_2 - q)^2\,.\label{eq:TwoLoopPropsSoft}
\end{alignat}
appear in the denominators (possibly with raised powers), and subleading corrections all appear in numerators. Such numerators are in turn written as linear combinations $\rho_i$.
The small-$|q|$ expansion consists of integrals of the form
\begin{equation}
G_{i_1, \, i_2, \dots , i_9} = \int \frac {\mathrm{d}^D \ell_1 \, e^{\EulerGamma\epsilon}}{\imath\pi^{D/2}} \int \frac {\mathrm{d}^D \ell_2 \, e^{\EulerGamma\epsilon}}{\imath\pi^{D/2}}
\frac{1}{\rho_1^{i_1} \rho_2^{i_2} \dots \rho_9^{i_9}} ,
\label{eq:twoloopGtilde}
\end{equation}
where negative indices represent numerators rather than denominators.
Note that, as in the previous subsection, we have removed a factor of \eqref{eq:loopfactor} per loop in the soft integrals.
There are a total of 10 master integrals for the $\RT$ topology\footnote{For reference, in the full equal-mass problem there are 23 master integrals \cite{Henn:2013woa}.} as shown in Figs.~\ref{fig:RTTopologiesEven} and \ref{fig:RTTopologiesOdd}.  
A pure basis is given by  
\begin{align}
f_{\RT,1}={}& \epsilon ^2 (-q^2) G_{0,0,0,0,1,2,2,0,0}\,,\label{eq:fRT1}\\ 
f_{\RT,2}={}&\epsilon ^4 \sqrt{\relfbar^2-1}\, G_{0,1,1,0,1,1,1,0,0}\,, \label{eq:fRT2}\\ 
f_{\RT,3}={}&\epsilon ^3 (-q^2) \sqrt{\relfbar^2-1}\, G_{0,1,1,0,2,1,1,0,0}\,,\\ 
f_{\RT,4}={}&-\epsilon ^2 (-q^2) G_{0,2,2,0,1,1,1,0,0} +\epsilon ^3 \relfbar \, (-q^2) G_{0,1,1,0,2,1,1,0,0}\,,\\ 
f_{\RT,5}={}&\epsilon ^3\sqrt{\relfbar^2-1}\,  (-q^2) G_{1,1,0,0,1,1,2,0,0}\,,\\ 
f_{\RT,6}={}&\epsilon ^3 (1-6 \epsilon)\, G_{1,0,1,0,1,1,1,0,0}\,, \label{eq:fRT6} \\ 
f_{\RT,7}={}&\epsilon ^4\left(\relfbar^2-1\right) (-q^2)  G_{1,1,1,1,1,1,1,0,0}\,,\\
f_{\mathrm{\RT,8}}={}&\epsilon ^3 \sqrtmQSq G_{1,0,0,0,1,1,2,0,0}\,,\\ 
f_{\mathrm{\RT,9}}={}&\epsilon ^3 \sqrtmQSq G_{0,2,1,0,1,1,1,0,0}\,,\\ 
f_{\mathrm{\RT,10}}={}&\epsilon ^4\sqrt{\relfbar^2-1} \sqrtmQSq G_{1,1,1,0,1,1,1,0,0}\,,\label{eq:fRT10}
\end{align}
where all the master integrals are normalized to be proportional to $(-q^2)^{-2\epsilon}$. The corresponding topologies are depicted in Figs.~\ref{fig:RTTopologiesEven} and~\ref{fig:RTTopologiesOdd}, where we have separated the integrals which are even and odd in $|q|$.  

We perform soft expansion and use IBP-reduction to write the results in terms of the master integrals. The double-box integral is given as the following small-$|q|$ expansion, 
\begin{align} \label{eq:rtsoftexpansion}
  I_{\RT} &= - \frac{(\bar\mu^2)^{2\epsilon}}{(4\pi)^4} \bigg\{\frac{1}{(-q^2)} \frac{1}{\bar m_1^2 \bar m_2^2 \left(y^2-1\right) \epsilon ^4} f_{\RT, 7} \nonumber \\
  & \quad + \frac{1} {\sqrtmQSq} \frac{2 (\bar m_1 + \bar m_2)}{\epsilon^3 (y-1) \sqrt{y^2-1} \, \bar m_1^3 \bar m_2^3} f_{\RT, 10} \nonumber \\
  & \quad -\frac{\bar{m}_1^2 \left(3 \left(y^2+1\right) \epsilon +1\right)+\bar{m}_2^2 \left(3 \left(y^2+1\right) \epsilon +1\right)+2 y (6 \epsilon +1) \bar{m}_2 \bar{m}_1}{24 \left(y^2-1\right)^2 \epsilon ^3 \bar{m}_1^4 \bar{m}_2^4}
  f_{\RT,1} \nonumber \\
  &\quad -\frac{ y \left(\bar{m}_1^2+\bar{m}_2^2\right) + 2 \bar{m}_1 \bar{m}_2}{\left(y^2-1\right)^{3/2} \epsilon ^2 \bar{m}_1^4 \bar{m}_2^4} f_{\RT,2} \nonumber \\
  &\quad - \frac{ y \left(\bar{m}_1^2+\bar{m}_2^2\right) + 2 \bar{m}_1 \bar{m}_2}{8 \left(y^2-1\right)^{3/2} \epsilon ^3 \bar{m}_1^4 \bar{m}_2^4} f_{\RT,3} \nonumber \\
  &\quad -\frac{\bar{m}_1^2 \left(3 \left(y^2+1\right) \epsilon +1\right)+\bar{m}_2^2 \left(3 \left(y^2+1\right) \epsilon +1\right)+2 y (6 \epsilon +1) \bar{m}_2 \bar{m}_1}{12 \left(y^2-1\right)^2 \epsilon ^3 \bar{m}_1^4 \bar{m}_2^4} f_{\RT,4} \nonumber \\
  &\quad +\frac{(2 \epsilon +3) \left(y \left(\bar{m}_1^2+\bar{m}_2^2\right)+2 \bar{m}_1 \bar{m}_2\right)}{12 \left(y^2-1\right)^{3/2} \epsilon ^3 \bar{m}_1^4 \bar{m}_2^4} f_{\RT,5} \nonumber \\
  &\quad -\frac{\bar{m}_1^2 \left(3 \left(y^2+1\right) \epsilon +1\right)+\bar{m}_2^2 \left(3 \left(y^2+1\right) \epsilon +1\right)+2 y (6 \epsilon +1) \bar{m}_2 \bar{m}_1}{12 \left(y^2-1\right)^2 \epsilon ^3 \bar{m}_1^4 \bar{m}_2^4} f_{\RT,6} \nonumber \\
&\quad -\frac{(4 \epsilon +3) \left(2 y \bar{m}_2 \bar{m}_1+\bar{m}_1^2+\bar{m}_2^2\right)}{12 \left(y^2-1\right)^2 \epsilon ^4 \bar{m}_1^4 \bar{m}_2^4} f_{\RT,7} \bigg\} \,,
  \end{align}
where the first line is of order $1/q^2$, the second line is of order $1/|q|$, and the remaining lines are of order $|q|^0$,
Since integration-by-parts will only produce analytic coefficients for master integrals, e.g.\ polynomials in $q^2$ but not $\sqrtmQSq$, 
the master integrals $f_{\RT,1}$ to $f_{\RT,7}$ appear in terms that are even in $|q|$ in the small-$|q|$ expansion of the amplitude, while $f_{\RT,8}$ to $f_{\RT,10}$ appear in expansion terms that are odd in $|q|$.

The differential equations for the master integrals are
\begin{equation}
  \mathrm{d} \vec {f}_{\RT}= \epsilon\left[A_{\RT,0}\,\dlog(x)+A_{\RT,+1}\,\dlog(x-1)+A_{\RT,-1}\,\dlog(x+1)\right]\vec{f}_{\RT}\,.\label{eq:RTSoftDE}
\end{equation}
The even- and odd-$|q|$ systems decouple and we can write 
\begin{align}
  A_{\RT,i}=\left(
\begin{array}{cc}
  A_{\RT, i}^{(\mathrm{e})} & 0 \\
  0 & A_{\RT,i}^{(\mathrm{o})} \\
\end{array}
\right)\,,
\end{align}
where the matrices are given by
\begin{align}\label{eq:RTmateven}
	A_{\RT,0}^{(\mathrm{e})}={}&\left(
	\begin{array}{ccccccc}
	0 & 0 & 0 & 0 & 0 & 0 & 0 \\
	-\frac{1}{2} & -6 & 0 & -1 & 0 & 0 & 0 \\
	-\frac{3}{2} & 0 & 2 & -2 & 0 & 0 & 0 \\
	0 & 12 & 2 & 0 & 0 & 0 & 0 \\
	-\frac{3}{4} & 0 & 0 & 0 & 0 & 0 & 0 \\
	0 & 0 & 0 & 0 & 0 & 0 & 0 \\
	0 & 0 & 1 & 0 & -2 & 0 & 0 \\
	\end{array}
	\right)\,,\qquad
A_{\RT,\pm1}^{(\mathrm{e})} = {}\left(
\begin{array}{ccccccc}
0 & 0 & 0 & 0 & 0 & 0 & 0 \\
0 & 6 & 0 & 0 & 0 & 0 & 0 \\
0 & 0 & -2 & 0 & 0 & 0 & 0 \\
0 & 0 & 0 & 0 & 0 & 0 & 0 \\
0 & 0 & 0 & 0 & 0 & 0 & 0 \\
0 & 0 & 0 & 0 & 0 & 0 & 0 \\
0 & 0 & 0 & 0 & 0 & 0 & 0 \\
\end{array}
\right)\,.
\end{align}
\begin{align} \label{eq:RTmatodd}
A_{\RT,0}^{(\mathrm{o})}={}\left(
\begin{array}{ccc}
0 & 0 & 0 \\
0 & -2 & 0 \\
0 & 1 & 0 \\
\end{array}
\right)\,,\quad
A_{\RT,+1}^{(\mathrm{o})}={}\left(
\begin{array}{ccc}
0 & 0 & 0 \\
-3 & -2 & 0 \\
0 & 0 & 0 \\
\end{array}
\right)\,,\quad
A_{\RT,-1}^{(\mathrm{o})}={}\left(
\begin{array}{ccc}
0 & 0 & 0 \\
3 & 6 & 0 \\
0 & 0 & 0 \\
\end{array}
\right)\,.
\end{align}
We make a technical observation here. Previously we found that at one loop $x$, is
the only symbol letter. As a consequence only
powers of $\log x$ will appear in the solutions to the differential equations.
In contrast, at two loops, there are multiple symbol letters appearing in the
differential equations in Eq.~\eqref{eq:RTSoftDE}:  $\{x,1\pm
x\}$, so the symbol alphabet is larger.  This generically results
in the solution of the differential equations being (harmonic \cite{Remiddi:1999ew,Gehrmann:2001pz})
polylogarithms, but we will see that at leading order in $\epsilon$ all
two-loop integrals only contain logarithms.

\subsubsection*{Re-expansion in the potential region}
As described in Section \ref{sec:int}, we obtain boundary conditions for the pure basis of soft integrals by re-expanding the integrals in the potential region following Eqs.~\eqref{eq:gravPotRegion} and \eqref{eq:matterPotRegion}, and then integrate over energy components of loop momenta using an appropriate prescription \cite{Bern:2019crd}. The energy components of $\ell_1$ and $\ell_2$ are written as $\omega_1$ and $\omega_2$, while the spatial components are written as $\vect \ell_1$ and $\vect \ell_2$.

For the double-box integral and non-planar variants with exactly three graviton propagators, we follow the prescription of Ref.~\cite{Bern:2019crd}, but with slight modifications to simplify the presentation. First, we symmetrize over $3!$ permutations of the energy components of the three gravitons, in a way that directly extends the one-loop prescription Eq.~\eqref{eq:oneloopsym},
\begin{equation}
  \int_{-\infty}^{\infty} \mathrm{d} \omega_1 \int_{-\infty}^{\infty} \mathrm{d} \omega_2 \, \mathcal I(\omega_1, \omega_2) \rightarrow
  \int_{-\infty}^{\infty} \mathrm{d} \omega_1 \int_{-\infty}^{\infty} \mathrm{d} \omega_2  \,
  \frac 1 {3!} \sum\limits_{\eta \in S_3} \mathcal I(\omega_{\eta(1)}, \omega_{\eta(2)})  \, , \label{eq:twoloopsym}
\end{equation}
with the definition $\omega_3 = -(\omega_1 + \omega_2)$,
and then proceed as usual, i.e.\ perform the $\omega_1$ and $\omega_2$ contour integrals one by one, closing the contour either above or below the real axis and always neglecting poles at infinity. As an example, we calculate the static limit of $G_{1,0,1,0,1,1,1,0,0}$, which appears in $f_{\RT, 6}$ in Eq.~\eqref{eq:fRT6} and is shown in Fig.~\ref{fig:RTTopologiesEven}(d). In the $y=1$ i.e.\ static limit, the graviton propagators are turned into $(D-1)$-dimensional propagators,
\begin{align}
  G_{1,0,1,0,1,1,1,0,0} \big|_{\relfbar=1} &= - \int \frac{\mathrm{d}^{D-1} \vect \ell_1 \, e^{\EulerGamma\epsilon}}{\imath\pi^{D/2}} \int \frac {\mathrm{d}^{D-1} \vect \ell_2 \, e^{\EulerGamma\epsilon}}{\imath\pi^{D/2}} \, \frac 1 {\vect \ell_1^2-\imath0}\, \frac 1 {\vect \ell_2^2-\imath0}\, \frac 1 {(\vect \ell_1 + \vect \ell_2 + \vect \ell_3)^2-\imath0} \nonumber \\
  &\quad \times \int_{-\infty}^\infty \mathrm{d} \omega_1 \int_{-\infty}^\infty \mathrm{d} \omega_2 \frac{1} {(2  \vect{\ell}_1 \cdot \vect u_1 -2 \omega_1 u_1^0 -\imath0)} \, \frac{1} {(-2 \vect \ell_2 \cdot \vect u_1 +2 \omega_2 u_1^0 -\imath0)}\,. \label{eq:Wstatic}
\end{align}
Again adopting the frame choice Eq.~\eqref{eq:u1u2framechoice} with $u_1 = (\omega_1,  \vect{u}_1) = (1,  \vect{0})$, the second line of the above equation becomes
\begin{align}
\frac 1 4 \int_{-\infty}^\infty \mathrm{d} \omega_1 \int_{-\infty}^\infty \mathrm{d} \omega_2 \frac{1} {-\omega_1 -\imath0} \, \frac{1} {\omega_2 -\imath0} \, .
\end{align}
By the prescription Eq.~\eqref{eq:twoloopsym}, this divergent integral is turned into
\begin{align}
  &\frac 1 4 \cdot \frac 1 {3!} \int_{-\infty}^\infty \mathrm{d} \omega_1 \int_{-\infty}^\infty \mathrm{d} \omega_2 \left(
  \frac{1} {-\omega_1 -\imath0} \, \frac{1} {\omega_2 -\imath0}
  + \frac{1} {-\omega_1 -\imath0} \, \frac{1} {-\omega_1-\omega_2 -\imath0} \right. \nonumber \\
  &\left.  + \frac{1} {-\omega_2 -\imath0} \, \frac{1} {-\omega_1-\omega_2  -\imath0}
  + \frac{1} {-\omega_2 -\imath0} \, \frac{1} {\omega_1 -\imath0}
  + \frac{1} {\omega_1+\omega_2 -\imath0} \, \frac{1} {\omega_1 -\imath0} \right. \nonumber \\
  & \left. + \frac{1} {\omega_1+\omega_2 -\imath0} \, \frac{1} {\omega_2 -\imath0}
  \right) \, . \label{eq:symmetrizedW}
\end{align}
Now let us perform the $\omega_1$ integral by picking up residues in the upper half plane. Only the 4th, 5th, and 6th terms in the bracket of Eq.~\eqref{eq:symmetrizedW} have $\omega_1$ poles in the upper half plane, and in fact the 5th term contributes two poles whose residues add to zero. The result of $\omega_1$ integration is
\begin{equation}
\frac 1 4 \cdot \frac 1 {3!} (2 \pi \imath) \int_{-\infty}^\infty \mathrm{d} \omega_2 \left( \frac 1 {-\omega_2 -\imath0} + \frac 1 {\omega_2 -\imath0} \right) \, . \label{eq:symmetrizedW1}
\end{equation}
Now we integrate over $\omega_2$ by picking up residues in either the upper or lower half plane, obtaining the same result
\begin{equation}
\frac 1 4 \cdot \frac 1 {3!} (2 \pi \imath)^2 = -\frac{\pi^2} 6 \, . \label{eq:symmetrizedW2}
\end{equation}
Putting it back into Eq.~\eqref{eq:Wstatic}, we obtain
\begin{equation}
G_{1,0,1,0,1,1,1,0,0} \big|_{\relfbar=1} = \frac \pi 6 \int \frac{\mathrm{d}^{D-1} \vect \ell_1\mathrm{d}^{D-1} \vect \ell_2 \, (e^{\EulerGamma\epsilon})^2}{(\imath\pi^{(D-1)/2})^2} \frac 1 {\vect \ell_1^2-\imath0}\, \frac 1 {\vect \ell_2^2-\imath0}\, \frac 1 {(\vect \ell_1 + \vect \ell_2 + \vect \ell_3)^2-\imath0}\,. 
\end{equation}
Now we check that the final result is also independent of the contour choice for $\omega_1$. If instead we perform the $\omega_1$ integral in Eq.~\eqref{eq:symmetrizedW} by picking up residues in the lower half plane, we obtain a result identical to Eq.~\eqref{eq:symmetrizedW1}, so the subsequent $\omega_2$ integration also gives the same result as Eq.~\eqref{eq:symmetrizedW2}. In conclusion, we have verified in this example that once the $S_3$ symmetrization over graviton energies are performed, the subsequent energy integration has no dependence on contour choice (in the sense of closing above or below the real axis).

Adopting the frame choice Eq.~\eqref{eq:u1u2framechoice}, and following this prescription, we find that in the static limit, the only non-vanishing master integrals are equal $f_{\RT,4}^{\pot}, f_{\RT,6}^{\pot}, f_{\RT,7}^{\pot}$ and $f_{\RT,10}^{\pot}$.
The computation of these integrals can be carried out by ordinary methods and is explained in Appendix~\ref{sec:3dintegrals}. By expanding up to $\mathcal{O}(\epsilon^4)$ they yield the following vector of boundary conditions
\begin{align}
  \left. \vec {f}^{\,\pot}_{\RT}\right|_{\relfbar=1}={}& (-q^2)^{-2\epsilon} \epsilon^2\pi^2 \left( 0,0,0, \frac 1 3 - \frac{7\pi^2 \epsilon^2} {18}, 0, -\frac 1 6 + \frac{7\pi^2 \epsilon^2} {36}, \frac 1 2 - \frac {\pi^2 \epsilon^2} {12}, \right.\nonumber\\
  &\qquad\qquad\qquad\quad\left. 0, 0, \frac{\imath\pi \epsilon} 4 -\frac{\imath\pi \log(2) \epsilon^2}{2} \right)^T+\mathcal{O}(\epsilon^5)\,.\label{eq:StaticBNDRT}
\end{align}
\subsubsection*{Result}
The solution of the differential equations~\eqref{eq:RTSoftDE} with the boundary conditions \eqref{eq:StaticBNDRT} and \eqref{eq:StaticBNDH} is presented in Eqs.~\eqref{eq:fRT2pot}--\eqref{eq:fRT10pot} in Appendix~\ref{sec:twoloopdetails}. Here we just note that all functions have an overall factor of $\pi^2\epsilon^2$ and therefore the transcendental weight of the solutions is effectively reduced by two. Consequently at the order considered, the only polylogarithmic function relevant is $\log(x)$ related to the $\operatorname{arcsinh}$ function characteristic of 3PM scattering \cite{Bern:2019nnu, Bern:2019crd} by the change of variable Eq.~\eqref{eq:xvar},
\begin{equation}
	\log (x)= -\log\left(\relfbar+\sqrt{\relfbar^2-1}\right)= -2\operatorname{arcsinh}\left(\sqrt{\frac{\relfbar-1}{2}}\right)
	=-2\operatorname{arcsinh}\left(\sqrt{\frac{\relf-1}{2}}\right)+\mathcal{O}(q^2)\,.
\end{equation} 
Going to $\mathcal{O}(\epsilon^4)$ we find an additional weight-two function
\begin{equation}
  \operatorname{Li}_2 (1-x^2)
\end{equation}
which has no singularity in the entire range $0<x<1$, so has no singularity in either the static limit $\relfbar\to 1$ or the high-energy limit $\relfbar \to \infty$. Barring cancellations, it is natural to expect that this function will be relevant at $\cO(G^4)$ (i.e. at the 4PM order).

Finally, inserting in Eq.~\eqref{eq:rtsoftexpansion} the values of the master integrals evaluated in the potential region, Eqs.~\eqref{eq:fRT2pot}--\eqref{eq:fRT10pot}, and changing variables according to Eqs.~\eqref{eq:mbar} and \eqref{eq:yGamma}, the end result for the double-box integrals is
\begin{align}\label{eq:RTres}
  I_{\RT}^{\pot} &= - \frac{1}{(4\pi)^4} \left(\frac{-q^2}{\bar\mu^2}\right)^{-2\epsilon} \Bigg \{ \frac{1}{(-q^2)} \frac{\pi^2}{2 m_1^2 m_2^2 (\relf^2-1)} \left[ \frac 1 {\epsilon^2} - \frac{\pi^2} 6 + \frac 2 3 \log^2(x) + \cO(\epsilon) \right] \nonumber \\
  &\quad + \frac{1}{\sqrtmQSq} \left[ \frac{\imath \pi ^3 \left(m_1+m_2\right)}{2 m_1^3 m_2^3 (\relf-1) \sqrt{\relf^2-1}} + \cO(\epsilon) \right] \nonumber \\
  &\quad +\bigg[ -\frac{\pi ^2 \left(2 m_2 m_1 \relf+m_1^2+m_2^2\right)}{8   m_1^4 m_2^4 \left(\relf^2-1\right)^2}\frac{1}{\epsilon} + \cO(\epsilon^0) \bigg] \Bigg \} \, .
\end{align}

\subsubsection{$\H$ and crossed $\H$ ($\xH$)}
\begin{figure}[tb!]  
	\centering
	\includegraphics{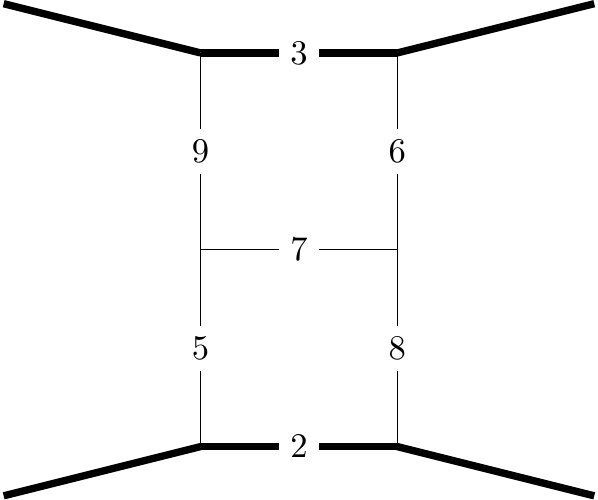}
	\caption{$\H$ topology. Indices correspond to the propagators listed in eq.~\eqref{eq:TwoLoopPropsFull}.}\label{fig:TopLevelH}
\end{figure}
\begin{figure}[tb!]  
	\centering
	\begin{subfigure}[b]{0.3\linewidth}
		\centering 
		\includegraphics{./figures/DE_SR}
		\caption{$f_{\H,1}$} 
	\end{subfigure}
	\begin{subfigure}[b]{0.3\linewidth}
		\centering
				\includegraphics{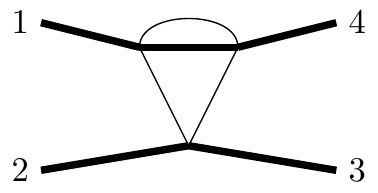}
		\caption{$f_{\H,2}$}
	\end{subfigure}
	\begin{subfigure}[b]{0.3\linewidth}
		\centering 
		\includegraphics{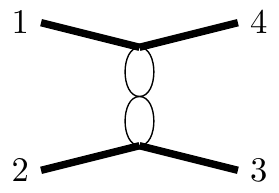}
		\caption{$f_{\H,3}$} 
	\end{subfigure}
	\begin{subfigure}[b]{0.3\linewidth}
		\centering
		\includegraphics{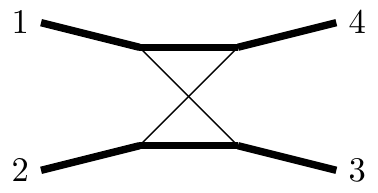}
		\caption{$f_{\H,4}$}
	\end{subfigure}
	\begin{subfigure}[b]{0.3\linewidth}
		\centering	\includegraphics{./figures/DE_SlashBox}
		\caption{$f_{\H,5},f_{\H,6},f_{\H,7}$}   
	\end{subfigure}
	\begin{subfigure}[b]{0.3\linewidth}
		\centering	\includegraphics{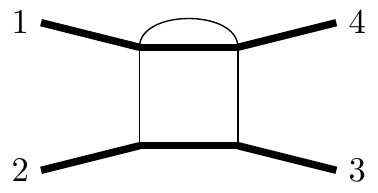}
		\caption{$f_{\H,8}$} 
	\end{subfigure}
	
	\begin{subfigure}[b]{0.3\linewidth}
		\centering	\includegraphics{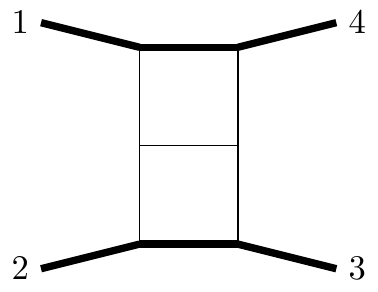}
		\caption{$f_{\H,9},f_{\H,10}$}
	\end{subfigure}
	\caption{Topologies relevant for the H master integrals.}\label{fig:HTopologies}
\end{figure}

  
Next we will consider the $\H$ topology, which can also be embedded in the family of indices in Eqs.~\eqref{eq:TwoLoopPropsFull} and \eqref{eq:TwoLoopPropsSoft} as shown in Fig.~\ref{fig:TopLevelH}, so that the scalar $\H$ integral is $I_{\H} = I_{0,1,1,0,1,1,1,1,1}$. We note that the case of equal masses has been evaluated in Ref.~\cite{Bianchi:2016yiq} without expansion in the soft or potential region.
For this topology we only need the leading contribution in $|q|$, which is even in $|q|$. Therefore we only give the pure basis of ten master integrals needed to express the even-$|q|$ terms\footnote{For reference, in the full equal-mass problem there are 25 master integrals  \cite{Bianchi:2016yiq}.},
\begin{align}
f_{\H,1}={}&\epsilon ^2 (-q^2) G_{0,0,0,0,0,0,1,2,2}\,, \label{eq:HpureBasis1} \\ 
f_{\H,2}={}&\epsilon ^2 (1-4 \epsilon )\, G_{0,0,2,0,1,0,1,1,0}\,,\\ 
f_{\H,3}={}&\epsilon ^2 (-q^2)^2 G_{0,0,0,0,2,1,0,1,2}\,,\\ 
f_{\H,4}={}&\epsilon ^4 (-q^2) G_{0,1,1,0,1,1,0,1,1}\,,\\ 
f_{\H,5}={}&\epsilon ^4 \sqrt{\relfbar^2-1}\, G_{0,1,1,0,0,0,1,1,1}\,,\\ 
f_{\H,6}={}&\epsilon ^3 \sqrt{\relfbar^2-1}\, (-q^2) G_{0,1,1,0,0,0,2,1,1}\,,\\ 
f_{\H,7}={}&-\epsilon ^2 (-q^2) G_{0,2,2,0,0,0,1,1,1} +\epsilon ^3 \relfbar \, (-q^2) G_{0,1,1,0,0,0,2,1,1}\,,\\  
f_{\H,8}={}&\frac{\epsilon ^2 (4 \epsilon -1)}{\sqrt{\relfbar^2-1}}\left[(2 \epsilon -1) G_{0,1,1,0,0,1,1,0,1}+\relfbar\, G_{0,2,0,0,0,1,1,0,1}\right] \,,\\
f_{\H,9}={}&\epsilon ^4\sqrt{\relfbar^2-1}\,  (-q^2)^2 G_{0,1,1,0,1,1,1,1,1}\,,\\ 
f_{\H,10}={}&-\epsilon ^4 (-q^2) G_{-1,1,1,-1,1,1,1,1,1} + \frac{1}{2} \epsilon^2 (2 \epsilon -1)\, G_{0,0,0,0,1,1,0,1,1}\nonumber\\
&+2\epsilon ^4 \relfbar\,  (-q^2) G_{0,1,1,0,1,1,0,1,1} + \epsilon  (3 \epsilon -2) (3 \epsilon -1)\, (-q^2)^{-1} G_{0,0,0,0,1,1,1,0,0}\,.
\label{eq:HpureBasis10}
\end{align}
The corresponding topologies are shown in Fig.~\ref{fig:HTopologies}. In terms of these the soft expansion of the $\H$ integral is simply given by
\begin{equation}\label{eq:Hsoftexp}
  I_{\H} = -\frac{1}{(4\pi)^4}  \left(\frac{1}{\bar\mu^2}\right)^{-2\epsilon} \left\{ \frac1{(-q^2)^2} \frac{1}{\epsilon^4 \bar m_1 \bar m_2  \sqrt{\relfbar^2-1}} f_{\H,9} + \cO\left( (-q^2)^{-3/2-2\epsilon} \right) \right\} \,.
\end{equation}
The differential equations for these master integrals are
\begin{equation}
  \mathrm{d} \vec {f}_{\H}= \epsilon\left[ A^{(e)}_{\H,0} \,\dlog(x)+A^{(e)}_{\H,+1}\,\dlog(x-1)+A^{(e)}_{\H,-1}\,\dlog(x+1) \right]\vec{f}_{\H}\,,
  \label{eq:2loopSoftDEH}
\end{equation}
where we have only kept the even-$|q|$ sector and the matrices are given by
\begin{equation}\label{eq:Hmateven}
  A_{\H,0}^{(\mathrm{e})}={}\left(
	\begin{array}{cccccccccc}
	0 & 0 & 0 & 0 & 0 & 0 & 0 & 0 & 0 & 0 \\
	0 & 0 & 0 & 0 & 0 & 0 & 0 & 0 & 0 & 0 \\
	0 & 0 & 0 & 0 & 0 & 0 & 0 & 0 & 0 & 0 \\
	0 & 0 & 0 & 0 & 0 & 0 & 0 & 0 & 0 & 0 \\
	-\frac{1}{2} & 0 & 0 & 0 & -6 & 0 & -1 & 0 & 0 & 0 \\
	-\frac{3}{2} & 0 & 0 & 0 & 0 & 2 & -2 & 0 & 0 & 0 \\
	0 & 0 & 0 & 0 & 12 & 2 & 0 & 0 & 0 & 0 \\
	0 & 2 & 0 & 0 & 0 & 0 & 0 & 2 & 0 & 0 \\
	2 & -4 & 0 & 0 & 0 & 4 & 2 & 4 & 2 & -2 \\
	-1 & 0 & -1 & 0 & 12 & 8 & 0 & 8 & 2 & -2 \\
	\end{array}
      \right)\,, \quad
	A_{\H,\pm1}^{(\mathrm{e})}={}\left(
	\begin{array}{cccccccccc}
	0 & 0 & 0 & 0 & 0 & 0 & 0 & 0 & 0 & 0 \\
	0 & 0 & 0 & 0 & 0 & 0 & 0 & 0 & 0 & 0 \\
	0 & 0 & 0 & 0 & 0 & 0 & 0 & 0 & 0 & 0 \\
	0 & 0 & 0 & 0 & 0 & 0 & 0 & 0 & 0 & 0 \\
	0 & 0 & 0 & 0 & 6 & 0 & 0 & 0 & 0 & 0 \\
	0 & 0 & 0 & 0 & 0 & -2 & 0 & 0 & 0 & 0 \\
	0 & 0 & 0 & 0 & 0 & 0 & 0 & 0 & 0 & 0 \\
	0 & 0 & 0 & 0 & 0 & 0 & 0 & -2 & 0 & 0 \\
	0 & 0 & 0 & 0 & 0 & -4 & 0 & -4 & -2 & 0 \\
	1 & 0 & 1 & \pm4 & 0 & 0 & 0 & 0 & 0 & 2 \\
	\end{array}
      \right)\,.
\end{equation}

We also need to consider the crossed $\H$, or $\xH$, integral, in Fig.~\ref{fig:twoloopvert}(b), which is just a crossing of the $\xH$ integral by $p_2 \leftrightarrow -p_3$. We note, however, that the $\H$ and $\xH$ integrals appear together in the amplitude, with the same coefficient\footnote{This is even true for the pure gravity amplitude \cite{Bern:2019nnu, Bern:2019crd} with an appropriate alignment of loop momentum labels across the two different diagrams, up to differences that only give quantum corrections.}. Thus we can directly evaluate their sum. Since the crossing $p_1 \leftrightarrow -p_4$ is equivalent to $p_2 \leftrightarrow -p_3$, this can be written in the symmetrized form
\begin{equation}
  I_{\H} + I_{\xH} = \frac 1 2 \left( I_{\H} + I_{\H} \big|_{p_2 \leftrightarrow -p_3} + I_{\H} \big|_{p_1 \leftrightarrow -p_4} + I_{\H} \big|_{p_2 \leftrightarrow -p_3, \, p_1 \leftrightarrow -p_4} \right) \, . \label{eq:HplusXH}
\end{equation}
As mentioned above, we only need to perform the soft expansion of $\H$ and $\xH$ to the leading order, due to the suppression by $t^2 = q^4$ factor in the numerator, and subleading corrections are not relevant classically. The leading soft expansion of Eq.~\eqref{eq:HplusXH} can be obtained from that of the H integral itself by the replacements
\begin{align}
\frac{1}{\rho_2 + \imath0} \rightarrow \frac{1}{\rho_2 + \imath0} + \frac{1}{-\rho_2 + \imath0} &= (-2 \pi \imath) \delta(\rho_2), \\
\frac{1}{\rho_3 + \imath0} \rightarrow \frac{1}{\rho_3 + \imath0} + \frac{1}{-\rho_3 + \imath0} &= (-2 \pi \imath) \delta(\rho_3),
\end{align}
followed by multiplying the resulting expression by $1/2$. Effectively we have ``cut'' the matter propagators and turned them into delta functions. However, we still need to define how to ``cut'' matter propagators raised to higher powers, because integrals with squared matter propagators appear when we construct differential equations, and also appear in our choice of a pure basis of master integrals Eqs.~\eqref{eq:HpureBasis1}--\eqref{eq:HpureBasis10}. An appropriate prescription is
\begin{equation}
  G_{i_1, i_2, i_3 \dots , i_9} \rightarrow G_{i_1,\cutindex{i_2}, \cutindex{i_3}, i_4, \dots , i_9}\,, \label{eq:GtoHatG}
\end{equation}
with the definition
\begin{align}
 G_{i_1,\cutindex{i_2}, \cutindex{i_3}, i_4, \dots , i_9} &= \frac 1 2 \int \frac{\mathrm{d}^D\ell_1}{\imath \pi^{D/2}}  \int \frac{\mathrm{d}^D\ell_2}{\imath \pi^{D/2}}  \frac{1}{\rho_1^{i_1} \rho_4^{i_4} \cdots \rho_9^{i_9}}  \nonumber \\
&\qquad \times \left[
         \frac 1 {(-\rho_2+ \imath0)^{i_2}} - \frac 1 {(-\rho_2 - \imath0)^{i_2}}
       \right]
       \left[  \frac 1 {(-\rho_3 + \imath0)^{i_3}} - \frac 1 {(-\rho_3 - \imath0)^{i_3}}
       \right] \, .
\label{eq:twoloopHtilde}
\end{align}
Here $ G_{i_1,\cutindex{i_2}, \cutindex{i_3}, i_4, \dots , i_9}$ vanishes whenever the integer $i_2$ or $i_3$ is non-positive, because the $\imath0$ prescription is of no relevance in the numerator, and the terms in one of the square brackets of Eq.~\eqref{eq:twoloopHtilde} add to zero. 
The advantage of this prescription is that it preserves IBP relations and the differential equations Eq.~\eqref{eq:2loopSoftDEH}. In particular, the pure basis of master integrals for the H topology, Eqs.~\eqref{eq:HpureBasis1}--\eqref{eq:HpureBasis10} can be mapped to the ``cut'' version
\begin{equation}
  f_{\H, n} \rightarrow f_{\rm c\H, n}, \quad 1 \leq n \leq 10, \label{eq:ftohatf}
\end{equation}
using Eqs.~\eqref{eq:GtoHatG} and \eqref{eq:twoloopHtilde}, and the resulting integrals satisfy differential equations
\begin{equation}\label{eq:diffeqcutH}
  \mathrm{d}{\vec{f}}_{\rm c \H}=\epsilon\left[A_{\rm c\H,0}\,\dlog(x)+A_{\rm c\H,+1}\,\dlog(x-1)+A_{\rm c\H,-1}\,\dlog(x+1)\right]\vec{ f}_{\rm c \H}\,,
\end{equation}
where the matrices, $A_{i,\rm c \H}$, are identical to the ones in Eqs.~\eqref{eq:2loopSoftDEH} and \eqref{eq:Hmateven} for the differential equations of the original uncut $\H$ topology. Hence the solution of the ``cut $\H$'' differential equations will only differ from the full $\H$ in the boundary conditions.

In order to obtain the boundary conditions for the ``cut $\H$'' integrals, we follow a prescription  for performing the energy integrals similar to that in the previous section. In this case, the prescription is simply to carry out the $\omega_1$ integral by residues, and then performing the $\omega_2$ integral by residues too. Each of the two integration steps is done by closing the contour either above or below the real axis, picking up residues from poles at finite values and discarding poles at infinity. We find that the only non-vanishing master integrals in the static limit are $ f_{\rm c \H,4}^{\pot}, f_{\rm c \H,7}^{\pot}$ and $f_{\rm c\H,10}^{\pot}$. The computation of these integrals is explained in Appendix~\ref{sec:3dintegrals}. By expanding up to $\mathcal{O}(\epsilon^4)$ they yield the following vector of boundary condition
\begin{align}
  \left. \vec {f}^{\,\pot}_{\mathrm{c}\H}\right|_{\relfbar=1}={}& (-q^2)^{-2\epsilon} \epsilon^2\pi^2 \left(0,0,0,\frac {\pi^2 \epsilon^2} 2, 0, 0, -\frac 1 2 + \frac{7 \pi^2 \epsilon^2} {12}, 0,0, \pi^2 \epsilon^2\right)^T+\mathcal{O}(\epsilon^5)\,.\label{eq:StaticBNDH}
\end{align}

The result of solving the differential equation in Eq.~\eqref{eq:diffeqcutH} with the boundary conditions in Eq.~\eqref{eq:StaticBNDH} is given in Eqs.~\eqref{eq:fH4pot}--\eqref{eq:fH10pot} in Appendix~\ref{sec:twoloopdetails}. The sum of $\H$ and $\xH$ is given by Eq.~\eqref{eq:Hsoftexp} with the replacement $f_{\H,9}\rightarrow f_{\rm c\H,9}$,
 which using the solution of the differential equation yields
\begin{equation}
  I_{\vphantom{\xH}\H}^{\pot} + I_{\xH}^{\pot} = -\frac{1}{(4\pi)^4} \left(\frac{-q^2}{\bar\mu^2}\right)^{-2\epsilon} \left\{
  \frac1{(-q^2)^2} \left[ \frac{2\pi^2}{ \epsilon} \frac{\text{arcsinh} \sqrt{\frac{\sigma -1}{2}}}{ m_1 m_2\sqrt{\sigma^2-1}} + \cO(\epsilon^0) \right]
  +\cO ((-q)^{-3/2} ) \right \}\,.
  \label{eq:HxHres}
\end{equation}
The part of Eq.~\eqref{eq:HxHres}, proportional to $\log(-q^2)$, which due to the $q^4$ suppression in the numerator is the only piece relevant for the classical dynamics, agrees with the result in Refs.~\cite{Bern:2019nnu,Bern:2019crd}.

\subsubsection{Non-planar double-box ($\IX$)}
\begin{figure}[tb!]  
	\centering
	\begin{subfigure}[b]{0.49\linewidth}
		\centering 
		\includegraphics{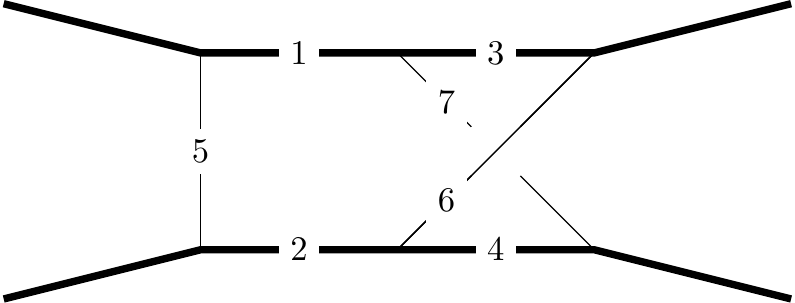}
		\caption{Non-planar double-box topology.} 
	\end{subfigure}
	\caption{Top-level topologies at two-loops. Indices correspond to the propagators listed in eq.~\eqref{eq:XBPropsFull}.}\label{fig:XBTopos}
\end{figure}

\begin{figure}[tb!]  
	\centering
	\begin{subfigure}[b]{0.3\linewidth}
		\centering 
		\includegraphics{./figures/DE_SR}
		\caption{$f_{\IX,1}$} 
	\end{subfigure}
	\begin{subfigure}[b]{0.3\linewidth}
		\centering	\includegraphics{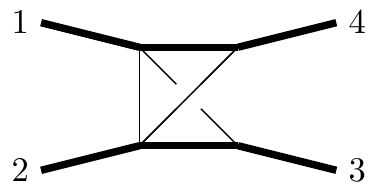}
		\caption{$f_{\IX,2},f_{\IX,3},f_{\IX,4}$}
	\end{subfigure}
	\begin{subfigure}[b]{0.3\linewidth}
		\centering	\includegraphics{./figures/DE_SlashBox}
		\caption{$f_{\IX,5},f_{\IX,6},f_{\IX,7}$}
	\end{subfigure}
	\begin{subfigure}[b]{0.35\linewidth}
		\centering
		\includegraphics{./figures/DE_TriTri}
		\caption{$f_{\IX,8}$}
	\end{subfigure}
	\begin{subfigure}[b]{0.2\linewidth}
		\centering 
		\includegraphics{./figures/DE_BubBoxRight}
		\caption{$f_{\IX,9}$} 
	\end{subfigure}\qquad\qquad
	\begin{subfigure}[b]{0.3\linewidth}
		\centering	\includegraphics{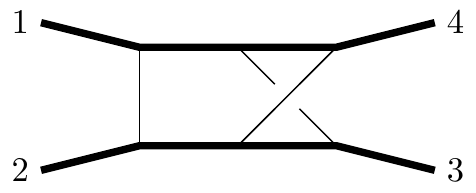}
		\caption{$f_{\IX,10}$}
	\end{subfigure}
	\caption{Even $|q|$ master integrals relevant for the non-planar double-box topology.}\label{fig:XBTopologiesEven}
\end{figure}

\begin{figure}[tbh!]  
	\centering
	\begin{subfigure}[b]{0.3\linewidth}
		\centering 
		\includegraphics{./figures/DE_BubTriRight}
		\caption{$f_{\IX,11}$} 
	\end{subfigure}
	\begin{subfigure}[b]{0.3\linewidth}
		\centering	\includegraphics{./figures/DE_SlashBox}
		\caption{$f_{\IX,12}$}
	\end{subfigure}
	\begin{subfigure}[b]{0.3\linewidth}
		\centering	\includegraphics{./figures/DE_TwistedSlashBox}
		\caption{$f_{\IX,13}$}
	\end{subfigure}
	\begin{subfigure}[b]{0.4\linewidth}
		\centering
		\includegraphics{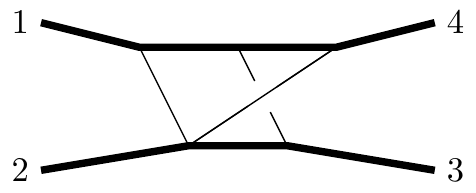}
		\caption{$f_{\IX,14}$}
	\end{subfigure}
	\begin{subfigure}[b]{0.4\linewidth}
		\centering
		\includegraphics{./figures/DE_BoxTriangle}
		\caption{$f_{\IX,15}$}
	\end{subfigure}
	\caption{Odd $|q|$ master integrals relevant for the non-planar double-box topology.}\label{fig:XBTopologiesOdd}
\end{figure}

Next we discuss the non-planar double-box topology. We only consider the $\IX$ topology, noting that the integral $\XI$ is identical. The full integral has been discussed in the equal mass case in Ref.~\cite{Heinrich:2004iq}. We first consider generic integrals of the form
\begin{equation}
I_{i_1,i_2,\dots, i_9}=\int\frac{\mathrm{d}^{D}\ell_1}{(2\pi)^D}\int\frac{\mathrm{d}^{D}\ell_2}{(2\pi)^D}\frac{1}{\rho_1^{i_1}\rho_2^{i_2}\cdots \rho_9^{i_9}}\,.
\end{equation}
Where the propagators are, as depicted in Fig.~\ref{fig:XBTopos}
\begin{alignat}{3}
&\tilde \rho_1=(\ell_1-p_1)^2-m_1^2\,,\qquad\quad &&\tilde \rho_2=(\ell_1+p_2)^2-m_2^2\,,\qquad\quad && \tilde \rho_3=(\ell_2-p_4)^2-m_1^2\,,\nonumber\\
&\tilde \rho_4=(\ell_1+\ell_2-q-p_3)^2-m_2^2\,,\qquad\quad&&\tilde \rho_5=\ell_1^2\,,\qquad\quad&&\tilde \rho_6=\ell_2^2 \,,\nonumber\\
&\tilde \rho_7=(\ell_1 + \ell_2 - q)^2\,,\qquad\quad&&\tilde \rho_8=(\ell_1 - q)^2\,,\qquad\quad&&\tilde \rho_9=(\ell_2 - q)^2\,,\label{eq:XBPropsFull}
\end{alignat}
and the scalar non-planar double-box integral is $I_{\IX} = I_{1,1,1,1,1,1,1,0,0}$.
The small-$|q|$ expansion consists of integrals of the form
\begin{equation}
G_{i_1, \, i_2, \dots , i_9} = \int \frac {\mathrm{d}^D \ell_1 \, e^{\EulerGamma \epsilon}}{\imath \pi^{D/2}} \int \frac {\mathrm{d}^D \ell_2 \, e^{\EulerGamma \epsilon}}{\imath \pi^{D/2}}
\frac{1}{\rho_1^{i_1} \rho_2^{i_2} \dots \rho_9^{i_9}} ,
\label{eq:XBtwoloopGtilde}
\end{equation}
where the leading order parts of the propagators are
\begin{alignat}{3}
&\rho_1=2\, \ell_1\cdot u_1\,,\qquad\qquad &&\rho_2=-2\, \ell_1 \cdot u_2\,,\qquad\qquad && \rho_3=-2\, \ell_2\cdot u_1\,,\nonumber\\
&\rho_4=-2\,(\ell_1+\ell_2)\cdot u_2\,,\qquad\qquad&&\rho_5=\ell_1^2\,,\qquad\qquad&&\rho_6=\ell_2^2 \,,\nonumber\\
&\rho_7=(\ell_1 + \ell_2 - q)^2\,,\qquad\qquad&&\rho_8=(\ell_1 -q)^2\,,\qquad\qquad&&\rho_9=(\ell_2 - q)^2\,.\label{eq:TwoLoopPropsSoftXI}
\end{alignat}
A pure basis of master integrals is given by
\begin{align}
f_{\IX,1}={}&\epsilon ^2(-q^2) G_{0,0,0,0,2,2,1,0,0}\,,\\ 
f_{\IX,2}={}&\epsilon ^4\sqrt{\relfbar^2-1}\,  G_{0,0,1,1,1,1,1,0,0}\,,\\ 
f_{\IX,3}={}&\epsilon ^3(-q^2) \sqrt{\relfbar^2-1}\,  G_{0,0,1,1,2,1,1,0,0}\,,\\ 
f_{\IX,4}={}&\epsilon ^2(-q^2) G_{0,0,2,2,1,1,1,0,0} + \epsilon ^3(-q^2) \relfbar  G_{0,0,1,1,2,1,1,0,0}\,,\\ 
f_{\IX,5}={}&\epsilon ^4\sqrt{\relfbar^2-1}\,   G_{0,1,1,0,1,1,1,0,0}\,,\\ 
f_{\IX,6}={}&\epsilon ^3(-q^2)\sqrt{\relfbar^2-1}\,   G_{0,1,1,0,1,1,2,0,0}\,,\\ 
f_{\IX,7}={}&\epsilon ^2(-q^2) G_{0,2,2,0,1,1,1,0,0} -\epsilon ^3(-q^2) \relfbar \,   G_{0,1,1,0,1,1,2,0,0}\,,\\ 
f_{\IX,8}={}&\epsilon^3(1-6\epsilon) G_{1,0,1,0,1,1,1,0,0}\,,\\ 
f_{\IX,9}={}&\epsilon ^3(-q^2)\sqrt{\relfbar^2-1}\,   G_{1,1,0,0,1,1,2,0,0}\,,\\ 
f_{\IX,10}={}&\epsilon ^4(-q^2) (\relfbar^2-1) G_{1,1,1,1,1,1,1,0,0}\, ,\\
f_{\IX,11}={}&\epsilon ^3\sqrtmQSq G_{1,0,0,0,1,1,2,0,0}\,,\\ 
f_{\IX,12}={}&\epsilon ^3\sqrtmQSq G_{0,2,1,0,1,1,1,0,0}\,,\\ 
f_{\IX,13}={}&\epsilon ^3\sqrtmQSq G_{0,0,2,1,1,1,1,0,0}\,,\\ 
f_{\IX,14}={}&\epsilon ^4\sqrtmQSq\sqrt{\relfbar^2-1}  G_{1,0,1,1,1,1,1,0,0}\,,\\ 
f_{\IX,15}={}&\epsilon ^4\sqrtmQSq\sqrt{\relfbar^2-1}  G_{1,1,1,0,1,1,1,0,0}\,,
\end{align}
where the corresponding topologies are shown in Fig.~\ref{fig:XBTopologiesEven} and Fig.~\ref{fig:XBTopologiesOdd}.
The functions $f_{\IX,1}$ to $f_{\IX,10}$ are even in $|q|$, while $f_{\IX,11}$ to $f_{\IX,15}$ are odd.
The soft expansion of the $\IX$ integral and subsequent IBP reduction gives
\begin{align}
  I_{\IX} &= - \frac{(\bar\mu^2)^{2\epsilon}}{(4\pi)^4} \bigg\{
  \frac{1} {-q^2} \frac{1}{\left(y^2-1\right) \epsilon ^4 \bar{m}_1^2 \bar{m}_2^2} f_{\IX, 10} \nonumber \\
  & \quad + \frac {1} {\sqrt{-q^2}} \bigg[
  \frac{-\bar{m}_1-\bar{m}_2}{2 \left(y^2-1\right) \epsilon ^3 \bar{m}_1^3 \bar{m}_2^3} f_{\IX, 12} \nonumber \\
  & \qquad + \frac{\bar{m}_1+\bar{m}_2}{2 \left(y^2-1\right) \epsilon ^3 \bar{m}_1^3 \bar{m}_2^3} f_{\IX, 13} \nonumber \\
  & \qquad + \frac{\left(\bar{m}_1+\bar{m}_2\right) (4 y \epsilon +y+2 \epsilon )}{2 \left(y^2-1\right)^{3/2} \epsilon ^4 \bar{m}_1^3 \bar{m}_2^3} f_{\IX, 14} \nonumber \\
  & \qquad + \frac{-\bar{m}_1-\bar{m}_2}{(y+1) \sqrt{y^2-1} \epsilon ^3 \bar{m}_1^3 \bar{m}_2^3} f_{\IX, 15} \bigg ] \nonumber \\
  & \quad + \frac{\epsilon  \bar{m}_1^2 \left(3 \left(y^2+1\right) \epsilon +1\right)+\epsilon  \bar{m}_2^2 \left(3 \left(y^2+1\right) \epsilon +1\right)+4 y \left(6 \epsilon ^2+7 \epsilon +1\right) \bar{m}_2 \bar{m}_1}{48 \left(y^2-1\right)^2 \epsilon ^4 \bar{m}_1^4 \bar{m}_2^4} f_{\IX,1} \nonumber \\
  & \quad + \frac{-y \epsilon  \bar{m}_1^2-y \epsilon  \bar{m}_2^2-4 (\epsilon +1) \bar{m}_2 \bar{m}_1}{4 \left(y^2-1\right)^{3/2} \epsilon ^3 \bar{m}_1^4 \bar{m}_2^4} f_{\IX, 2} \nonumber \\
  & \quad + \frac{y \left(12 \epsilon ^2+4 \epsilon -1\right) \bar{m}_1^2+y \left(12 \epsilon ^2+4 \epsilon -1\right) \bar{m}_2^2+2 (2 \epsilon -1) \bar{m}_2 \bar{m}_1}{16 \left(y^2-1\right)^{3/2} \epsilon ^3 (2 \epsilon -1) \bar{m}_1^4 \bar{m}_2^4} f_{\IX,3} \nonumber \\
  & \quad + \frac 1 {48 \left(y^2-1\right)^2 \epsilon ^4 (2 \epsilon -1) \bar{m}_1^4 \bar{m}_2^4}
  \Big( \epsilon  \bar{m}_2^2 \left(\left(12-24 y^2\right) \epsilon ^2+4 \epsilon +1\right) \nonumber \\
  &\qquad +\bar{m}_1^2 \left(\left(12-24 y^2\right) \epsilon ^3+4 \epsilon ^2+\epsilon \right)-4 y \left(12 \epsilon ^3+8 \epsilon ^2-5 \epsilon -1\right) \bar{m}_2 \bar{m}_1 \Big)  f_{\IX, 4} \nonumber \\
  & \quad + \frac{y \epsilon  \bar{m}_1^2+y \epsilon  \bar{m}_2^2+4 (\epsilon +1) \bar{m}_2 \bar{m}_1}{4 \left(y^2-1\right)^{3/2} \epsilon ^3 \bar{m}_1^4 \bar{m}_2^4} f_{\IX,5} \nonumber \\
  & \quad + \frac{y \left(6 \epsilon ^2+3 \epsilon -1\right) \bar{m}_1^2+y \left(6 \epsilon ^2+3 \epsilon -1\right) \bar{m}_2^2+2 \left(4 \epsilon ^2-1\right) \bar{m}_2 \bar{m}_1}{8 \left(y^2-1\right)^{3/2} \epsilon ^3 (2 \epsilon -1) \bar{m}_1^4 \bar{m}_2^4} f_{\IX, 6} \nonumber \\
  & \quad + \frac 1 {48 \left(y^2-1\right)^2 \epsilon ^4 (2 \epsilon -1) \bar{m}_1^4 \bar{m}_2^4}
  \Big( \epsilon  \bar{m}_1^2 \left(12 \left(y^2-2\right) \epsilon ^2+\left(6 y^2-2\right) \epsilon +1\right) \nonumber \\
  & \qquad +\epsilon \bar{m}_2^2 \left(12 \left(y^2-2\right) \epsilon ^2+\left(6 y^2-2\right) \epsilon +1\right)-4 y \left(12 \epsilon ^3+8 \epsilon ^2-5 \epsilon -1\right) \bar{m}_2 \bar{m}_1 \Big) f_{\IX, 7} \nonumber \\
  & \quad + \frac{\epsilon  \bar{m}_1^2 \left(6 \left(y^2-2\right) \epsilon -1\right)+\epsilon  \bar{m}_2^2 \left(6 \left(y^2-2\right) \epsilon -1\right)-4 y \left(6 \epsilon ^2+7 \epsilon +1\right) \bar{m}_2 \bar{m}_1}{48 \left(y^2-1\right)^2 \epsilon ^4 \bar{m}_1^4 \bar{m}_2^4} f_{\IX, 8} \nonumber \\
  & \quad + \frac{-y (2 \epsilon +3) \bar{m}_1^2-y (2 \epsilon +3) \bar{m}_2^2+2 (2 \epsilon -1) \bar{m}_2 \bar{m}_1}{24 \left(y^2-1\right)^{3/2} \epsilon ^3 \bar{m}_1^4 \bar{m}_2^4} f_{\IX,9} \nonumber \\
  & \quad -\frac{4 y (\epsilon +1) \bar{m}_2 \bar{m}_1+(4 \epsilon +3) \bar{m}_1^2+(4 \epsilon +3) \bar{m}_2^2}{12 \left(y^2-1\right)^2 \epsilon ^4 \bar{m}_1^4 \bar{m}_2^4} f_{\IX, 10} \bigg\} \, . \label{eq:IXIBPresult}
\end{align}

The differential equations are
\begin{equation}
  \mathrm{d}\vec{f}_\IX=\epsilon\left[A_{\IX,0}\,\dlog(x)+A_{\IX,+1}\,\dlog(x-1)+A_{\IX,-1}\,\dlog(x+1)\right]\vec{f}_\IX.\label{eq:XB2loopSoftDE}
\end{equation}
The even- and odd-$|q|$ systems decouple and we can write 
\begin{align}
  A_{\IX,i}=\left(
	\begin{array}{cc}
	  A_{\IX,i}^{(\mathrm{e})} & 0 \\
	  0 & A_{\IX,i}^{(\mathrm{o})} \\
	\end{array}
	\right)\,,
\end{align}
where the matrices for the even integrals are given by
\begin{align} \label{eq:IXmateven}
  A_{\IX,0}^{(\mathrm{e})}={}&
\left(
\begin{array}{cccccccccc}
0 & 0 & 0 & 0 & 0 & 0 & 0 & 0 & 0 & 0 \\
\frac{1}{2} & -6 & 0 & -1 & 0 & 0 & 0 & 0 & 0 & 0 \\
\frac{3}{2} & 0 & 2 & -2 & 0 & 0 & 0 & 0 & 0 & 0 \\
0 & 12 & 2 & 0 & 0 & 0 & 0 & 0 & 0 & 0 \\
-\frac{1}{2} & 0 & 0 & 0 & -6 & 0 & 1 & 0 & 0 & 0 \\
-\frac{3}{2} & 0 & 0 & 0 & 0 & 2 & 2 & 0 & 0 & 0 \\
0 & 0 & 0 & 0 & -12 & -2 & 0 & 0 & 0 & 0 \\
0 & 0 & 0 & 0 & 0 & 0 & 0 & 0 & 0 & 0 \\
-\frac{3}{4} & 0 & 0 & 0 & 0 & 0 & 0 & 0 & 0 & 0 \\
0 & 0 & -\frac{1}{2} & 0 & 0 & -1 & 0 & 0 & 1 & 0 \\
\end{array}
\right),
A_{\IX, \pm1}^{(\mathrm{e})}={}\left(
\begin{array}{cccccccccc}
0 & 0 & 0 & 0 & 0 & 0 & 0 & 0 & 0 & 0 \\
0 & 6 & 0 & 0 & 0 & 0 & 0 & 0 & 0 & 0 \\
0 & 0 & -2 & 0 & 0 & 0 & 0 & 0 & 0 & 0 \\
0 & 0 & 0 & 0 & 0 & 0 & 0 & 0 & 0 & 0 \\
0 & 0 & 0 & 0 & 6 & 0 & 0 & 0 & 0 & 0 \\
0 & 0 & 0 & 0 & 0 & -2 & 0 & 0 & 0 & 0 \\
0 & 0 & 0 & 0 & 0 & 0 & 0 & 0 & 0 & 0 \\
0 & 0 & 0 & 0 & 0 & 0 & 0 & 0 & 0 & 0 \\
0 & 0 & 0 & 0 & 0 & 0 & 0 & 0 & 0 & 0 \\
0 & 0 & 0 & 0 & 0 & 0 & 0 & 0 & 0 & 0 \\
\end{array}
\right)\,.
\end{align}
The matrices for the odd integrals are 
\begin{align} \label{eq:IXmatodd}
  A_{\IX,0}^{(\mathrm{o})}=\left(
	\begin{array}{ccccc}
	0 & 0 & 0 & 0 & 0 \\
	0 & -2 & 0 & 0 & 0 \\
	0 & 0 & -2 & 0 & 0 \\
	0 & -1 & 1 & 0 & 0 \\
	0 & 1 & 0 & 0 & 0 \\
	\end{array}
	\right)\,,
	A_{\IX,+1}^{(\mathrm{o})}=\left(
	\begin{array}{ccccc}
	0 & 0 & 0 & 0 & 0 \\
	-3 & -2 & 0 & 0 & 0 \\
	3 & 0 & 6 & 0 & 0 \\
	0 & 0 & 0 & 0 & 0 \\
	0 & 0 & 0 & 0 & 0 \\
	\end{array}
	\right)\,,
	A_{\IX,-1}^{(\mathrm{o})}=\left(
	\begin{array}{ccccc}
	0 & 0 & 0 & 0 & 0 \\
	3 & 6 & 0 & 0 & 0 \\
	-3 & 0 & -2 & 0 & 0 \\
	0 & 0 & 0 & 0 & 0 \\
	0 & 0 & 0 & 0 & 0 \\
	\end{array}
	\right)\,.
\end{align}
We proceed by computing the boundary condition in the static limit analogously to the planar double-box discussed above. As before the integrals in this limit are evaluated using the residue method, yielding three-dimensional integrals tabulated in Appendix~\ref{sec:3dintegrals}. 
Only the functions $f_4,f_7,f_8$ and $f_{15}$ are non-vanishing on the boundary and we have
\begin{align}
\left. \vec {f}^{\,\pot}_{\IX}\right|_{\relfbar=1}={}& (-q^2)^{-2\epsilon} \epsilon^2\pi^2 \left( 0,0,0, -\frac{1}{6} + \frac{7\pi^2 \epsilon^2} {36}, 0,0,\frac{1}{3}-\frac{7\pi^2\epsilon^2}{18},-\frac{1}{6}+\frac{7\pi^2\epsilon^2}{36},0,0,\right.\\
&\qquad\qquad\qquad\left. 0,  0, 0, 0, \frac{\imath \pi \epsilon} 4 -\frac{\imath \pi \log(2) \epsilon^2} {2} \vphantom{\frac{1} {1}}\right)^T+\mathcal{O}(\epsilon^5)\,.\label{eq:StaticBNDIX}
\end{align}
Solving the differential equation
\eqref{eq:XB2loopSoftDE} with the boundary conditions \eqref{eq:StaticBNDIX} up to $\mathcal{O}(\epsilon^4)$ gives the results in Eqs.~\eqref{eq:fIX2}--\eqref{eq:fIX15} in Appendix \ref{sec:twoloopdetails}. Using these results in Eq.~\eqref{eq:IXIBPresult} yields the following result for the non-planar double-box integral $I_{\IX}$,
\begin{align}\label{eq:IXres}
  I_{\IX}^{\pot} = I_{\XI}^{\pot} &= - \frac{1}{(4\pi)^4} \left(\frac{-q^2}{\bar\mu^2}\right)^{-2\epsilon} \Bigg \{ \frac{1}{(-q^2)} \frac{\pi^2}{2 m_1^2 m_2^2 (\relf^2-1)} \left[ - \frac 5 6 \log^2(x) + \cO(\epsilon) \right] \nonumber \\
  &\quad + \frac{1}{\sqrtmQSq} \left[ -\frac{\imath \pi ^3 \left(m_1+m_2\right)}{2 m_1^3 m_2^3 (\relf+1) \sqrt{\relf^2-1}} + \cO(\epsilon) \right] \nonumber \\
  &\quad + (-q^2)^0 \bigg[0 + \cO(\epsilon^0) \bigg] \Bigg \} \, .
\end{align}

\subsubsection{Crossed integrals}
\label{sec:xIntegrals}
In order to evaluate the integrand \eqref{eq:twoloopintegrand} we also need the integrals that that are obtained from $\RT$ and $\IX$  by $p_2\to p_3$ crossing (denoted $\xRT$ and $\xIX$). Since the energy integration step produces non-analytic behavior, these integrals cannot be directly obtained from analytic continuation, and we have to solve the differential equations again. From Eq.~\eqref{eq:xvar}, we can see that $x \rightarrow -x$ corresponds to the change $y \rightarrow -y, \, \sqrt{y^2-1} \rightarrow -\sqrt{y^2-1}$. The differential equations for the crossed integrals are thus obtained from the differential equations for the original integrals, Eqs.~\eqref{eq:RTSoftDE} and \eqref{eq:XB2loopSoftDE}, when we change the LHS by
\begin{equation}
\vec {f}_{\rm T} \rightarrow \vec {f}_{\bar{\rm T}}, \quad u_1^\mu \rightarrow u_1^\mu, \, u_2^\mu \rightarrow -u_2^\mu, \, y \rightarrow -y, \, \sqrt{y^2-1} \rightarrow -\sqrt{y^2-1},
\end{equation}
where $\rm T \in {\RT,\IX}$ denotes the topology, and change the RHS by
\begin{equation}
\log(x)\to\log(x) \,,\quad\log(1-x)\to\log(1+x)\,,\quad \log(1+x)\to\log(1-x)\,.
\end{equation}
The static boundary conditions for $\xRT$ and $\xIX$ integrals are obtained by the same energy integration method covered before, and are explicitly given by
\begin{align}
\left. \vec {f}^{\,\pot}_{\xRT}\right|_{\relfbar=1}={}& (-q^2)^{-2\epsilon} \epsilon^2\pi^2 \left( 0,0,0, -\frac {1}{6} + \frac{7\pi^2 \epsilon^2} {36}, 0, -\frac 1 6 + \frac{7\pi^2 \epsilon^2} {36}, 0, 0, 0, 0 \right)^T+\mathcal{O}(\epsilon^5)\,,\label{eq:StaticBNDxRT}\\
\left. \vec {f}^{\,\pot}_{\xIX}\right|_{\relfbar=1}={}& (-q^2)^{-2\epsilon} \epsilon^2\pi^2 \left( 0,0,0, \frac {1}{3} - \frac{7\pi^2 \epsilon^2} {18}, 0,0, -\frac 1 6 + \frac{7\pi^2 \epsilon^2} {36}, -\frac 1 6 + \frac{7\pi^2 \epsilon^2} {36}, 0, 0,\right.\nonumber\\
&\qquad\qquad\qquad\left.0,0,0,0,0 \vphantom{\frac{1} {1}}\right)^T+\mathcal{O}(\epsilon^5)\,.\label{eq:StaticBNDxIX}
\end{align}
Solving the differential equations obtained by crossing of ~Eqs.~\eqref{eq:RTSoftDE} and \eqref{eq:XB2loopSoftDE} with the boundary conditions \eqref{eq:StaticBNDxRT} and \eqref{eq:StaticBNDxIX} gives the result in Eqs.~\eqref{eq:firstxRT}--\eqref{eq:leadingxRT} and \eqref{eq:firstxRT}--\eqref{eq:leadingxRT} in Appendix \ref{sec:twoloopdetails}. These can be used in the soft expansion of the crossed double-box and non-planar double-box integrals provided in the ancillary file, which gives especially simple final results,
\begin{align}
  I_{\xRT}^{\pot} &= - \frac{1}{(4\pi)^4} \left(\frac{-q^2}{\bar\mu^2}\right)^{-2\epsilon} \Bigg \{ \frac{1}{(-q^2)} \frac{\pi^2}{2 m_1^2 m_2^2 (\relf^2-1)} \left[ - \frac 1 3 \log^2(x) + \cO(\epsilon) \right] \nonumber \\
  &\quad + \frac{1}{\sqrtmQSq} \left[ 0 + \cO(\epsilon) \right] \nonumber \\
  &\quad + (-q^2)^0 \bigg[0 + \cO(\epsilon^0) \bigg] \Bigg \} \, ,
  \label{eq:xRTres}
\end{align}
and
\begin{align}
  I_{\xIX}^{\pot} =  I_{\xXI}^{\pot} &= - \frac{1}{(4\pi)^4} \left(\frac{-q^2}{\bar\mu^2}\right)^{-2\epsilon} \Bigg \{ \frac{1}{(-q^2)} \frac{\pi^2}{2 m_1^2 m_2^2 (\relf^2-1)} \left[ \frac 2 3 \log^2(x) + \cO(\epsilon) \right] \nonumber \\
  &\quad + \frac{1}{\sqrtmQSq} \left[ 0 + \cO(\epsilon) \right] \nonumber \\
  &\quad + (-q^2)^0 \bigg[0 + \cO(\epsilon^0) \bigg] \Bigg \} \, .
\label{eq:xIXres}
\end{align}

 \section{Scattering amplitudes in the potential region} 
\label{sec:amplitudes}

In the previous section we calculated the integrals necessary to evaluate the one- and two-loop conservative amplitudes in the potential region, which we will denote by $M_{4,\pot}$. In this section we will put together the integrals to construct such scattering amplitudes. 

\subsection{Tree-level amplitude}
For completeness, let us start by considering the tree-level amplitude  in Eq.~\eqref{eq:treeresmasslessexchange}. In this case the restriction to the potential region is trivial and we simply have
\begin{equation}
  M_{4,\pot}^{\tree} =  32\pi Gm_1^2m_2^2 (\sigma - \cos\phi)^2 \frac{1}{-q^2}\,,
  \label{eq:treeAmplitude}
\end{equation}
which we have written in a form which will be convenient later.

\subsection{One-loop amplitude}

The one-loop integrand for the conservative black-hole amplitude in $\mathcal N=8$ supergravity is given in terms of the sum of the box and crossed box integrals in the potential region,
\begin{equation}
M_{4,\pot}^{\oneloop} = -\imath(8\pi G)^2\,(s-|m_1+m_2e^{\imath\phi}|^2)^4\, \big(\, I^{\pot}_{\rm II} + I^{\pot}_{\rm X}\,) \,.
  \label{eq:1lintegrandpot}
\end{equation}
From the results in Eqs.~\eqref{eq:IIres} and \eqref{eq:Xres} we find
\begin{align}
I_{\II}^{\pot} + I_{\X}^{\pot} &=\frac{\imath}{(4\pi)^2} \left(\frac{-q^2}{\bar\mu^2}\right)^{-\epsilon}
\bigg\{
\frac{1}{(-q^2)} \frac {\imath \pi} {2 m_1 m_2 \sqrt{\relf^2-1}}
\frac{ e^{\epsilon \EulerGamma} \Gamma (-\epsilon )^2 \Gamma (1 + \epsilon)}{\Gamma (-2 \epsilon )}\nonumber \\
&\quad -  \epsilon \frac{1}{\sqrtmQSq} \frac{ \sqrt{\pi}(m_1 + m_2)}{m_1^2 m_2^2(\relf^2-1) }
\frac{ e^{\epsilon \EulerGamma} \Gamma \left(\frac{1}{2}-\epsilon \right)^2 \Gamma \left(\epsilon+\frac{1}{2}\right)} { \Gamma(1-2 \epsilon )}\nonumber \\
&\quad -  \epsilon\frac{ \imath \pi  \left(m_1^2+m_2^2+2 m_1 m_2 \relf\right)}{8 m_1^3 m_2^3 \left(\relf^2-1\right)^{3/2}}
\frac{ e^{\epsilon \EulerGamma} \Gamma (-\epsilon )^2 \Gamma (1 + \epsilon)}{\Gamma (-2 \epsilon )}
\nonumber \\
&\quad + \mathcal O \left( \sqrtmQSq \right)
\bigg\}  \, . \label{eq:boxplusxbox}
\end{align}
Note that this formula is valid in arbitrary dimension. In particular it agrees with the soft-integrals in Eqs.~(B.36) and (B.40) of Ref.~\cite{Cristofoli:2020uzm}. This reference also calculated the contribution in the potential region at leading order in velocity, which, as expected, did not match the full soft integrals away from the static limit. It is well known that  the contributions of soft and potential region coincide at one loop in $D = 4$, up to differences that are suppressed in the classical limit. Our result shows that this is also true in arbitrary dimensions. As a cross-check we have also calculated the result directly in the soft region, by solving the differential equations for the soft integrals subject to their full boundary conditions without restricting to the potential region, and found agreement to $\mathcal{O}(\epsilon^0)$ for both the $1/(-q^2)$ coefficient and the $1/\sqrtmQSq$ coefficient. Details will be given elsewhere.

With the sum of the boxes at hand we can evaluate the one-loop amplitude \eqref{eq:1lintegrand} with the result 
\begin{align}
  M_{4,\pot}^{\oneloop}={} &64G^2\,m_1^3m_2^3(\sigma-\cos\phi)^4\left(\frac{-q^2}{\bar \mu^2}\right)^{-\epsilon}
\bigg\{
\frac{1}{(-q^2)} \frac {\imath \pi} {2\sqrt{\relf^2-1}}
\frac{e^{\epsilon \EulerGamma}\Gamma (-\epsilon )^2 \Gamma (1 + \epsilon)}{\Gamma (-2 \epsilon )}\nonumber \\
&\qquad\qquad -  \epsilon\frac{1}{\sqrtmQSq} \frac{\sqrt{\pi} (m_1 + m_2)}{m_1 m_2(\relf^2-1) }
\frac{e^{\epsilon \EulerGamma} \Gamma \left(\frac{1}{2}-\epsilon \right)^2 \Gamma \left(\epsilon+\frac{1}{2}\right)} {\Gamma(1-2 \epsilon )}\nonumber \\
&\qquad\qquad -  \epsilon\frac{ \imath \pi  \left(m_1^2+m_2^2+2 m_1 m_2 \relf\right)}{8 m_1^2 m_2^2 \left(\relf^2-1\right)^{3/2}}
\frac{ e^{\epsilon \EulerGamma}\Gamma (-\epsilon )^2 \Gamma (1 + \epsilon)}{\Gamma (-2 \epsilon )} + \mathcal O \left( \sqrtmQSq \right)
\bigg\}\,. \label{eq:1loopAmplitude}
\end{align}

\subsection{Two-loop amplitude}

Next we use the integrals in Section~\ref{sec:twoloop} to assemble the two-loop amplitude.
The two-loop amplitude in the potential region is given by
\begin{align} 
  &M_{4,\pot}^{\twoloop} = (8\pi G)^3 (s-|m_1+m_2e^{\imath\phi}|^2)^4\,  \\
  & \hspace{1cm} \times \bigg[ (s-|m_1+m_2e^{\imath\phi}|^2)^2(I^{\pot}_{\RT} +I^{\pot}_{\XI} + I^{\pot}_{\IX} + I^{\pot}_{\xRT} +I^{\pot}_{\xXI} + I^{\pot}_{\xIX})  + (-q^2)^2 (I^{\pot}_{\rm H} +  I^{\pot}_{\xH}) \bigg]\,.  \nonumber
\label{eq:twoloopintegrandpot}
\end{align}
where the remaining integrals are suppressed in the classical limit. Naively, the double-boxes and crossed double-boxes appear with different prefactor in \eqref{eq:twoloopintegrand}. 
We have 
\begin{equation}
	u-|m_1-m_2e^{\imath\phi}|^2=-s+|m_1+m_2e^{\imath\phi}|^2-q^2\,,
\end{equation}
so the $\mathcal{O}(|q|^2)$ mismatch could in principle combine with the leading order of the crossed double-boxes which are of $\mathcal{O}(|q|^{-2})$. 
The explicit results for this integrals in Eqs.~\eqref{eq:xRTres} and \eqref{eq:xIXres} shows however that these do not contribute to the classical part of the amplitude and all the double-boxes contribute with the same coefficient.
Using Eqs.~\eqref{eq:RTres}, \eqref{eq:IXres}, \eqref{eq:xRTres}, and \eqref{eq:xIXres} the relevant combination of double-box integrals is then
\begin{align}
  &\quad I_{\RT}^{\pot} + I_{\XI}^{\pot} + I_{\IX}^{\pot} + I_{\xRT}^{\pot} + I^{\pot}_{\xXI} + I^{\pot}_{\xIX} \nonumber \\
  &= -\frac{1}{(4\pi)^4}\left(\frac{-q^2}{\bar\mu^2}\right)^{-2\epsilon}
\bigg\{
\frac{1}{(-q^2)} \frac {\pi^2} {2 m_1^2 m_2^2 (\relf^2-1)}
\left[ \frac 1 {\epsilon^2} - \frac{\pi^2} 6 + \mathcal O(\epsilon^1) \right] \nonumber \\
&\quad + \frac{1}{\sqrtmQSq} \frac{\imath \pi^3 (m_1 + m_2)}{m_1^3 m_2^3(\relf^2-1)^{3/2} }
\left[ 1 + \mathcal O(\epsilon^1) \right] \nonumber \\
&\quad - \frac {\pi^2 (m_1^2+m_2^2+2\relf m_1 m_2) } {8 m_1^4 m_2^4 (\relf^2-1)^2}
\left[ \frac 1 {\epsilon} + \mathcal O(\epsilon^0) \right] \nonumber \\
&\quad + \mathcal O \left( \sqrtmQSq \right)
\bigg\}  + \text{analytic terms}\, , \label{eq:twoloopladderstotal}
\end{align}
where ``analytic terms'' stand for terms with polynomial (including constant) dependence on $q^2$, with or without poles in $\epsilon$. Such analytic terms give contact terms after Fourier transform to impact parameter space, and are irrelevant for long-range classical physics. Note that the classical $\log(-q^2)$ arises from the Taylor expansion of $(-q^2)^{-2 \epsilon}$.
With these, together with the $\H$-type integrals in Eq.~\eqref{eq:HxHres}, we can evaluate the conservative two-loop amplitude 
\begin{align}
  M_{4,\pot}^{\twoloop}={} &-32 \pi G^3\,m_1^4m_2^4(\sigma-\cos\phi)^4\left(\frac{-q^2}{\bar\mu^2}\right)^{-2\epsilon}
\bigg\{
\frac{1}{(-q^2)} \frac {2(\sigma-\cos\phi)^2} { (\relf^2-1)}
\left[ \frac 1 {\epsilon^2} - \frac{\pi^2} 6 \right] \nonumber \\
&\quad + \frac{1}{\sqrtmQSq} \frac{4\imath \pi (m_1 + m_2)(\sigma-\cos\phi)^2}{m_1 m_2(\relf^2-1)^{3/2} }
 \nonumber \\
&\quad -\frac{1}{\epsilon}\left[ \frac { (m_1^2+m_2^2+2\relf m_1 m_2)(\sigma-\cos\phi)^2 } {2 m_1^2 m_2^2 (\relf^2-1)^2}
  -2\frac{\operatorname{arcsinh}\left(\sqrt{\frac{\relf-1}{2}}\right)}{m_1m_2\sqrt{\sigma^2-1}}
\right] \nonumber \\
&\quad + \mathcal O \left( \sqrtmQSq \right)
\bigg\}+ \text{analytic terms}\,.  \label{eq:2loopAmplitude}
\end{align}

 \section{Eikonal phase, scattering angle and graviton dominance}
\label{sec:eikonal}

In this section we will study eikonal exponentiation of the conservative amplitudes directly in momentum space. We will check the exponentiation of the leading and subleading eikonal in the two-loop amplitude. Then we will use the eikonal phase to evaluate the scattering angle in $\mathcal{N}=8$ supergravity. Finally we will compare the high-energy limit of our result to that of Einstein gravity.

\subsection{The eikonal phase in $\cN = 8$ supergravity}

In traditional treatments of eikonal exponentiation, it is customary to Fourier transform the scattering amplitudes to impact parameter space in order to extract the eikonal phase. Here we will take a slightly different approach and study eikonal exponentiation directly in momentum space. There is a simple reason why we prefer this approach: First, in the presence of a Coulomb-like tree-level interaction, such as graviton exchange, the Fourier transform has the side effect of introducing an additional infrared divergence, which in dimensional regularization gives the appearance that one needs to carefully analyze the scattering amplitude at $\cO(\epsilon)$ and keep track of $\epsilon/\epsilon$ contributions to extract the eikonal phase at a fixed order. The momentum space approach has the advantage that the Coulomb-like singularities directly cancel, making clear that the $\cO(\epsilon)$ pieces of the $L$-loop amplitude cannot  contribute to the $L$-loop phase.\footnote{Unfortunately, one still needs to calculate $\cO(\epsilon)$ parts of the lower loop amplitudes to extract the phase at a given order.}  Working in momentum space comes at a cost nevertheless: simple products in impact parameter space become convolutions in momentum space. However, all convolutions can be easily evaluated as they are equivalent to iterated bubble integrals.

As usual in the eikonal approach, we will consider the amplitude as a function of a $D-2$-dimensional vector, $\vect q_{\perp}$, transverse to the scattering plane, which has the same magnitude as the four-momentum exchange, i.e., $\vect q_{\perp}^2 = - q^2$ (see e.g. Ref.~\cite{Amati:1990xe}). The conservative amplitude only depends on powers of $q$, so this poses no problem. The statement of eikonal exponentiation is that one can write the scattering amplitude in the potential region as a convolutional exponential of the eikonal phase
\begin{align}
  \imath M_{\pot}(\sigma,\vect q_\perp) &= \text{cexp}\left(\imath \delta(\sigma,\vect q_\perp)\right) - 1  \\
  &\coloneqq  \imath \delta(\sigma,\vect q_\perp) - \frac{1}{2!}  \delta(\sigma,\vect q_\perp) \otimes  \delta(\sigma,\vect q_\perp) -\imath  \frac{1}{3!} \delta(\sigma,\vect q_\perp) \otimes  \delta(s,\vect q_\perp) \otimes  \delta(\sigma,\vect q_\perp) + \cdots \nonumber
\end{align}
where we defined the convolution as integral over the $D-2$ dimensional transverse space
\begin{equation}
  f_1(\vect q_\perp) \otimes f_2(\vect q_\perp) = \frac1N\int \frac{\mathrm{d}^{D-2}\vect\ell_\perp}{(2\pi)^{D-2}}\, f_1(\vect\ell_\perp)\,f_2(\vect q_\perp- \vect\ell_\perp)\,,
  \label{eq:convolution}
\end{equation}
with a normalization factor $N=4 m_1 m _2 \sqrt{\sigma^2-1}$. Equivalently, one can write the inverse relation,
\begin{align}
  \delta(\sigma,\vect q_\perp) &= -\imath \, \text{c}\!\log(1+\imath M_{\pot}(\sigma,\vect q_\perp))  \nonumber\\
  &= M_{\pot}(\sigma,\vect q_\perp)  -\frac {\imath}{2} M_{\pot}(\sigma,\vect q_\perp) \otimes M_{\pot}(\sigma,\vect q_\perp) \\
  &\hspace{1cm}-\frac13 M_{\pot}(\sigma,\vect q_\perp) \otimes M_{\pot}(\sigma,\vect q_\perp) \otimes M_{\pot}(\sigma,\vect q_\perp) + \cdots \nonumber
\end{align}
We expand $\delta$  perturbatively
\begin{align}
  \delta &= \delta^{(0)} + \delta^{(1)} + \delta^{(2)} + \cdots
\end{align}
where $\delta^{(L)}$ is $\cO(G^{L+1})$. Then, from the discussion above we can write the phase in terms of the amplitudes 
\begin{align}\label{eq:Mtodelta0}
  \delta^{(0)} &= M^\tree_{4,\pot}\,,\\
  \delta^{(1)} &= M^\oneloop_{4,\pot} - \frac{\imath}{2} M^\tree_{4,\pot}\otimes  M^\tree_{4,\pot}\,,\\
  \delta^{(2)} &= M^\twoloop_{4,\pot} - \imath M^\tree_{4,\pot}\otimes  M^\oneloop_{4,\pot} - \frac1{3} M^\tree_{4,\pot}\otimes  M^\tree_{4,\pot}\otimes  M^\tree_{4,\pot}\,. \label{eq:Mtodelta2}
\end{align}
Looking at Eqs.~\eqref{eq:treeAmplitude}-\eqref{eq:2loopAmplitude}, we see that to calculate the right-hand side of these equations we need the following convolutions
\begin{align}
  &\frac{1}{\vect q_\perp^2} \otimes \frac{1}{\vect q_\perp^2} 
  =  \frac{1}{N} \frac{1}{4\pi \vect q_\perp^2} \left(\frac{\vect q_\perp^2}{\bar\mu^2}\right)^{-\epsilon} \frac{e^{\epsilon \EulerGamma}\Gamma(-\epsilon)^2\Gamma(1+\epsilon)}{\Gamma(-2\epsilon)} \,,
  \label{eq:conv1}\\
  &\frac{1}{\vect q_\perp^2} \otimes \frac{1}{(\vect q_\perp^2)^{1+\epsilon}} 
  = \frac{1}{N} \left(\frac{\vect q_\perp^2}{\bar\mu^2}\right)^{-2\epsilon} \left(\frac{e^{\EulerGamma}}{4\pi}\right)^\epsilon \left[-\frac1\epsilon \frac{3}{8\pi \vect q_\perp^2} + \cO(\epsilon)\right]\,,
  \label{eq:conv2}\\
  &\frac{1}{\vect q_\perp^2} \otimes \frac{1}{\vect q_\perp^2} \otimes \frac{1}{\vect q_\perp^2} 
  = \frac{1}{N^2} \left(\frac{\vect q_\perp^2}{\bar\mu^2}\right)^{-2\epsilon}\left[\frac1{\epsilon^2} \frac{3}{16\pi^2 \vect q_\perp^2} - \frac1{32  \vect q_\perp^2}+ \cO(\epsilon)\right] \,,
  \label{eq:conv3}\\
  &\frac{1}{\vect q_\perp^2} \otimes \frac{1}{(\vect q_\perp^2)^{\frac12+\epsilon}} 
  = \frac{1}{N} \left(\frac{\vect q_\perp^2}{\bar\mu^2}\right)^{-2\epsilon} \left(\frac{e^{\EulerGamma}}{4\pi}\right)^\epsilon \left[-\frac1\epsilon \frac{1}{4\pi |\vect q_\perp|} + \frac{\log (2)}{\pi |\vect q_\perp|} + \cO(\epsilon)\right]\,,
  \label{eq:conv4}\\
  &\frac{1}{\vect q_\perp^2} \otimes \frac{1}{(\vect q_\perp^2)^{\epsilon}} 
  = \frac{1}{N} \left(\frac{\vect q_\perp^2}{\bar\mu^2}\right)^{-2\epsilon} \left(\frac{e^{\EulerGamma}}{4\pi}\right)^\epsilon \left[-\frac1\epsilon \frac{1}{8\pi} + \cO(\epsilon)\right]\,,
  \label{eq:conv5}
\end{align}
which can all be evaluated by using Eq.~\eqref{eq:SmirnovMasterTriangle} in Appendix~\ref{sec:3dintegrals} with $c=0$ and $\epsilon \rightarrow \epsilon - 1/2$.
Using the first convolution, Eq.~\eqref{eq:conv1}, we find
\begin{equation}\label{eq:firstMconv}
  - \frac{\imath}{2} M^\tree_{4,\pot}\otimes  M^\tree_{4,\pot} =  -\imath 32\pi G^2 m_1^3m_2^3\left(\frac{\vect q_\perp^2}{\bar\mu^2}\right)^{-\epsilon} \frac1{\vect q_\perp^2} \frac{(\sigma - \cos\phi)^4}{\sqrt{\sigma^2-1}} \frac{e^{\epsilon \EulerGamma}\Gamma(-\epsilon)^2\Gamma(1+\epsilon)}{\Gamma(-2\epsilon)}\,,
\end{equation}
which exactly cancels the $\cO(|q|^{-2})$ of the one-loop amplitude in Eq.~\eqref{eq:1loopAmplitude} to all orders in $\epsilon$.
Similarly, using Eqs.~\eqref{eq:conv2} and \eqref{eq:conv3} we find\footnote{Note that exponentiation at one loop implies $M^\oneloop_{4,\pot}\Big|_{\cO(|q|^{-2})} =  \frac{\imath}{2} M^\tree_{4,\pot} \otimes  M^\tree_{4,\pot}$, so the first line can also be written as $ \frac1{3!} M^\tree_{4,\pot} \otimes  M^\tree_{4,\pot} \otimes  M^\tree_{4,\pot}$.}
\begin{align}
  & - \imath M^\tree_{4,\pot}\otimes  M^\oneloop_{4,\pot}\Big|_{\cO(|q|^{-2})} - \frac1{3} M^\tree_{4,\pot}\otimes  M^\tree_{4,\pot}\otimes  M^\tree_{4,\pot} 
  \\& \hspace{4cm}= 64 \pi G^3 m_1^4m_2^4 \frac{ (\sigma - \cos\phi)^4}{\sigma^2-1} \left(\frac{\vect q_\perp^2}{\bar\mu^2}\right)^{-2\epsilon} \frac1{\vect q_\perp^2} \left[ \frac1{\epsilon^2} - \frac{\pi^2}{6} + \cO(\epsilon)\right]\,. \nonumber
\end{align}
Using Eq.~\eqref{eq:conv4}, we find
\begin{align}
  - \imath M^\tree_{4,\pot}\otimes  M^\oneloop_{4,\pot}\Big|_{\cO(|q|^{-1})} &=  - 128\imath \pi^2 G^3 m_1^3m_2^3(m_1+m_2) \\
  &\hspace{2cm}\times \frac{ (\sigma - \cos\phi)^6}{(\sigma^2-1)^{3/2}} \left(\frac{\vect q_\perp^2}{\bar\mu^2}\right)^{-2\epsilon} \frac1{|\vect q_\perp|} \bigg[ 1 + \cO(\epsilon)\bigg] \,. \nonumber
\end{align}
Using Eq.~\eqref{eq:conv5}, we find
\begin{align}\label{eq:lastMconv}
  - \imath M^\tree_{4,\pot}\otimes  M^\oneloop_{4,\pot}\Big|_{\cO(|q|^0)} &=  - 16 \pi G^3 m_1^2m_2^2 (2m_1m_2\sigma + m_1^2+m_2^2) \\
  &\hspace{2cm}\times \frac{ (\sigma - \cos\phi)^6}{(\sigma^2-1)^{2}} \left(\frac{\vect q_\perp^2}{\bar\mu^2}\right)^{-2\epsilon} \bigg[ \frac{1}{\epsilon} + \cO(\epsilon)\bigg]\,. \nonumber
\end{align}
These expressions respectively cancel the $\cO(|q|^{-2})$, the $\cO(|q|^{-1})$ and the $\cO(|q|^0)$ contributions to the two-loop amplitude Eq.~\eqref{eq:2loopAmplitude}, which arise from the double-box-type diagrams. Therefore, the double-box-type diagrams at two loops give exactly zero contribution to the eikonal exponent, up to the order of $q$ relevant for classical dynamics at $\cO(G^3)$. This cancellation is a check of the exponentiation of the leading and subleading eikonal phase in the two-loop amplitude. Henceforth we will assume exponentiation of the two-loop phase and leave a proof for further work. We note that this zero result relies on delicate cancellations between \emph{all six} double-box diagrams which leave only the contributions of the $\H$-type diagrams to the two-loop eikonal phase.

In summary, putting together Eqs.~\eqref{eq:treeAmplitude}--\eqref{eq:2loopAmplitude} and \eqref{eq:firstMconv}--\eqref{eq:lastMconv} in \eqref{eq:Mtodelta0}--\eqref{eq:Mtodelta2} the result of calculation our of the eikonal phase is
\begin{align}
  \delta^{(0)}(\sigma, \vect q_\perp) &= 32\pi Gm_1^2m_2^2 (\sigma - \cos\phi)^2 \frac{1}{\vect q_\perp^2}\,, \\
\delta^{(1)}(\sigma, \vect q_\perp) &= 0 + \cO(\epsilon|\vect q_\perp|^0)\,,\\
\delta^{(2)}(\sigma, \vect q_\perp) &=  -64\pi (Gm_1m_2)^3 \frac{(\sigma - \cos\phi)^4}{\sqrt{\sigma^2-1}}  \text{arcsinh} \sqrt{\frac{\sigma -1}{2}}  \frac{1}{\epsilon} \left(\frac{\vect q_\perp^2}{\bar\mu^2}\right)^{-2\epsilon} + \cO(\epsilon^0 |\vect q_\perp|) \,.
\end{align}
Note that $\delta^{(2)}$ includes an $\cO(\epsilon^0 |\vect q_\perp|)$ which we have not calculated. This however goes beyond the classical power counting and so is a quantum correction to the phase.
Finally, we can readily perform the Fourier transform to obtain the more familiar eikonal phase in impact parameter space
\begin{equation}
  \delta(\sigma, \vect b_\mathrm{e}) = \frac1N\int \frac{\mathrm{d}^{D-2}\vect q_\perp}{(2\pi)^{D-2}}\, e^{\imath\vect b_\mathrm{e} \cdot \vect q_\perp}\, \delta(\sigma, \vect q_\perp)\,,
\end{equation}
with the result
\begin{align}
  \delta^{(0)}(\sigma, \vect b_\mathrm{e}) &= -2 Gm_1m_2 \frac{(\sigma - \cos\phi)^2}{\sqrt{\sigma^2-1}} \left(\frac1{\epsilon} + \log \vect b_\mathrm{e}^2\right)\,,  \\
   \delta^{(1)}(\sigma, \vect b_\mathrm{e}) &= 0\,, \\
   \delta^{(2)}(\sigma, \vect b_\mathrm{e}) &= -32 G^3m_1^2m_2^2 \frac{(\sigma - \cos\phi)^4}{\sigma^2-1} \text{arcsinh} \sqrt{\frac{\sigma -1}{2}} \frac{1}{\vect b_\mathrm{e}^2}\,,
\end{align}
where we have dropped $\cO(\epsilon)$ and quantum parts. As a cross-check we have verified that the same result is obtained by using the more common approach in which one directly transforms the amplitudes to impact parameter space.

\subsubsection*{Soft vs. potential and exponentiation}
Let us stress that it was very important that we evaluated the amplitude in the potential region to extract the conservative piece.
For the one-loop amplitude, the expansion in the soft region differs from that of the potential region at $\cO(\epsilon \, |\bm q|^0)$, which, in addition to the $\epsilon$ suppression, is a quantum correction since the classical dynamics arises from $\cO(1/|\bm q|)$ terms.
For the two-loop amplitude, however, the two expansions still differ from each other at $\cO(|\bm q|^0)$, which is at the same order as the terms responsible for the classical dynamics at two loops, and the difference is also no longer suppressed by $\epsilon$.
In fact, when we directly evaluate the integrals in the soft region at two loops we find non-exponentiating effects which cause infrared divergences that are not canceled by either matching to non-relativistic EFT or by extracting the eikonal exponent, signaling the appearance of contributions that cannot be interpreted as arising from a conservative potential.\footnote{This is reminiscent of the situation in the EFT formulation of the Regge limit of massless scattering \cite{Rothstein:2016bsq}, where contributions from the Glauber region exponentiate while the full soft regions contain non-exponentiating effects.} The evaluation of the soft integrals using the differential equations above and a detailed discussion of this point will be presented elsewhere.

\subsection{Scattering angle from eikonal phase}

Let us now calculate the gravitational scattering angle from the eikonal phase. The formula relating the two can be derived from the stationary phase approximation of the Fourier transform of the exponentiated impact-parameter amplitude back to momentum space \cite{Amati:1987uf}, which yields the relation
\begin{equation}
  \vect q   =-  \frac{\partial }{\partial \vect b_\mathrm{e}} \delta(\sigma,\vect b_\mathrm{e})\,.
  \label{eq:qdelta}
\end{equation}
The magnitude of $\vect q$ is related to the scattering angle $\chi$ and the magnitude of the three-momentum $\vect p$ in the center of mass by
\begin{equation}
  |\vect q| = 2 |\vect p| \sin\frac\chi 2\,,
\label{eq:qangle}
\end{equation}
where in terms of the center of mass energy $E=\sqrt{s}$ and/or $\sigma$
\begin{equation}
  |\vect p| = \frac{m_1m_2\sqrt{\sigma^2-1}}{E} = \frac{m_1m_2\sqrt{\sigma^2-1}}{\sqrt{m_1^2 + m_2^2 + 2m_1m_2\sigma}}\,.
\end{equation}
From Eqs.~\eqref{eq:qdelta} and \eqref{eq:qangle} we can derive the formula for the scattering angle
\begin{equation}
  \sin\frac\chi 2 = - \frac{1}{2 |\vect p|} \frac{\partial }{\partial |\vect b_\mathrm{e}|} \delta(\sigma,\vect b_\mathrm{e})\,.
\end{equation}
Using this formula we find the following result for the scattering angle
\begin{equation}
  \sin\frac\chi 2 = { G m_1 m_2 \over  |\vect p| |\vect b_\mathrm{e}| }   \frac{2 (\sigma - \cos \phi)^2}{\sqrt{\sigma^2-1}} -\frac{G^3 m_1^3 m_2^3}{|\vect p|^3 |\vect b_\mathrm{e}|^3} \frac{32 m_1 m_2 (\sigma - \cos \phi)^4}{m_1^2 + m_2^2 + 2m_1m_2\sigma} \text{arcsinh} \sqrt{\frac{\sigma -1}{2}}\,.
\end{equation}
or separating the different orders
\begin{align}
  \chi^{\rm 1PM}_{\rm eik} &=  { G m_1 m_2 \over  |\vect p| |\vect b_\mathrm{e}| }  \frac{4 (\sigma - \cos \phi)^2}{\sqrt{\sigma^2-1}}  \,,\\
  \chi^{\rm 2PM}_{\rm eik} &= 0\,,\\
  \chi^{\rm 3PM}_{\rm eik} &=   -\frac{G^3 m_1^3 m_2^3}{|\vect p|^3 |\vect b_\mathrm{e}|^3} 16 
\biggl[-\frac{(\sigma - \cos \phi)^6}{6(\sigma^2-1)^{3/2}}  + \frac{4m_1m_2 (\sigma - \cos \phi)^4}{m_1^2 + m_2^2 + 2m_1m_2\sigma} \text{arcsinh} \sqrt{\frac{\sigma -1}{2}}        
\biggr] \,.
\end{align}
Looking ahead, in order to more easily to compare with the results from EFT in the next section, we will write the formula in terms of the angular momentum, $J$. The angular momentum is defined as 
\begin{equation}
  J = | \vect b \times \vect p | =|\vect b| |\vect p|~,
  \label{eq:lb}
\end{equation}
where $\vect b$ is an impact parameter perpendicular the incoming center of mass momentum $\vect p$. This is however not the impact parameter, $\vect b_\mathrm{e}$, which arises naturally from the eikonal phase. Eq.~\eqref{eq:qdelta} shows that $\vect b_\mathrm{e}$ points in the direction of the momentum transfer. The magnitude of $\vect b$ and $\vect b_\mathrm{e}$ are then related by 
\begin{equation}
  |\vect b| = |\vect b_\mathrm{e}| \cos\frac{\chi}{2}\,,
\end{equation}
so that the angular momentum is
\begin{equation}
  J =|\vect b_\mathrm{e}| |\vect p|  \cos\frac{\chi}{2}\,.
  \label{eq:lbe}
\end{equation}
For small angle scattering $|\vect b| \sim |\vect b_\mathrm{e}|$, and the difference is unimportant at leading order. Our results, however, go beyond the leading order and the difference matters. Using the relation \eqref{eq:lbe} we find the scattering angle in terms of the angular momentum
\begin{align}
  \chi^{\rm 1PM}_{\rm eik} &=  { G m_1 m_2 \over  J }  \frac{4 (\sigma - \cos \phi)^2}{\sqrt{\sigma^2-1}}  \,,\\
  \chi^{\rm 2PM}_{\rm eik} &= 0~,\\
  \chi^{\rm 3PM}_{\rm eik} &=   -\frac{G^3 m_1^3 m_2^3}{J^3} 16 
\biggl[\frac{(\sigma - \cos \phi)^6}{3(\sigma^2-1)^{3/2}}  + \frac{4m_1m_2 (\sigma - \cos \phi)^4}{m_1^2 + m_2^2 + 2m_1m_2\sigma} \text{arcsinh} \sqrt{\frac{\sigma -1}{2}}        
\biggr] \,.
\end{align}
For later convenience we can rewrite this in terms of the total mass, $m$, and symmetric mass ratio, $\nu$,
\begin{align}
  m &= m_1 + m_2\,,\quad \nu = \frac{m_1m_2}{(m_1 + m_2)^2}\,,
\end{align}
as follows,
\begin{align}
  \chi^{\rm 1PM}_{\rm eik} &=  { G m^2 \nu \over  J }  \frac{4 (\sigma - \cos \phi)^2}{\sqrt{\sigma^2-1}}  \,, \label{eikangle1}\\
  \chi^{\rm 2PM}_{\rm eik} &= 0~, \label{eikangle2}\\
  \chi^{\rm 3PM}_{\rm eik} &=   -\frac{G^3 m^6 \nu^3}{J^3} 16 
\biggl[\frac{(\sigma - \cos \phi)^6}{3(\sigma^2-1)^{3/2}}  +  \nu \frac{4 (\sigma - \cos \phi)^4}{2(\sigma-1)\nu+1} \text{arcsinh} \sqrt{\frac{\sigma -1}{2}}        
  \biggr] \,. \label{eikangle3}
\end{align}

\subsubsection*{Probe limit}
As a cross-check we can compare the probe limit $\nu\rightarrow 0$ of our result with the scattering angle of a particle of mass $\mu$ moving along geodesics in the background of the half-BPS black hole of mass $M$ \cite{Andrianopoli:1997wi,Arcioni:1998mn}. Ref.~\cite{Caron-Huot:2018ape} studied the precession of the periastron, which is given by 
\begin{equation}
  \frac12 \Delta\Phi = \int_{r_{\mathrm{min}}}^{r_{\mathrm{max}}} \mathrm{d}r \, \frac{\mathrm{d}\chi}{\mathrm{d}r}  =  J \int_{r_{\mathrm{min}}}^{r_{\mathrm{max}}} \frac{\mathrm{d}r}{r^2  \sqrt{p_r(r)^2}}\,,
\end{equation}
where $p_r$ is the radial momentum of the probe particle, related to its three-momentum by $\vect p_{\rm p}^2 = p_r^2 + J^2/r^2$. The scattering angle is given by the same integral with different limits
\begin{equation}
  \frac12 (\chi + \pi) =   J \int_{r_{\mathrm{min}}}^{\infty}  \frac{\mathrm{d}r}{r^2  \sqrt{p_r(r)^2}}\,,
\end{equation}
so their calculation can be easily adapted to obtain this quantity. Let us spare the details to the reader and just give the result
\begin{align}
  \frac12 \chi_{\rm p}  &= \operatorname{arctan} \left[\frac{G M^2 \nu_{\rm p} }{J}\frac{2 (\relf_{\rm p} - \cos\phi_{\rm p})^2}{(\relf_{\rm p}^2-1)^{1/2}}\right] \\
  &\hspace{1.5cm}= \frac{G M^2 \nu_{\rm p} }{J}\frac{ 4 (\relf_{\rm p} - \cos\phi_{\rm p})^2}{(\relf_{\rm p}^2-1)^{1/2}} - \frac{G^3 M^6 \nu_{\rm p}^3 }{J^3}\frac{ 16 (\relf_{\rm p} - \cos\phi_{\rm p})^6}{3(\relf_{\rm p}^2-1)^{3/2}}\,, \nonumber
\end{align}
where $\relf_{\rm p}$ and $\phi_{\rm p}$ are the relativistic factor and charge misalignment of the probe particle respectively, and $\nu_{\rm p} = \mu/M$. Interestingly the structure of the result is the same of that for a Newtonian potential (see e.g Ref.~\cite{Kalin:2019rwq}, Eq.~(4.34)), which could be expected from the fact that Ref.~\cite{Caron-Huot:2018ape} found no precession. Finally, it is easy to check that with the identifications
\begin{equation}
  \sigma \leftrightarrow \sigma_{\rm p}\,,\quad \phi \leftrightarrow \phi_{\rm p}\,,\quad M\leftrightarrow m\,, \quad \nu \leftrightarrow \nu_{\rm p} \,,
\end{equation}
this matches Eqs.~\eqref{eikangle1}--\eqref{eikangle3} in the limit $\nu\rightarrow 0$, in which the term with the $\operatorname{arcsinh}$ is suppressed by its coefficient, thus providing a check of our result.

\subsection{High-energy limit and graviton dominance}

At this point we would like to compare our result for the scattering angle with that of Einstein gravity obtained in Refs.~\cite{Bern:2019nnu,Bern:2019crd}. Famously, the high-energy limit of scattering amplitudes in a theory with gravity is dominated by the exchange of gravitons \cite{tHooft:1987vrq}. This is proven at leading order in $G m_1 m_2 /J$ but not beyond that. Recently, in Ref.~\cite{Bern:2020gjj}, a similar result was found by explicit calculation at order $G^3$ for the case of massless scattering. Although a general proof of graviton dominance at this order is lacking, this reference calculated from first principles the scattering angle for $\cN \leq 4$ supergravity and Einstein gravity using eikonal and partial wave techniques, and found that it coincides in all such theories.\footnote{In massless theories the classical limit and the high-energy limit are not distinct, so the full classical angle agrees.} In addition, the result for Einstein gravity was found to agree with an earlier result by Amati, Ciafaloni and Veneziano \cite{Amati:1990xe} and contradicts a modified proposal by Damour \cite{Damour:2019lcq}. 

Motivated by the universality in the massless case, we will study the high-energy limit of our result by taking $\sigma\rightarrow \infty$ in our result for the scattering angle at order $G^3$, which yields 
\begin{equation}
  \chi^{\rm 3PM}_{\cN=8} \overset{\sigma\rightarrow \infty}{=} -\frac{16 G^3 m^6 \nu ^3 \sigma ^3 \log (\sigma )}{J^3} + \cdots\,.
\end{equation}
This can be compared with the high-energy limit of the Einstein gravity result in Eq.~(11.32) of Ref.~\cite{Bern:2019crd},
\begin{equation}
  \chi^{\rm 3PM}_{\rm EG} \overset{\sigma\rightarrow \infty}{=} -\frac{16 G^3 m^6 \nu ^3 \sigma ^3 \log (\sigma )}{J^3} + \cdots\,,
\end{equation}
finding perfect agreement. This strongly suggests that the coefficient of the $\text{arcsinh}$ term features graviton dominance, and universality also holds in the case of massive scattering. Note that this result does not trivially follow from the massless one since here we impose the limits $J\gg 1$ and then $\sigma\gg1$ in this order (or equivalently $|\bm q| \ll m$). The limits do not commute, so the high energy limit of classical massive scattering is distinct from the Regge limit of massless scattering. 
Admittedly, our calculation provides is only one point of comparison with Einstein gravity, so the question of graviton dominance merits further investigation, either by calculating the scattering angle in other supergravity theories or by directly proving universality. We leave this for future work.
 \section{Consistency check from effective field theory}
\label{sec:eft}
In this section we will calculate the conservative amplitudes using the non-relativistic integration method of Refs.~\cite{Cheung:2018wkq,Bern:2019nnu,Bern:2019crd}, which is optimized for EFT matching. This method avoids explicit computation of infrared divergent integrals in dimensional regularization, by canceling such integrals between the full theory and the effective field theory using a four-dimensional matching procedure. We will use the EFT Hamiltonian to calculate the scattering angle solving the classical dynamics. Finally, we will compare to our predictions for the amplitude and the angle from the previous section.

The EFT is defined in the center of mass frame 
\begin{align}
  p_1 = (-E_1,   \vect p )\,, \quad
  p_2 = (-E_2,  -\vect p )\,, \quad
  p_3 = (E_2 ,   \vect p')\,, \quad
  p_4 = (E_1 ,  -\vect p')\,, \quad
\end{align}
where the magnitude of the three-momenta, $|\vect p|=|\vect p'|$, is unchanged in the scattering and the energies are $E_i = \sqrt{m_i^2+\vect p^2}$. In this frame the momentum transfer is purely spatial and given by $\vect q = \vect p  - \vect p' $, and the usual Mandelstam invariants are
\begin{align}
s &= (p_1 + p_2)^2 = (E_1+ E_2)^2 = E^2\,, \\
\label{eq:mandlestamt}
t &= (p_1 + p_4)^2 = -(\bm p  - \bm p' )^2 = -\bm q^2 = -4\bm p^2 \frac{1-\cos\chi}{2} = - 4\bm p^2 \sin^2\frac{\chi}{2} \,, \\
u &= (p_1 + p_3)^2 = (E_1 -  E_2)^2 -(\bm p  + \bm p' )^2 = E^2(1-4\xi) -4\bm p^2 \cos^2\frac{\chi}{2} \,,
\end{align}
where $\chi$ is the scattering angle and we introduced the total center of mass energy, $E$, and the symmetric energy ratio, $\xi$,
defined as
\begin{align}
  E &= E_1 + E_2,\;\quad \xi = \frac{E_1E_2}{(E_1 + E_2)^2}~.
\end{align}
We will use these variables throughout this section.

\subsection{Scattering amplitude with IR subtractions optimized for EFT matching}

Here we will use the method of Refs.~\cite{Bern:2019nnu, Bern:2019crd},  which first expand in the small-velocity limit in the potential region to produce three-dimensional integrals, and then expand in the limit of small $q$. Divergent integrals will be kept unevaluated, to be canceled against EFT amplitudes in the matching procedure.

First let us calculate the scattering amplitudes optimized for EFT matching. At tree level the relevant piece comes from the $1/t$ pole
\begin{equation}
  \cM_1 = \frac{8\pi G m_1^2 m_2^2}{E_1E_2} \frac{(\sigma - \cos\phi)^2}{\vect q^2} \,,
\end{equation}
where we have divided by the non-relativistic normalization $4E_1E_2$. We will use the notation in Refs.~\cite{Bern:2019nnu,Bern:2019crd} and denote the conservative amplitudes in this section with calligraphic $\cM$ to distinguish them from those evaluated in dimensional regularization in previous sections.
The one-loop amplitude can be easily obtained from the one-loop integrand in Eq.~\eqref{eq:1lintegrand}. As explained in Ref.~\cite{Bern:2019crd}, Sec.~7.2.2 and 7.3.3, the scalar crossed box gives a vanishing contribution in the potential region (in strictly four dimensions), and the box yields the following three dimensional integral
\begin{equation}
  I_{\II}^{(p)} =  \int \frac{\mathrm{d}^{D-1} \vect \ell }{(2\pi)^{D-1}} \, {1 \over 2E \vect \ell^2 ( \vect \ell + \vect q)^2 (\vect \ell^2 + 2 \vect p \vect \ell)} + \text{evanescent terms}.
\end{equation}
Here \emph{evanescent terms} refer to two classes of terms: (1) terms that are suppressed in $\epsilon$ or $|q|$ after loop integration, (2) terms that arise from EFT diagrams with insertions of EFT operators suppressed by $\epsilon$ or $|q|$ omitted from Eq.~\eqref{eq:eftPotential}. Due to divergences associated with loop integration, terms of class (2) may be naively of the same order of $\epsilon$ and $|q|$ as terms that directly correspond to four-dimensional classical dynamics, but nevertheless such evanescent terms cancel in the EFT matching procedure and do not contribute to the final results. The one-loop amplitude optimized for EFT matching is then
\begin{align}
\cM_2 &= \frac{(16\pi G)^2 m_1^4 m_2^4}{2E_1E_2(E_1+E_2)} (\sigma - \cos\phi)^4 \int \frac{\mathrm{d}^{D-1} \vect \ell }{(2\pi)^{D-1}} \, \frac{1}{ \vect \ell^2 ( \vect \ell + \vect q)^2 (\vect \ell^2 + 2 \vect p \vect \ell)} + \text{evanescent terms.}\end{align}

Finally, we extract the two-loop conservative amplitude optimized for EFT matching from the two-loop integrand in Eq.~\eqref{eq:twoloopintegrand}. Let us first consider the integrals in the first line of such an equation. As explained in Ref.~\cite{Bern:2019crd}, when using the non-relativistic integration method all the non-planar scalar double-boxes vanish. Intuitively this is  because the energy flow would require the propagation of an antiparticle, which is not allowed, so only the planar double-box contributes in the potential region as \cite{Bern:2019crd}
\begin{align}
  I_{\RT}^{(p)}  &=  \frac{1}{4 E^2} \int \frac{\mathrm{d}^{D-1} \vect \ell_1 }{(2\pi)^{D-1}} \frac{\mathrm{d}^{D-1} \vect \ell_2 }{(2\pi)^{D-1}}  \frac{1}{  \vect \ell_1 ^2 (  \vect \ell_2 - \vect \ell_1)^2 (\vect \ell_2 + \vect q)^2 ( \vect \ell_1^2 + 2 \vect p \vect \ell_1 ) ( \vect \ell_2^2 + 2 \vect p \vect \ell_2 )} \nonumber \\
  &\quad + \text{evanescent terms}\,.
\end{align}
We must note that the vanishing of the non-planar integrals is a consequence of the loop-by-loop integration procedure used in Ref.~\cite{Bern:2019crd}, which at every stage drops evanescent contributions. In a two-loop integral these can hit at $1/\epsilon$ or $1/|q|$ pole coming from a different loop and generate finite contributions with classical power-counting such as those calculated in Section~\ref{sec:twoloop}. These contributions arising from evanescent terms are scheme dependent, and, as mentioned above, their ultimate fate is to cancel in the EFT matching procedure. In particular, they will not affect any physical quantity. Thus, as long as the integration in full theory and EFT is done consistently one might drop such evanescent terms. This effectively gives us a four-dimensional regularization method which, in contrast to our eikonal calculation based on dimensional regularization, does not need quantum corrections of $\cO(|q|^0)$, and $\cO(\epsilon)$ contributions at one-loop in order to extract the classical dynamics at two loops.

Next we consider the integrals in the second line of Eq.~\eqref{eq:twoloopintegrand}. As explained in previous sections only $I_{\H}$ and $I_{\xH}$ contribute with value given by Eq.~\eqref{eq:HxHres}, which we reprint here
\begin{equation}
  I^{\pot}_{\vphantom{\xH}\H} + I^{\pot}_{\xH}  = \frac{\log \vect q^2}{ 64\pi^2 m_1 m_2 \vect q^4} \frac{\text{arcsinh} \sqrt{\frac{\sigma -1}{2}}}{ \sqrt{\sigma^2 -1}}  + \text{evanescent terms}\,,
\end{equation}
where we dropped $1/\epsilon$ pole terms that do not generate non-analytic dependence on $q^2$.
Putting the pieces together we find the full two-loop amplitude optimized for EFT matching
\begin{align}
  \cM_3 &=  \frac{32\pi G^3 m_1^3 m_2^3 \,(\sigma - \cos\phi)^4 }{E_1E_2}  \left[ \log \vect q^2 \frac{\text{arcsinh} \sqrt{\frac{\sigma -1}{2}}}{ \sqrt{\sigma^2 -1}}  + \frac{64 \pi^2 m_1^3 m_2^3 (\sigma - \cos\phi)^2}{(E_1+E_2)^2} \right. \nonumber\\
  & \left. \hspace{2cm} \times \int \frac{\mathrm{d}^{D-1} \vect \ell_1 }{(2\pi)^{D-1}} \frac{\mathrm{d}^{D-1} \vect \ell_2 }{(2\pi)^{D-1}}  \frac{1}{  \vect \ell_1 ^2 (  \vect \ell_2 - \vect \ell_1)^2 (\vect \ell_2 + \vect q)^2 ( \vect \ell_1^2 + 2 \vect p \vect \ell_1 ) ( \vect \ell_2^2 + 2 \vect p \vect \ell_2 )}   \right] \nonumber \\
  & \hspace{2cm} + \text{evanescent terms}.
\end{align}

For later convenience we rewrite the conservative amplitudes in terms of the total energy, mass and cross ratios as
\begin{align}
  \cM_1 &= \frac{8\pi G \nu^2 m^4}{E^2 \xi} \frac{(\sigma - \cos\phi)^2}{\vect q^2}~,\\
 \cM_2 &= \frac{(16\pi G)^2\nu^4 m^8}{2E^3 \xi} (\sigma - \cos\phi)^4 \int \frac{\mathrm{d}^{D-1} \vect \ell }{(2\pi)^{D-1}} \, {1 \over  \vect \ell^2 ( \vect \ell + \vect q)^2 (\vect \ell^2 + 2 \vect p \vect \ell)} + \text{evanescent terms}~,\\
  \cM_3 &=  \frac{32\pi G^3 \nu^3 m^6 \,(\sigma - \cos\phi)^4 }{\xi E^2}  \left[ \log \vect q^2\frac{\text{arcsinh} \sqrt{\frac{\sigma -1}{2}}}{ \sqrt{\sigma^2 -1}}  + \frac{64 \pi^2 \nu^3 m^6 (\sigma - \cos\phi)^2}{E^2} \right. \nonumber\\
  & \left. \hspace{2cm} \times \int \frac{\mathrm{d}^{D-1} \vect \ell_1 }{(2\pi)^{D-1}} \frac{\mathrm{d}^{D-1} \vect \ell_2 }{(2\pi)^{D-1}}  \frac{1}{  \vect \ell_1 ^2 (  \vect \ell_2 - \vect \ell_1)^2 (\vect \ell_2 + \vect q)^2 ( \vect \ell_1^2 + 2 \vect p \vect \ell_1 ) ( \vect \ell_2^2 + 2 \vect p \vect \ell_2 )}   \right] \nonumber \\
  & \hspace{2cm} + \text{evanescent terms}.
\end{align}

\subsection{EFT matching and classical Hamiltonian}
Following Ref.~\cite{Cheung:2018wkq}, we want to match the amplitudes above to an EFT with an ordinary Hamiltonian with a potential, which we later will use to solve for the classical dynamics. The EFT describes two massive scalars interacting with momentum space Lagrangian given by
\begin{align}
  {\cal L} =& \int {\mathrm{d}^{D-1} \bm p \over (2\pi)^{D-1}}   \, \phi_1^\dagger(-\bm p) \left(\imath\partial_t - \sqrt{\bm p^2 + m_1^2}\right) \phi_1(\bm p) \nonumber\\
  &+ \int  {\mathrm{d}^{D-1} \bm p \over (2\pi)^{D-1}}    \, \phi_2^\dagger(-\bm p) \left(\imath\partial_t - \sqrt{\bm p^2 + m_2^2}\right) \phi_2(\bm p) \nonumber\\
&
-\int  {\mathrm{d}^{D-1} \bm p \over (2\pi)^{D-1}}   {\mathrm{d}^{D-1} \bm p' \over (2\pi)^{D-1}}   \, V(\bm p ,\bm p') \, \phi_1^\dagger(\bm p') \phi_1(\bm p) \phi_2^\dagger(-\bm p') \phi_2(-\bm p) \,,
\end{align}
where the form of the kinetic term manifests the absence of anti-particles.  The potential is given by
\begin{align}
V(\bm p,\bm p - \bm q) &=\sum_{n=1}^\infty { (G/2)^n (4\pi)^{( D-1)/2} \over |\bm q|^{D-1-n}}      \frac{\Gamma\left[ (D-1-n)/2 \right] }{\Gamma\left[ n/2 \right]} \, c_n \hspace{-0.12cm} \left( \vect p^2 \right) \\[3pt]
&= \frac{4 \pi G }{\vect q^2}\, c_1\hspace{-0.12cm}\left( {\bm p^2 } \right) + \frac{2 \pi^2 G^2 }{|\bm q|} \, c_2\hspace{-0.12cm}\left( \bm p^2  \right) -2 \pi G^3 \log \bm q^2  \, c_3\hspace{-0.12cm}\left( \vect p^2 \right) + \cdots \, ,
\label{eq:eftPotential}
\end{align}
where for conciseness we have put the external legs on-shell. As in the full theory, here we have also dropped evanescent terms suppressed by $\epsilon$ or $q^2$ at each order in $G$, which can affect the scattering amplitudes but do not have physical effects. If we Fourier transform $\vect q$ back to position space this yields the more familiar potential with an expansion in $G/|\vect r|$.

The EFT amplitudes calculated with the Lagrangian above are very simple. Due to the absence of anti-particles they are given by iterated bubble diagrams. The results up to order $G^3$ are given by Ref.~\cite{Bern:2019crd},
\begin{align}
{\cM}^{\rm EFT}_{1}  & = -\frac{4 \pi G  c_1}{\vect q^2}  \,, \nonumber\\
{\cM}^{\rm EFT}_{2}  &=  -\frac{2\pi^2 G^2
   c_2}{|\vect q|} + {\pi^2 G^2 \over  E \xi |\vect q|} \bigg[ (1-3 \xi ) c_1^2+4 \xi ^2 E^2 c_1 c_1' \bigg] +   \int {\mathrm{d}^{D-1} {\vect \ell}  \over (2\pi)^{D-1} } { 32 E \xi  \pi^2 G^2 c_1^2 \over \vect \ell^2 (\vect \ell + \vect q)^2  (\vect \ell^2 + 2 \vect p \vect \ell)}  \,, 
   \nonumber\\
{\cM}^{\rm EFT}_{3} 
&=   2 \pi G^3
   \log \vect q^2 c_3   - { \pi G^3 \log \vect q^2 \over  E^2 \xi }  \bigg[ (1-4 \xi ) c_1^3-8 \xi ^3 E^4 c_1  c_1' {}^2-4 \xi ^3 E^4 c_1^2 c_1'' +4 \xi ^2 E^3 c_2  c_1'  \nonumber\\
&\quad\hskip 3 cm  +  4\xi ^2 E^3 c_1  c_2' -2 (3-9 \xi ) \xi  E^2 c_1^2 c_1' +2E(1-3\xi)c_1  c_2   \bigg] \nonumber\\
&\quad  + \int {\mathrm{d}^{D-1} {\vect \ell}  \over (2\pi)^{D-1} } {16 \pi^3 G^3 c_1 \left[ 2E \xi c_2 - (1- 3 \xi) c_1^2 - 4 \xi ^2 E^2 c_1 c_1' \right] \over \vect \ell^2 |\vect \ell + \vect q|  (\vect \ell^2 + 2 \vect p \vect \ell)} \nonumber\\
&\quad -    \int {\mathrm{d}^{D-1} {\vect \ell_1}  \over (2\pi)^{D-1} } {\mathrm{d}^{D-1} {\vect \ell_2}  \over (2\pi)^{D-1} } { 256 E^2 \xi^2 \pi^3 G^3 c_1^3 \over \vect \ell_1^2 (\vect \ell_1 + \vect \ell_2)^2 (\vect \ell_2 + \vect q)^2  (\vect \ell_1^2 + 2 \vect p \vect \ell_1) (\vect \ell_2^2 + 2 \vect p \vect \ell_2)} \,,
\label{eq:M_EFT}
\end{align}
where $c_i = c_i(\vect p^2)$ and the primes denote derivatives. The EFT matching is performed by requiring ${\cM}_{n} = {\cM}^{\rm EFT}_{n}$, which yields the following coefficients for the potential
\begin{align}
  c_1(\vect p^2) &= -\frac{m^4\nu^2}{E^2\xi} 2 (\sigma - \cos\phi)^2 \,,\\
  c_2(\vect p^2) &= \frac{m^6\nu^3}{E^3\xi^2} \left[ -8 (\sigma - \cos\phi)^3 + \frac{2 \nu  (\sigma - \cos\phi)^4}{E^2\xi} \right]\,, \\
  c_3(\vect p^2) &= \frac{m^6\nu^3}{E^2\xi} \left[ \frac{16(\sigma - \cos\phi)^4\text{arcsinh} \sqrt{\frac{\sigma -1}{2}}}{ \sqrt{\sigma^2 -1}}  -\frac{40 m^2 \nu  (\sigma - \cos\phi )^4}{E^2 \xi ^2} \right. \nonumber \\ 
  &\hspace{2cm} \left. +\frac{8 m^4 \nu ^2 (3 -4 \xi) (\sigma - \cos\phi)^5}{E^4 \xi ^3} 
  -\frac{4 m^6 \nu ^3 (1-2 \xi) (\sigma - \cos\phi )^6}{E^6 \xi ^4} \right]\,.
\end{align}
A simple check is that for $\phi=0$ the potential should vanish in the static limit, $\sigma \rightarrow 1$ because the black holes are extremal. At higher loops this will continue to hold because the amplitude is proportional $stuM^{\rm tree}$, which vanishes as $(1-\cos\phi)^4$. Note that at one loop there are no triangles so the result is pure iteration
\begin{equation}
  c_2(\vect p^2) = \left[ \frac{ (1-3\xi)}{2E \xi} +  E \xi \partial_{ \vect p^2} \right] c_1( \vect p^2)^2\,.
\end{equation}
We note that in the high-energy limit $\sigma \rightarrow \infty$, the potential also matches the result from Einstein gravity in Refs.~\cite{Bern:2019nnu,Bern:2019crd}.

\subsection{Scattering angle from the classical Hamiltonian}
The scattering angle can be calculated from the Hamiltonian 
  \begin{equation}
H(\bm p, \bm r) = \sqrt{\bm p^2 + m_1^2}+ \sqrt{\bm p^2 + m_2^2} 
+ V(\bm p, \bm r) \, ,
\end{equation} 
by solving the classical equations of motion. As shown in Ref.~\cite{Bern:2019nnu}, this yields a formula that expresses the scattering angle directly in terms the IR finite part of the PM amplitudes, ${\cal M}'_i$, which are defined by dropping the unevaluated integrals in the expressions above
\begin{align}
2\pi \chi =  {d_1 \over  J } + {d_2  \over J^2  }  + \frac{1}{J^3}
 \biggl( \vphantom{\frac{1}{\pi^2}} {  -4  d_3     }   
 + {  d_1 d_2 \over \pi^2    }  
-{ d_1^3 \over 48 \pi^2  }  \biggr) \,,
\label{angle_ito_amplitudes}
\end{align}
where  $d_i$ are defined in terms of  ${\cal M}'_i$ as
\begin{equation}
d_1 = E \xi {\vect q}^2 {\cal M}_{1}'/|{\vect p}| \,,
\hskip 1 cm 
d_2 = E \xi |\vect q| {\cal M}_{2}' \,,
\hskip 1 cm 
d_3 =  E \xi |{\vect p}| {\cal M}_{3}' / \log {\vect q}^2 \,.
\end{equation}
Using our results for $\cN=8$ supergravity we find
\begin{align}
  d_1 &= 8\pi G m^2 \nu \frac{(\sigma - \cos \phi)^2}{\sqrt{\sigma^2-1}} \,,
\hskip 1 cm 
d_2 = 0 \,,\\
\hskip 1 cm 
d_3 &=  \frac{32 \pi G^3 m^6 \nu^4(\sigma - \cos \phi)^4}{2(\sigma-1)\nu+1} \text{arcsinh} \sqrt{\frac{\sigma -1}{2}} \,,
\end{align}
so the scattering angle calculated from the EFT is
\begin{align}
  \chi^{\rm 1PM} &=  { G m^2 \nu \over  J }  \frac{4 (\sigma - \cos \phi)^2}{\sqrt{\sigma^2-1}}  \,,\\
  \chi^{\rm 2PM} &= 0~,\\
  \chi^{\rm 3PM} &=   -\frac{G^3 m^6 \nu^3}{J^3} 16 
\biggl[\frac{(\sigma - \cos \phi)^6}{3(\sigma^2-1)^{3/2}}  +  \nu \frac{4 (\sigma - \cos \phi)^4}{2(\sigma-1)\nu+1} \text{arcsinh} \sqrt{\frac{\sigma -1}{2}}        
  \biggr] \,,
\end{align}
which precisely matches our results in Eqs.~\eqref{eikangle1}--\eqref{eikangle3} from the eikonal analysis.

One might be tempted to use the Hamiltonian to also calculate the precession of the periastron, $\Delta\Phi$, but as explained in Refs.~\cite{Kalin:2019rwq,Kalin:2019inp}, there is a simple relation between this quantity and the scattering angle
\begin{equation}
  \Delta\Phi = \chi(J) + \chi(-J)\,,
\end{equation}
which implies that odd orders in $G$ (i.e. odd PM orders), which are also odd in $J$, do not produce a precession, which can be confirmed by explicit calculation using the Hamiltonian. This means that the absence of precession observed in Ref.~\cite{Caron-Huot:2018ape} extends to $\cO(G^3)$, although for trivial reasons, and a calculation at the next order will be needed to test their conjecture of no precession to all orders. The precise statement of the conjecture in Ref.~\cite{Caron-Huot:2018ape} is that the quantum energy levels of the bound system, which we have not explored in this work, remain exactly degenerate. However, the fact that there is a correction to the classical scattering angle at $\cO(G^3)$, although suppressed in the probe limit, makes us less optimistic about the possibility of the orbits remaining integrable at this and higher orders. 

\section{Conclusion}
\label{sec:conclusions}
In this paper we computed the conservative classical dynamics of scattering of two spinless extremal black holes in $\mathcal N=8$ supergravity at $\cO(G^3)$. In Refs.~\cite{Bern:2019nnu,Bern:2019crd} the $\cO(G^3)$ (or 3rd-post-Minkowskian) conservative potential in Einstein gravity was calculated using an EFT matching procedure that avoids evaluation of infrared divergent integrals and provides a velocity expansion to high orders.  Here, in contrast, we have directly calculated the IR-divergent scattering amplitude in dimensional regularization, and have directly obtained exact velocity dependence using differential equations, without the need to resum a series expansion.

This has allowed us to probe the delicate IR structure of eikonal exponentiation, where terms that vanish in four dimensions or vanish in the classical limit have to be evaluated explicitly at one loop, in order to construct IR subtraction terms at two loops to isolate genuine classical contributions at $O(G^3)$. Our novel integration method paves the way to a rigorous verification of the velocity resummation of Refs.~\cite{Bern:2019nnu,Bern:2019crd}, and to streamline further calculations. The ability to evaluate the divergent two-loop amplitudes in dimensional regularization also opens the door to applying the method of Refs.~\cite{Kosower:2018adc, Maybee:2019jus} which computes classical observables directly from appropriate phase space integrations of the S-matrix. Our differential equations method is highly flexible as the only difference between the soft region and the potential region is in the boundary conditions. The evaluation of the amplitude in the soft region at two loops and the emergence of non-exponentiating terms will be discussed elsewhere.

By computing the classical gravitational scattering angle in both the eikonal approximation and EFT formalism, we have explicitly established their equivalence at $\cO(G^3)$ for the scattering of massive particles for the first time. While the EFT formalism gives a more direct connection to the classical Hamiltonian, the eikonal approximation provides a more direct relation between two gauge-invariant quantities, the scattering amplitude and the scattering angle. It would be interesting to prove the all-order eikonal exponentiation structure for massive scattering from first principles beyond the one-loop case \cite{Bjerrum-Bohr:2018xdl}, perhaps by generalizing the partially massive case studied at two loops in Ref.~\cite{Akhoury:2011kq}, and to prove the validity of the eikonal angle formula beyond two loops.

Remarkably, we found that the classical scattering angle of two extremal black holes in $\cN=8$ supergravity coincides in the limit of high energy with that of two Schwarzschild black holes in Einstein gravity \cite{Bern:2019nnu,Bern:2019crd}. Since the classical limit satisfies $|\bm q| \ll M$ and does not commute with the massless limit $M \to 0$, our result is reminiscent of, but not a direct consequence of, the universality of massless gravitational scattering in the Regge limit recently unveiled in Ref.~\cite{Bern:2020gjj}, and strongly suggests graviton dominance, whose mechanism still needs to be understood, is generic at this order.

Beyond universality, several aspects of the scattering of black holes in $\cN=8$ supergravity deserve further study. For instance, it would be very interesting to re-analyze the two-loop calculation for dyonic black holes with generic charge misalignments. This might require an improved understanding of the structure of the S-matrix for mutually non-local particles. Furthermore, it would be interesting to calculate the exact quantum energy levels of the bound system and their decay rates to explore the precise integrability conjecture of Ref.~\cite{Caron-Huot:2018ape}. More generally, this conjecture should be investigated at the next order, where precession can arise.
Given the simplicity of loop integrands in $\mathcal N=8$ supergravity, we expect this highly symmetric theory to be an excellent theoretical laboratory for other aspects of black hole binary dynamics, such as spin-dependent scattering at $\cO(G^3)$ and spinless scattering at $\cO(G^4)$, both of which are unexplored frontiers in post-Minkowskian expansion of black hole binary dynamics, but are amenable to treatment by our techniques.\footnote{See Refs.~\cite{Bini:2020wpo, Levi:2020kvb, Levi:2020uwu} for some related recent results in the post-Newtonian expansion. Also see Refs.~\cite{Vines:2017hyw, Vines:2018gqi, Guevara:2018wpp, Chung:2018kqs, Maybee:2019jus, Guevara:2019fsj, Arkani-Hamed:2019ymq, Damgaard:2019lfh, Siemonsen:2019dsu, Aoude:2020onz, Bern:2020buy} for spin-dependence in the post-Minkowskian expansion up to $\cO(G^2)$.} We hope to explore some of these questions in the near future.

 \section*{Acknowledgments}
\label{sec:acknowledgements}
We thank Harald Ita for collaboration during initial stages of the project. We thank Zvi Bern, Clifford Cheung, Thibault Damour, Harald Ita, Chia-Hsien Shen, and Mikhail P.\ Solon for helpful comments and discussions.
Special thanks go to Simon Caron-Huot and Zahra Zahraee for enlightening discussions and for sharing results for some of the integrals.
J.P-.M.\ is supported by the US Department of State through a Fulbright scholarship and by the Mani L. Bhaumik
Institute for Theoretical Physics.
M.S.R.’s work is funded by the German Research Foundation (DFG) within the Research Training Group GRK 2044.
M.Z.~is supported by the Swiss National Science Foundation under contract SNF200021 179016
and the European Commission through the ERC grant pertQCD.
 \appendix
\section{Dimensionally regularized integrals for the potential region}
\label{sec:3dintegrals}
In this appendix we present results for dimensionally regularized Feynman integrals in $D-1 = 3-2\epsilon$ spatial dimensions, needed for re-expanding the ``soft integrals'' in the potential region. All of these integrals are the result of evaluating the energy integrals using the residue prescriptions explained in the main text.

Following widely used conventions in the literature on Feynman integrals, the integrals are presented with the following normalization,
\begin{equation}
\frac {\mathrm{d}^{D-1} \vect{\ell}}{\pi^{(D-1)/2}} = 8\pi^{3/2} \, (4\pi)^{-\epsilon} \frac {\mathrm{d}^{D-1} \vect{\ell} \, }{(2\pi)^{(D-1)}} \, .
\end{equation}
In the frame chosen the external three-momentum transfer $\bm q$ is in the transverse $(x,y)$ direction, while some integrals have linear propagators of the form $1/\ell_z = 1/(\vect{\ell} \cdot \vect{n}_z)$, where $\vect{n}_z$ is the unit vector in the $z$-direction. The final results are fully relativistic and functions of $\vect q^2 = -q^2$.
Unless otherwise shown, we will consider the $-\imath 0$ prescription to be implicitly present in every propagator.

\subsection{One-loop integrals}
At one loop we need to evaluate the linearized triangle and bubble integrals in Eqs.~\eqref{eq:lintri} and \eqref{eq:linbub}.
These can evaluated using traditional methods. Concrete the general linearized triangle integral is given by \cite{Smirnov:2012gma} 
\begin{align}
&\int\!\frac{\mathrm{d}^{D-1}\vect{\ell}}{\pi^{(D-1)/2}}\frac{1}{(\vect{\ell}^2-\imath 0)^a[(\vect{\ell}-\vect{q})^2-\imath 0]^b(2\ell^z - \imath 0 )^c}\label{eq:SmirnovMasterTriangle}\\
={}& e^{\frac {\imath \pi c} {2}} (\vect{q}^2)^{\frac{3}{2}-a-b-\frac{c}{2}-\epsilon} \frac{\Gamma \left(\frac{c}{2}\right) \Gamma \left(\frac{3}{2}-a-\frac{c}{2}-\epsilon\right) \Gamma \left(\frac{3}{2}-b-\frac{c}{2}-\epsilon \right)
	\Gamma \left(a+b+\frac{c}{2}+\epsilon -\frac{3}{2}\right)}{2\Gamma (a) \Gamma (b) \Gamma (c) \Gamma (3-a-b-c-2\epsilon)} \,. \nonumber
\end{align}
The usual bubble integrals with $c=0$ can be recovered by
\begin{equation}
\lim_{c \rightarrow 0} \frac{\Gamma (c/2) } {2 \Gamma (c)} = 1 \, .
\end{equation}
In particular, for $a=b=1$, $c\rightarrow 0$, Eq.~\eqref{eq:SmirnovMasterTriangle} gives
\begin{align} \label{eq:Bub3D}
\int \frac{\mathrm{d}^{D-1}\vect{\ell}} {\pi^{(D-1)/2}}
\frac{1}{\vect{\ell}^2(\vect{\ell} - \vect{q})^2}
&=\left(-q^2 \right)^{-\epsilon} \frac{1}{\sqrtmQSq}
\frac{\Gamma\left(\frac 1 2 - \epsilon \right)^2 \Gamma\left(\frac 1 2 + \epsilon\right)} {\Gamma(1-2\epsilon)}\,.
\end{align}
Setting $a=b=c=1$ in Eq.~\eqref{eq:SmirnovMasterTriangle} gives
\begin{align} \label{eq:Tri3D}
\int\!\frac{\mathrm{d}^{D-1}\vect{\ell}}{\pi^{(D-1)/2}}\frac{1}{\vect{\ell}^2(\vect{\ell} - \vect{q})^2(2 \ell^z)}
&= \left(-q^2\right)^{-\epsilon}\frac{1}{-q^2}
\frac {\imath \sqrt \pi \, \Gamma(-\epsilon)^2 \Gamma(1+\epsilon)} {2\Gamma(-2\epsilon)} 
\,.
\end{align}
Another way to evaluate this integral is by using symmetrization over the possible assignments of loop momenta
\begin{align}
&\int\!\frac{\mathrm{d}^{D-1}\vect{\ell}}{\pi^{(D-1)/2}}\frac{1}{\vect{\ell}^{2}(\vect{\ell} - \vect{q})^2(2 \ell_1^z - \imath 0)}\\
={}& \!\!\int\!\frac{\mathrm{d}^{D-1}\vect{\ell}_1}{\pi^{(D-1)/2}} \mathrm{d}^{D-1}\vect{\ell}_2 \frac{1}{(2\ell_1^z - \imath 0)\prod_i\vect{\ell}_i^{\,2}}\delta(\sum \ell_i^z)\delta^{(D-2)}(\sum\vect{\ell}_i^\perp - \vect q^\perp)\nonumber\,.
\end{align}
Where the $\vect{\ell}_i^\perp$ and $\vect{q}^\perp$ are the components of $\vect{\ell}_i$ and  $\vect{q}$ in the plane orthogonal to $\vect{n}_z$, respectively.
Now we symmetrize over the two loop momenta, using 
\begin{align}
	\frac{1}{2!}\left[\frac{1}{2\ell_1^z - \imath 0}+\frac{1}{2\ell_2^z - \imath 0}\right]\delta(\sum \ell_i^z)
={}\frac{\imath\pi}{2}\delta(\ell^z_1)\delta(\ell^z_2)\,.\label{eq:1loopSymmetrization}
\end{align}
Using $q^z=0$, we can trivially preform the $z$-integration to obtain a $(D-2)$-dimensional bubble integral
\begin{align}
\int\!\frac{\mathrm{d}^{D-1}\vect{\ell}}{\pi^{(D-1)/2}}\frac{1}{\vect{\ell}^{2}(\vect{\ell}-\vect{q})^2(2 \ell_1^z - \imath 0)}
={}& \frac{\imath\sqrt{\pi}}{2}\!\!\int\!\frac{\mathrm{d}^{D-2}\vect{\ell}_1}{\pi^{(D-2)/2}}\frac{1}{\vect{\ell}^{2}(\vect{\ell}-\vect{q})^2}\nonumber\\
={}&\left(-q^2\right)^{-\epsilon}\frac{1}{-q^2}\frac{\imath \sqrt{\pi } \Gamma (-\epsilon )^2 \Gamma (\epsilon +1)}{2 \Gamma (-2 \epsilon )}\,,
\end{align}
in agreement with Eq.~\eqref{eq:Tri3D}.
\subsection{Two-loop integrals}

\subsubsection*{Double box ($\RT$)}

Adopting the frame choice Eq.~\eqref{eq:u1u2framechoice}, and after energy integration, we find that in the static limit, the pure basis of master integrals, Eqs.~\eqref{eq:fRT1}--\eqref{eq:fRT10}, for the double-box family are equal to
\begin{align}
f_{\RT,4}^{\pot} \big|_{\relfbar=1} ={}&\frac{\pi}{6}\epsilon ^2(1+2\epsilon) (-q^2)\int\!\frac{\mathrm{d}^{D-1}\vect{\ell}_1\mathrm{d}^{D-1}\vect{\ell}_2 \left( e^{\EulerGamma\epsilon}\right)^2}
  {(\imath\pi^{(D-1)/2})^2 \,(\vect {\ell}_1^{\,2})^2\vect{\ell}_2^{\,2}(\vect{\ell}_1+\vect{\ell}_2-\vect{q})^2}\,,\label{eq:g4staticReduced}\\ 
f_{\RT,6}^{\pot} \big|_{\relfbar=1} ={}& \frac{\pi}{6}\epsilon^3(1-6\epsilon)\int\!\frac{\mathrm{d}^{D-1}\vect{\ell}_1\mathrm{d}^{D-1}\vect{\ell}_2 \left( e^{\EulerGamma\epsilon}\right)^2}
{(\imath\pi^{(D-1)/2})^2 \, \vect {\ell}_1^{\,2}\vect{\ell}_2^{\,2}(\vect{\ell}_1+\vect{\ell}_2-\vect{q})^2}\,,\label{eq:g12staticReduced}\\ 
f_{\RT,7}^{\pot} \big|_{\relfbar=1} ={}& \pi \epsilon ^4  (-q^2) \!\int\!\frac{\mathrm{d}^{D-1}\vect{\ell}_1\mathrm{d}^{D-1}\vect{\ell}_2 \left( e^{\EulerGamma\epsilon}\right)^2}
{(\imath\pi^{(D-1)/2})^2 \, \vect{\ell}_1^{\,2}\vect{\ell}_2^{\,2}(\vect{\ell}_1+\vect{\ell}_2-\vect{q})^2(2\ell_1^z)(-2\ell_2^z)}\label{eq:g13staticReduced}\,,\\
f_{\RT,10}^{\pot} \big|_{\relfbar=1} ={}& -\frac {\epsilon^4} 8 \sqrtmQSq \!\int\!\frac{\mathrm{d}^{D-1}\vect{\ell}_1\mathrm{d}^{D-1}\vect{\ell}_2 \left( e^{\EulerGamma\epsilon}\right)^2}
{(\imath\pi^{(D-1)/2})^2 \, \vect{\ell}_1^{\,2} \vect{\ell}_2^{\,2}(\vect{\ell}_1+\vect{\ell}_2-\vect{q})^2 (2\ell_1^z)}\,,
\end{align}
where we have omitted the other integrals in the basis which vanish in the static limit.
The first, second and fourth of these integrals can be evaluated by first performing a sub-loop integral over $\vect{\ell}_2$ using Eq.~\eqref{eq:SmirnovMasterTriangle}, and then evaluating the resulting $\vect{\ell}_1$ integral again using Eq.~\eqref{eq:SmirnovMasterTriangle} with non-integer propagator powers,
\begin{align}
	& \int\!\frac{\mathrm{d}^{D-1}\vect{\ell}_1}{\pi^{(D-1)/2}}\frac{\mathrm{d}^{D-1}\vect{\ell}_2}{\pi^{(D-1)/2}}\frac{1}{(\vect{\ell}_1^{\,2})^2\vect{\ell}_2^{\,2}(\vect{\ell}_1+\vect{\ell}_2-\vect{q})^2} \nonumber \\
	& \hspace{1cm}=
        \left(-q^2\right)^{-2\epsilon}\frac{1}{(-q^2)}
        \frac{\Gamma \left(-\epsilon -\frac{1}{2}\right) \Gamma \left(\frac{1}{2}-\epsilon \right)^2 \Gamma (2 \epsilon +1)}{\Gamma \left(\frac{1}{2}-3 \epsilon \right)}\,, \\
& \int\!\frac{\mathrm{d}^{D-1}\vect{\ell}_1}{\pi^{(D-1)/2}}\frac{\mathrm{d}^{D-1}\vect{\ell}_2}{\pi^{(D-1)/2}}\frac{1}{\vect{\ell}_1^{\,2}\vect{\ell}_2^{\,2}(\vect{\ell}_1+\vect{\ell}_2-\vect{q})^2} \nonumber \\
	& \hspace{1cm}=
        \left(-q^2\right)^{-2\epsilon}
        \frac{\Gamma \left(\frac{1}{2}-\epsilon \right)^3 \Gamma (2 \epsilon )}{\Gamma \left(\frac{3}{2}-3 \epsilon \right)}\,, \\
& \int\!\frac{\mathrm{d}^{D-1}\vect{\ell}_1}{\pi^{(D-1)/2}}\frac{\mathrm{d}^{D-1}\vect{\ell}_2}{\pi^{(D-1)/2}}\frac{1}{\vect{\ell}_1^{\,2}\vect{\ell}_2^{\,2}(\vect{\ell}_1+\vect{\ell}_2-\vect{q})^2(2\ell_1^z)} \nonumber \\
	& \hspace{1cm}=
	\left(-q^2\right)^{-2\epsilon}\frac{1}{\sqrtmQSq}
        \frac{\imath \sqrt{\pi } \Gamma \left(\frac{1}{2}-2 \epsilon \right) \Gamma \left(\frac{1}{2}-\epsilon \right)^2 \Gamma (-\epsilon ) \Gamma \left(2 \epsilon +\frac{1}{2}\right)}{2 \Gamma \left(\frac{1}{2}-3 \epsilon \right) \Gamma (1-2 \epsilon )}\,.
\end{align}
The evaluation of the remaining integral follows closely the evaluation of the one-loop triangle integral by symmetrization. We first rewrite
\begin{align}
	&\int\!\frac{\mathrm{d}^{D-1}\vect{\ell}_1}{\pi^{(D-1)/2}}\frac{\mathrm{d}^{D-1}\vect{\ell}_2}{\pi^{(D-1)/2}}\frac{1}{\vect{\ell}_1^{\,2}\vect{\ell}_2^{\,2}(\vect{\ell}_1+\vect{\ell}_2-\vect{q})^2(2 \ell_1^z - \imath 0)(-2\ell_2^z-\imath 0)}\\
	={}& \!\!\int\!\frac{\mathrm{d}^{D-1}\vect{\ell}_1}{\pi^{(D-1)/2}}\frac{\mathrm{d}^{D-1}\vect{\ell}_2}{\pi^{(D-1)/2}} \mathrm{d}^{D-1}\vect{\ell}_3 \frac{1}{(2\ell_1^z - \imath 0)(-2 \ell_2^z-\imath 0)\prod_i\vect{\ell}_i^{\,2}}\delta(\sum \ell_i^z)\delta^{(D-2)}(\sum\vect{\ell}_i^\perp - \vect q^\perp)\nonumber\,.
\end{align}
Symmetrizing over all loop momenta, results in the identity similar to Eq.~\eqref{eq:1loopSymmetrization},
\begin{align}
\frac{1}{3!}\left[ \frac{1}{(2\ell_1^z - \imath 0) (-2 \ell_2^z - \imath 0) } \, +\text{perms.} \right]
\delta\left(\sum \ell_i^z\right)
=
-\frac{\pi^2}{6}\delta(\ell_1^z)\delta(\ell_2^z)\delta(\ell_3^z)\,.
\end{align}
Using $q^z=0$, we can trivially preform the $z$-integration to obtain a $(D-2)$-dimensional integral
\begin{align}
&\int\!\frac{\mathrm{d}^{D-1}\vect{\ell}_1}{\pi^{(D-1)/2}}\frac{\mathrm{d}^{D-1}\vect{\ell}_2}{\pi^{(D-1)/2}}\frac{1}{\vect{\ell}_1^{\,2}\vect{\ell}_2^{\,2}(\vect{\ell}_1+\vect{\ell}_2-\vect{q})^2(2\ell_1^z - \imath 0)(-2\ell_2^z - \imath 0)}\label{eq:DoubleTriange3D}
\\
={}&
-\frac{\pi}{6}\int\!\frac{\mathrm{d}^{D-2}\vect{\ell}_1}{\pi^{(D-2)/2}}\frac{\mathrm{d}^{D-2}\vect{\ell}_2}{\pi^{(D-2)/2}}\frac{1}{\vect{\ell}_1^{\,2}\vect{\ell}_2^{\,2}(\vect{\ell}_1 + \vect{\ell}_2 - \vect{q})^2} ={}
- \frac \pi 6
\left(-q^2\right)^{-2\epsilon}\frac{1}{(-q^2)}
\frac{\Gamma (-\epsilon )^3 \Gamma (2 \epsilon +1)}{\Gamma (-3 \epsilon )}\,. \nonumber \end{align}
Therefore, the integrals with nonzero values in the static limit are
\begin{align}
  f_{\RT,4}^{\pot} \big|_{\relfbar=1} ={}& -2  f_{\RT,6}^{\pot} \big|_{\relfbar=1} =\frac{2\pi}{3}\epsilon ^3  (-q^2)^{-2\epsilon}e^{2\EulerGamma\epsilon}  \frac{\Gamma \left(\frac{1}{2}-\epsilon \right)^3 \Gamma (2 \epsilon )}{\Gamma \left(\frac{1}{2}-3 \epsilon \right)}\,,
  \label{eq:StaticBNDg7}\\
f_{\RT,7}^{\pot} \big|_{\relfbar=1} ={}&\frac{\pi^2}{6}\epsilon ^4  (-q^2)^{-2\epsilon}e^{2\EulerGamma\epsilon}  \frac{\Gamma (-\epsilon )^3 \Gamma (2 \epsilon +1)}{\Gamma (-3 \epsilon )}\,,\label{eq:StaticBNDg13} \\
f_{\RT,10}^{\pot} \big|_{\relfbar=1} ={}& -\frac{\imath \epsilon^4 \pi^{3/2}}{4}  (-q^2)^{-2\epsilon}e^{2\EulerGamma\epsilon} 
\frac{\Gamma \left(\frac{1}{2}-2 \epsilon \right) \Gamma \left(\frac{1}{2}-\epsilon \right)^2 \Gamma (-\epsilon ) \Gamma \left(\frac{1}{2} + 2 \epsilon \right)}{\Gamma \left(\frac{1}{2}-3 \epsilon \right) \Gamma (1-2 \epsilon )}
\,.
\end{align}
By expanding in $\epsilon$ one can check that such boundary conditions  \eqref{eq:StaticBNDg4}--\eqref{eq:StaticBNDg13} are of uniform transcendental weight, and yield the boundary vector \eqref{eq:StaticBNDRT} used in the text.

\subsubsection*{$\H$ and $\xH$}
The integrals for the sum of $\H$ and $\xH$ topologies with non-vanishing static limits are
\begin{align}
  f_{\rm c\H,4}^{\pot} \big|_{\relfbar=1} ={}&-\frac{\pi}{2} \epsilon ^4 (-q^2) \int\!\frac{\mathrm{d}^{D-1}\vect{\ell}_1\mathrm{d}^{D-1}\vect{\ell}_2\left( e^{\EulerGamma\epsilon}\right)^2}
   {(\imath\pi^{(D-1)/2})^2\vect{\ell}_1^{\,2}\vect{\ell}_2^{\,2}(\vect{\ell}_1-\vect{q})^2(\vect{\ell}_2-\vect{q})^2}\,,\label{eq:g4HstaticReduced}\\ 
 f_{\rm c\H,7}^{\pot} \big|_{\relfbar=1} ={}&-\frac{\pi}{4}\epsilon ^2(1+2\epsilon)\int\!\frac{\mathrm{d}^{D-1}\vect{\ell}_1\mathrm{d}^{D-1}\vect{\ell}_2\left( e^{\EulerGamma\epsilon}\right)^2}
   {(\imath\pi^{(D-1)/2})^2(\vect{\ell}_1^{\,2})^2\vect{\ell}_2^{\,2}(\vect{\ell}_1+\vect{\ell}_2-\vect{q})^2}\,,\label{eq:g7HstaticReduced}\\
   f_{\rm c\H,10}^{\pot} \big|_{\relfbar=1} ={}&-\pi \epsilon^4 (-q^2)
\int\!\frac{\mathrm{d}^{D-1}\vect{\ell}_1\mathrm{d}^{D-1}\vect{\ell}_2\left( e^{\EulerGamma\epsilon}\right)^2}
   {(\imath\pi^{(D-1)/2})^2\vect{\ell}_1^2\vect{\ell}_2^2(\vect{\ell}_1-\vect{q})^2(\vect{\ell}_2-\vect{q})^2}
   \,.\label{eq:g10HstaticReduced}
\end{align}
The second integral has already been evaluated, and equals to
\begin{equation}
f_{\rm c\H,7}^{\pot} \big|_{\relfbar=1} ={}-\frac{3}{2}f_{\RT,4}^{\pot} \big|_{\relfbar=1}\,.
\end{equation}
The remaining integrals are proportional to a two-loop double-bubble integral which factorizes and is trivially the square of the one-loop bubble integral \eqref{eq:Bub3D}
\begin{align}
\int\!\frac{\mathrm{d}^{D-1}\vect{\ell}_1}{\pi^{(D-1)/2}}\frac{\mathrm{d}^{D-1}\vect{\ell}_2}{\pi^{(D-1)/2}}\frac{1}{\vect{\ell}_1^2\vect{\ell}_2^2(\vect{\ell}_1-\vect{q})^2(\vect{\ell}_2-\vect{q})^2}
&=\left(-q^2 \right)^{-2\epsilon} \frac{1}{(-q^2)}
\frac{\Gamma\left(\frac 1 2 - \epsilon \right)^4 \Gamma\left(\frac 1 2 + \epsilon\right)^2} {\Gamma(1-2\epsilon)^2}\,. \label{eq:FactBubBube3D}
\end{align}
In summary we find the following result for the static integrals
\begin{align}
f_{\rm c\H,4}^{\pot} \big|_{\relfbar=1} ={}&\frac{1}{2}f_{\rm c\H,10}^{\pot} \big|_{\relfbar=1} =\frac{\pi}{2}\epsilon ^4 (-q^2)^{-2\epsilon}e^{2\EulerGamma\epsilon} \left[\frac{   \Gamma \left(\frac{1}{2}-\epsilon \right)^2\!\Gamma \left(\epsilon +\frac{1}{2}\right)}{\Gamma (1-2 \epsilon )}\right]^2,\label{eq:StaticBNDg4}\\ 
f_{\rm c\H,7}^{\pot} \big|_{\relfbar=1} =&-\pi\epsilon ^3 (-q^2)^{-2\epsilon} e^{2\EulerGamma\epsilon} \frac{\Gamma \left(\frac{1}{2}-\epsilon \right)^3 \Gamma (2 \epsilon )}{\Gamma \left(\frac{1}{2}-3 \epsilon \right)}
\,, \label{eq:StaticBNDH710}
\end{align}
which yields the boundary vector in Eq.~\eqref{eq:StaticBNDH}.
\subsubsection*{$\IX$ and crossed integrals}
The evaluation of the boundary vector for the $\IX$ and crossed integrals proceeds analogously to the computations in the previous subsections\footnote{The explicit values for the boundary conditions can be obtained from the solutions inside the ancillary files accompanying this paper.}. In particular all three-dimensional integrals necessary have already been computed therein. 
 \section{Solution of the differential equations}
\label{sec:twoloopdetails}
Having the canonical form of the differential equations at hand the systems can be straightforwardly solved order-by-order in $\epsilon$, yielding harmonic polylogarithms.
In this appendix we present the solution of the differential equations for two-loop master integrals in the potential region up to
$\mathcal O(\epsilon^4)$. All the functions not shown vanish.
The solution of the differential equations in Eq.~\eqref{eq:RTSoftDE} with the matrices in Eqs.~\eqref{eq:RTmateven} and \eqref{eq:RTmatodd} and boundary conditions in Eq.~\eqref{eq:StaticBNDRT} is
\begin{align}
  f_{\RT,2}^{\pot}= (-q^2)^{-2\epsilon} {}& \epsilon^2 \pi^2 \left[ - \frac 1 3 \epsilon \log(x) + \epsilon^2 \left( \operatorname{Li}_2(1-x^2) + \log^2(x) \right)  \right] \label{eq:fRT2pot}
  \,,\\
  f_{\RT,3}^{\pot}= (-q^2)^{-2\epsilon} {}& \epsilon^2 \pi^2 \left[ -\frac 2 3 \epsilon \log(x) -\frac 2 3 \epsilon^2 \left( \operatorname{Li}_2(1-x^2) + \log^2(x) \right)  \right]
  \,,\\
  f_{\RT,4}^{\pot}= (-q^2)^{-2\epsilon} {}& \epsilon^2 \pi^2 \left[ \frac 1 3 + \frac 1 {18} \epsilon^2 \left( -7 \pi^2 -48 \log^2(x) \right) \right]
  \,,\\
  f_{\RT,6}^{\pot}= (-q^2)^{-2\epsilon} {}& \epsilon^2 \pi^2 \left[ -\frac 1 6 + \frac{7 \epsilon^2 \pi^2}{36} \right]
  \,,\\
  f_{\RT,7}^{\pot}= (-q^2)^{-2\epsilon} {}& \epsilon^2 \pi^2 \left[ \frac 1 2 - \frac 1 {12}\epsilon^2 \left( 4 \log^2(x) + \pi^2 \right) \right]
  \,, \label{eq:fRT7pot} \\
  f_{\RT,10}^{\pot}= (-q^2)^{-2\epsilon} {}& \epsilon^2 \pi^2 \left[ \frac {\imath \pi\epsilon }{4} -\frac{\imath\pi \log(2) \epsilon^2}{2} \right]
  \,. \label{eq:fRT10pot} 
\end{align}
The solution of the differential equations in Eq.~\eqref{eq:2loopSoftDEH} with the matrices in Eq.~\eqref{eq:Hmateven} and boundary conditions in Eq.~\eqref{eq:StaticBNDH} is
\begin{align}
  f_{\rm c\H,4}^{\pot}= (-q^2)^{-2\epsilon} {}& \epsilon^2 \pi^2 \left[ \frac{\epsilon^2 \pi^2} 2  \right]
  \,, \label{eq:fH4pot}\\
  f_{\rm c\H,5}^{\pot}= (-q^2)^{-2\epsilon} {}& \epsilon^2 \pi^2 \left[ \frac 1 2 \epsilon \log(x) - \frac 3 2 \epsilon^2 \left( \operatorname{Li}_2(1-x^2) + \log^2(x) \right)  \right]
  \,,\\
  f_{\rm c\H,6}^{\pot}= (-q^2)^{-2\epsilon} {}& \epsilon^2 \pi^2 \left[ \epsilon \log(x) + \epsilon^2 \left( \operatorname{Li}_2(1-x^2) + \log^2(x) \right)  \right]
  \,,\\
  f_{\rm c\H,7}^{\pot}= (-q^2)^{-2\epsilon} {}& \epsilon^2 \pi^2 \left[ -\frac 1 2 + \epsilon^2 \left( \frac{7\pi^2} {12} + 4 \log^2(x) \right)  \right]
  \,,\\
  f_{\rm c\H,9}^{\pot}= (-q^2)^{-2\epsilon} {}& \epsilon^2 \pi^2 \left[ - \epsilon \log(x) + \epsilon^2 \left( \operatorname{Li}_2(1-x^2) + \log^2(x) \right)  \right]
  \,,\\
  f_{\rm c\H,10}^{\pot}= (-q^2)^{-2\epsilon} {}& \epsilon^2 \pi^2 \left[\epsilon^2 \left( \pi^2 + 6 \log^2(x) \right)  \right] \label{eq:fH10pot}\, .
\end{align}
The solution of the differential equations in Eq.~\eqref{eq:XB2loopSoftDE} with the matrices in Eqs.~\eqref{eq:IXmateven} and \eqref{eq:IXmatodd} boundary conditions in Eq.~\eqref{eq:StaticBNDIX} is
\begin{align} \label{eq:fIX2}
f_{\IX,2}^{\pot}= (-q^2)^{-2\epsilon} {}& \epsilon^2 \pi^2 \left[ \frac 1 6 \epsilon \log(x) -\frac{1}{2} \epsilon^2 \left( \operatorname{Li}_2(1-x^2) + \log^2(x) \right)  \right]\,,\\
f_{\IX,3}^{\pot}= (-q^2)^{-2\epsilon} {}& \epsilon^2 \pi^2 \left[ \frac 1 3 \epsilon \log(x) +\frac 1 3 \epsilon^2 \left( \operatorname{Li}_2(1-x^2) + \log^2(x) \right)  \right]\,,\\
f_{\IX,4}^{\pot}= (-q^2)^{-2\epsilon} {}& \epsilon^2 \pi^2 \left[ -\frac 1 6 + \frac 1 {36} \epsilon^2 \left( 7 \pi^2 +48 \log^2(x) \right) \right]\,,\\
f_{\IX,5}^{\pot}= (-q^2)^{-2\epsilon} {}& \epsilon^2 \pi^2 \left[ \frac 1 3\epsilon\log(x) -\epsilon^2 \left( \operatorname{Li}_2(1-x^2) + \log^2(x) \right) \right]\,,\\
f_{\IX,6}^{\pot}= (-q^2)^{-2\epsilon} {}& \epsilon^2 \pi^2 \left[ \frac 2 3\epsilon\log(x) +\frac{2}{3}\epsilon^2 \left( \operatorname{Li}_2(1-x^2) + \log^2(x) \right) \right]\,,\\
f_{\IX,7}^{\pot}= (-q^2)^{-2\epsilon} {}& \epsilon^2 \pi^2 \left[\frac{1}{3}-\frac{1}{18}\epsilon^2 \left( 7 \pi^2 +48 \log^2(x) \right)\right]\,,\\
f_{\IX,8}^{\pot}= (-q^2)^{-2\epsilon} {}& \epsilon^2 \pi^2 \left[ -\frac{1}{6}+\frac{7\pi^2\epsilon^2}{36}\right]\,,\\
f_{\IX,10}^{\pot}= (-q^2)^{-2\epsilon} {}& \epsilon^2 \pi^2 \left[ -\frac{5}{12}\epsilon^2\log^2(x)\right]\,,\\
f_{\IX,15}^{\pot}= (-q^2)^{-2\epsilon} {}& \epsilon^2 \pi^2 \left[ \frac {\imath\pi \epsilon }{4} -\frac{\imath \pi \log(2) \epsilon^2}{2} \right]
\,. \label{eq:fIX15}
\end{align} 
The differential equation for the $\xRT$ topology is obtained by crossing from the differential equation for the $\RT$ topology in Eq.~\eqref{eq:RTSoftDE} with the matrices in Eqs.~\eqref{eq:RTmateven} and \eqref{eq:RTmatodd}. Using the boundary conditions in Eq.~\eqref{eq:StaticBNDxRT}, we find the solutions
\begin{align} \label{eq:firstxRT}
f_{\xRT,3}^{\pot}= (-q^2)^{-2\epsilon} {}& \epsilon^2 \pi^2 \left[ -\frac {1}{6} \epsilon \log(x) +\frac{1}{2} \epsilon^2 \left( \operatorname{Li}_2(1-x^2) + \log^2(x) \right)  \right]
\,,\\
f_{\xRT,4}^{\pot}= (-q^2)^{-2\epsilon} {}& \epsilon^2 \pi^2 \left[ -\frac 1 3 \epsilon\log(x) - \frac{1}{3} \epsilon^2 \left( \operatorname{Li}_2(1-x^2) + \log^2(x) \right) \right]
\,,\\
f_{\xRT,5}^{\pot}= (-q^2)^{-2\epsilon} {}& \epsilon^2 \pi^2 \left[ -\frac 1 6 + \frac 1 {36} \epsilon^2 \left( 7 \pi^2 +48 \log^2(x) \right) \right]
\,,\\
f_{\xRT,6}^{\pot}= (-q^2)^{-2\epsilon} {}& \epsilon^2 \pi^2 \left[ -\frac 1 6 + \frac{7 \epsilon^2 \pi^2}{36} \right]
\,,\\
f_{\xRT,7}^{\pot}= (-q^2)^{-2\epsilon} {}& \epsilon^2 \pi^2 \left[ - \frac{1}{6}\epsilon^2\log^2(x) \right]
\,.\label{eq:leadingxRT}
\end{align}
The differential equation for the $\xIX$ topology is obtained by crossing from the differential equation for the $\IX$ topology in Eq.~\eqref{eq:RTSoftDE} with the matrices in Eqs.~\eqref{eq:IXmateven} and \eqref{eq:IXmatodd}. Using the boundary conditions in Eq.~\eqref{eq:StaticBNDxIX}, we find the solutions
\begin{align} \label{eq:firstxIX}
f_{\xIX,2}^{\pot}= (-q^2)^{-2\epsilon} {}& \epsilon^2 \pi^2 \left[ -\frac {1}{3} \epsilon \log(x) +\epsilon^2 \left( \operatorname{Li}_2(1-x^2) + \log^2(x) \right)  \right]
\,,\\
f_{\xIX,3}^{\pot}= (-q^2)^{-2\epsilon} {}& \epsilon^2 \pi^2 \left[ -\frac {2}{3} \epsilon \log(x) -\frac{2}{3} \epsilon^2 \left( \operatorname{Li}_2(1-x^2) + \log^2(x) \right)  \right]
\,,\\
f_{\xIX,4}^{\pot}= (-q^2)^{-2\epsilon} {}& \epsilon^2 \pi^2 \left[ \frac 1 3 - \frac 1 {18} \epsilon^2 \left( 7 \pi^2 +48 \log^2(x) \right) \right]
\,,\\
f_{\xIX,5}^{\pot}= (-q^2)^{-2\epsilon} {}& \epsilon^2 \pi^2 \left[ -\frac {1}{6} \epsilon \log(x) +\frac{1}{2} \epsilon^2 \left( \operatorname{Li}_2(1-x^2) + \log^2(x) \right)  \right]
\,,\\
f_{\xIX,6}^{\pot}= (-q^2)^{-2\epsilon} {}& \epsilon^2 \pi^2 \left[ -\frac {1}{3} \epsilon \log(x) -\frac{1}{3} \epsilon^2 \left( \operatorname{Li}_2(1-x^2) + \log^2(x) \right)  \right]
\,,\\
f_{\xIX,7}^{\pot}= (-q^2)^{-2\epsilon} {}& \epsilon^2 \pi^2 \left[ -\frac 1 6 + \frac 1 {36} \epsilon^2 \left( 7 \pi^2 +48 \log^2(x) \right) \right]
\,,\\
f_{\xIX,8}^{\pot}= (-q^2)^{-2\epsilon} {}& \epsilon^2 \pi^2 \left[ -\frac 1 6 + \frac{7\pi^2\epsilon^2} {36} \right]
\,,\\
f_{\xIX,10}^{\pot}= (-q^2)^{-2\epsilon} {}& \epsilon^2 \pi^2 \left[ \frac 1 3\epsilon^2\log^2(x) \right]
\,.\label{eq:leadingxIX}
\end{align}

\bibliography{main.bib}
\end{document}